\documentclass[fleqn,usenatbib,useAMS]{mnras}

\usepackage{graphicx}	% Including figure files
\usepackage{amsmath}	% Advanced maths commands
\usepackage{amssymb}	% Extra maths symbols
\usepackage{multicol}   % Multi-column entries in tables
\usepackage{bm}		    % Bold maths symbols, including upright Greek
\usepackage{tabularx}

\def\dr{{\rm d}}

\def\kms{{\rm km/s}}
\def\yr{{\rm yr}}
\def\myr{{\rm Myr}}
\def\gyr{{\rm Gyr}}
\def\pc{{\rm pc}}

\def\MK{{\rm MK}}

\def\outer{0}

\def\Lgas{L_{\rm g}}
\def\Lgasout{L_{\rm g, \outer}}

\def\Sigmagas{\Sigma_{\rm g}}

\def\tdyn{\tau_{\rm g}}
\def\tdynout{\tau_{\rm g,\outer}}

\def\Rgas{R_{\rm g}}
\def\Rgasout{R_{\rm g,\outer}}

\def\Mgas{M_{\rm g}}
\def\Mgasout{M_{\rm g, \outer}}

\def\vrmsgas{\sigma_{\rm g}}
\def\vrmsgasout{\sigma_{\rm g,0}}

\def\tp{t^\prime}

\def\tf{t_{\rm f}}
\def\tfmin{t_{\rm f, min}}
\def\tfmax{t_{\rm f, max}}

\def\mf{m_{\rm f}}
\def\mfmax{m_{\rm f, max}}

\def\mdotin{\dot{m}_{\rm in}}
\def\mdotinmax{\dot{m}_{\rm in,max}}

\def\mdotacc{\dot{m}_{\rm acc}}

\def\mdotaccmax{\dot{m}_{\rm acc,max}}

\def\epsout{\epsilon_{\rm out}}

\def\mwind{{m}_{\rm wind}}
\def\mdotw{\dot{m}_{\rm wind}}
\def\mdwh{\dot{m}_{\rm wind,\odot,100}}
\def\mdwhz{\dot{m}_{\rm wind,100}(Z)}

\def\Mdotlow{\dot{M}_{\rm low}}

\def\vems{{\rm V/EMS}}
\def\vemss{V/EMSs}

\def\MGC{{M}_{\rm GC}}

\def\MGCi{{M}_{\rm GC, 0}}

\def\trh{\tau_{\rm rh}}
\def\rhoh{\rho_{\rm h}}

\def\DYw{\langle \Delta Y\rangle_{\rm wind}}
\def\DYmax{\Delta Y_{\rm max}}
\def\DYobs{\Delta Y_{\rm P2, P1}}
\def\DYnor{\Delta Y_{\rm no\,rej}}

\def\mvms{m_{\rm V/EMS}}
\def\mvmsw{\langle \mvms\rangle_{\rm wind}}

\def\Teff{T_{\rm eff}}
\def\Tc{T_{\rm c}}

\def\Tcw{\langle \Tc\rangle_{\rm wind}}

\def\Flin#1#2{F_{{\rm #1,#2}}}

\def\psif{\psi_{\rm f}}
\def\psil{\psi_{\rm low}}

\def\ms{1\,\msun}
\def\minf{m_{\infty}}
\def\tinf{\tau_{\infty}}

\def\fdil{f_{\rm dil}}
\def\fdilinf{f_{{\rm dil},\infty}}

\def\mfdil{\langle f_{\rm dil}\rangle}
\def\tform{t_{\rm form}}

\def\Yinf{Y_{\infty}}
\def\Ydot{\dot{Y}}
\def\tnuc{\tau_{\rm nuc}}

\def\Mwind{{M}_{\rm wind}}
\def\Mdwind{\dot{M}_{\rm wind}}
\def\Mdwindpoll{\dot{M}_{\rm wind,poll}}
\def\Mdprist{\dot{M}_{\rm prist}}

\def\fpoll{f_{\rm P2}}

\def\sfrv{{\rm SFR}_{\rm V}}
\def\rsn{\mathcal{R}_{\rm SN}}

\def\rhogas{\rho_{\rm g}}
\def\sfe{{\rm SFE}}

\def\zsun{{\rm Z}_\odot}

\def\msun{{\rm M}_\odot}
\def\msunyr{{\rm M}_\odot\,{\rm yr}^{-1}}
\def\msunmyr{{\rm M}_\odot\,{\rm Myr}^{-1}}
\def\feh{{\rm [Fe/H]}}
\def\ElFe{{\rm [El/Fe]}}
\def\NaFe{{\rm [Na/Fe]}}
\def\AlFe{{\rm [Al/Fe]}}
\def\AlFemax{{\rm [Al/Fe]}_{\rm max}}
\def\NaFemax{{\rm [Na/Fe]}_{\rm max}}
\def\NeFe{{\rm [Ne/Fe]}}
\def\OFe{{\rm [O/Fe]}}
\def\MgFe{{\rm [Mg/Fe]}}

\def\OFemin{{\rm [O/Fe]_{\rm min}}}
\def\OFemax{{\rm [O/Fe]_{\rm max}}}
\def\LaEu{{\rm [La/Eu]}}
\def\ALi{{\rm A(Li)}}

\def\arctanh{{\rm arctanh}}

\usepackage[T1]{fontenc}
\usepackage{ae,aecompl}

\usepackage{txfonts}
\usepackage{soul}
\usepackage[normalem]{ulem}
\usepackage{soul}
\usepackage[dvipsnames]{xcolor}

\definecolor{orange}{rgb}{1, 0.35, 0.}

\title[GC formation with extremely massive stars]
{Globular cluster formation from inertial inflows: accreting extremely massive stars as the origin of  abundance anomalies}
\author[]{Mark Gieles$^{1,2}$\thanks{Contact e-mail:\href{mailto:mgieles@icc.ub.edu}{mgieles@icc.ub.edu}}\href{https://orcid.org/0000-0002-9716-1868}{\includegraphics[scale=0.5]{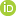}},
Paolo Padoan$^{2,3}$, 
Corinne  Charbonnel$^{4,5}$,
Jorick  S. Vink$^{6}$,\newauthor
Laura  Ramírez-Galeano$^4$\\
$^1$ ICREA, Pg. Llu\'{i}s Companys 23, E08010 Barcelona, Spain,\\
$^2$ Institut de Ci\`{e}ncies del Cosmos (ICCUB), Universitat de Barcelona (UB), c. Mart\'{i} i Franqu\`{e}s, 1, 08028 Barcelona, Spain\\
$^3$ Department of Physics and Astronomy, Dartmouth College, 6127 Wilder Laboratory, Hanover, 03755, NH, USA\\
$^4$ Department of Astronomy, University of Geneva, Chemin des Maillettes 51, CH-1290 Versoix, Switzerland\\
$^5$ IRAP, Universit\'{e} de Toulouse, CNRS, Observatoire Midi Pyr\'{e}n\'{e}s, 57 Avenue d’Azereix, 65008 Tarbes, France\\
$^6$ Armagh Observatory, College Hill, Armagh, BT61 9DG, Northern Ireland, UK
}

\date{Accepted 2025 July 09. Received 2025 July 07 in original form 2025 January 21}

\pubyear{2025}

\begin{document}
\label{firstpage}
\pagerange{\pageref{firstpage}--\pageref{lastpage}}
\maketitle

\begin{abstract}  % MAX 250 words
We use the inertial-inflow model of massive star formation to describe the formation of globular clusters (GCs) in turbulent molecular clouds. A key aspect of this model is that the maximum stellar mass scales linearly with cloud mass, such that extremely massive stars (EMSs, $10^{3-4}\,\msun$) form in massive GCs ($\gtrsim10^5\,\msun$). 
The total wind mass loss is dominated by accreting EMSs (aEMSs), whose wind mass-loss rates have become comparable to their accretion rates ($\gtrsim10^{-2}\,\msun\,\yr^{-1}$). These winds pollute the intra-cluster medium with hot-hydrogen burning yields during  GC formation. We propose a parameterised model for the evolution of the stellar mass function during GC formation  ($\sim 1-2\,\myr$), accounting for  gas inflow, wind mass loss and mixing of aEMS yields with pristine gas that has initial proto-GC abundances. Low-mass stars  ($\lesssim1\,\msun$) form continuously from this mixed gas and their abundances resemble observed abundance trends with GC mass and metallicity, specifically: (i) the helium spread in a typical GC is small ($\Delta Y \simeq 0.01$) and increases with GC mass; (ii) the fraction of polluted stars increases with GC mass and metallicity; (iii) the extent of the Mg-Al anticorrelations is more pronounced in  metal-poor and massive GCs. We conclude that  GCs formed  with a population of EMSs from gas with surface densities $\gtrsim10^3\,\msun\,\pc^{-2}$ and that nitrogen-rich galaxies discovered by the James Webb Space Telescope ({\it JWST}) are dominated by EMS-rich GCs that formed in the earliest phases of galaxy formation. These EMSs  may have left behind intermediate-mass black holes with masses above the pair-instability gap ($\gtrsim120\,\msun$) that could be found with ongoing gravitational wave experiments. 
\end{abstract}

%%%%%%%%%%%%%%%%%%%%%%%%%%%%%%%%%%%%%%%%%%
\section{Introduction}
The problem of globular cluster (GC) formation is one with a rich history. Their old ages have led to models invoking special conditions in the early Universe \citep*[for example,][]{1968ApJ...154..891P, 1985ApJ...298...18F, 2015ApJ...808L..35T}, or specific environments, such as dwarf galaxies \citep[for example,][]{2002ApJ...566L...1B}. In recent years, a consensus has emerged that the physics of GCs formation  is similar at all cosmic times \citep{1994ApJ...429..177H,1997ApJ...480..235E} and massive star clusters forming in high-redshift galaxies can evolve into old clusters matching the properties of observed GC systems \citep{2018MNRAS.480.2343C,2018MNRAS.475.4309P,2019MNRAS.486.3180K, 2023MNRAS.521..124R, 2023MNRAS.522.5340G}. Hydrodynamical  simulations are  able to resolve certain aspects of GC formation \citep[for example,][]{2016ApJ...823...52K,2023MNRAS.519.1366G}, but a complete (magneto-)hydrodynamic model of GC formation in which stars of all masses form self-consistently with realistic boundary conditions is still out of reach. We  address the problem of GC formation by scaling up the inertial-inflow model for massive star formation \citep{Padoan+20massive, Pelkonen+21} in which massive stars ($\sim10-100\,\msun$) form accompanied by  a fully sampled stellar initial mass function (IMF) in star clusters with masses of $\sim 10^4\,\msun$. By extrapolating this model with scaling relations to higher masses, we can address certain aspects of GC formation. In this  study, we focus on the problem of the anomalous abundances of GCs, also referred to as the `multiple populations' (MPs) problem. The ubiquity of MPs in GCs made \citet{2015AJ....149...91P} conclude that  ``GC formation and the origin of their MPs have now become one and
the same problem.''

GCs are characterised by anomalous light-element (HeCNONaMgAl) abundances. They display spreads of these abundances, but there is also evidence for discrete  MPs \citep[see for example,][]{2002ASPC..265...87A, 2009A&A...505..117C, 2017MNRAS.464.3636M,2017A&A...601A.112P,2019A&ARv..27....8G}. The observed He spread and the specific anticorrelations between C-N, O-Na and Mg-Al resemble the ashes of hydrogen burning at high temperatures ($\sim50-80\,\MK$), well above  the central temperatures of the GC stars ($\sim15\,\MK$). Combined with the ubiquity of these abundance patterns in GCs and the rarity of these stars in the field\footnote{N-rich  and Al-rich stars have been found in the Milky Way halo \citep[for example,][]{2016ApJ...825..146M,2021MNRAS.500.5462H} and these are believed to have originated from dissolved GCs \citep{2023MNRAS.522.5340G,2023MNRAS.525.4456B}.}, the working hypothesis is that GC stars formed out of proto-GC gas polluted by material produced during or soon after GC formation. Several  polluters have been proposed, such as asymptotic giant branch stars \citep[AGB, $6-8\,\msun$;][]{2001ApJ...550L..65V}, fast-rotating massive stars \citep[FRMSs, $\gtrsim20\,\msun$;][]{2006A&A...448L..37M,2006A&A...458..135P,2007A&A...475..859D}, interacting massive binaries \citep[$\gtrsim20\,\msun$;][]{2009A&A...507L...1D}, very massive stars \citep[VMSs, $\gtrsim100\,\msun$;][]{2018A&A...615A.119V} and supermassive stars \citep*[SMSs, $\gtrsim10^4\,\msun$;][]{2014MNRAS.437L..21D}, with associated GC formation (toy) models (\citealt{2007A&A...464.1029D, 2008MNRAS.391..825D, 2009A&A...507L...1D,2013MNRAS.436.2398B, 2017ApJ...836...80E, 2018MNRAS.478.2461G}; \citealt*{2022MNRAS.513.2111R}). Despite all these ongoing efforts, no consensus picture exists that satisfies all observational constraints. The yields of stars $\gtrsim10^3\,\msun$ resemble the observed abundance patterns  \citep{2017A&A...608A..28P}. Nevertheless, a GC formation model in which MPs follow naturally from VMSs and/or SMSs\footnote{\citet{2018MNRAS.478.2461G} proposed a GC formation model in which SMSs form from stellar collisions, which we discuss in Section~\ref{ssec:collisions}.} still lacks and is needed to explain the ubiquity of MPs in GCs with masses $\gtrsim10^5\,\msun$ and ages $\gtrsim2\,\gyr$ \citep{2018MNRAS.473.2688M}. In this study we focus on hydrogen-burning stars with masses in between the hitherto considered VMSs \citep[few$\times10^2\,\msun$, for example,][]{2023A&A...679L...9V} and SMSs \citep[few$\times10^4\,\msun$, for example,][]{2015MNRAS.448.3314D}. We refer to stars in the mass range $10^3-10^4\,\msun$ as extremely massive stars (EMSs)\footnote{Following the naming convention of stellar metallicities \citep{2005ARA&A..43..531B}. Some studies \citep{2007ARA&A..45..481Z} refer to them as ultramassive stars and others as SMSs \citep{2017A&A...608A..28P}.}. 
The motivation to have a distinguishing name for this mass decade is twofold: firstly, the detailed abundance patterns vary for each order of magnitude variation of stellar mass \citep{2017A&A...608A..28P}, with the abundances of EMSs being superior to that of VMSs and SMSs in reproducing observing GC abundances (Section~\ref{ssec:anticorrs}). Secondly, above $\sim10^3\,\msun$ stars are expected to be fully convective while accreting (Section~\ref{ssec:mixing}), making them physically different from VMSs.

We will discuss a scenario in which VMSs and EMSs (but not SMSs) form as part of the IMF of GCs and in which EMSs are the dominant source of pollution. Because we consider a fully sampled IMF up to EMSs, we will refer to  stars in the mass range $10^{2-4}\,\msun$ as \vems.

In the inertial-inflow model  for massive star formation \citep{Padoan+20massive, Pelkonen+21}, massive stars  form in turbulent molecular clouds and grow in mass from turbulence driven inertial inflows on the scale of the giant molecular cloud. The mass of the most massive star is $\sim0.1$ per cent of the mass of the parent cloud, such that stars $\gtrsim10^3\,\msun$  are expected in clouds of $\gtrsim10^6\,\msun$, resulting in stars clusters of $\gtrsim10^5\,\msun$ for a star formation efficiency of 10 per cent. This theory prediction for the existence of EMSs in proto-GCs is a key motivation to consider their role as the origin of MPs. 

Another motivation to consider \vemss\ as polluters is that VMSs have been observed in star clusters. Individual VMSs have been  found in the young massive clusters (YMCs) R136 in the Large Magellanic Cloud (LMC) and NGC\,3603 in the Milky Way disc \citep[][]{2010MNRAS.408..731C}. Spectral features of VMSs have been found in the central star clusters of the blue compact dwarf galaxy NGC\,5253 \citep{2016ApJ...823...38S}; in the super star cluster NGC\,3125-A1 \citep{2023MNRAS.523.3949W}; in the starburst region Mrk 71 in the local dwarf galaxy NGC\,2366  \citep{2023ApJ...958..194S} and in a YMC in the Sunburst Arc galaxy at redshift $z\sim2.4$ \citep[][]{2023A&A...673A..50M}, thanks to the James Webb Space Telescope ({\it JWST}). The presence of VMSs appears to be  ubiquitous in UV-bright galaxies at Cosmic Noon
 \citep{2024A&A...686A.185U}. Although individual EMSs have not been observed yet, the Extremely Large Telescope \citep[{\it ELT},][]{2007Msngr.127...11G} may be able to resolve them. For example, the young clusters in NGC\,5253 ($\sim10^6\,\msun$) have metallicities similar to R\,136. The VMSs in R136 are separated by $\gtrsim0.2\arcsec$ \citep{2010MNRAS.408..731C}. The gain in spatial resolution with {\it ELT} approximately compensates the larger distance of NGC\,5253, such that EMSs, if present, could be resolved. 

\begin{figure}
\includegraphics[width=\columnwidth]{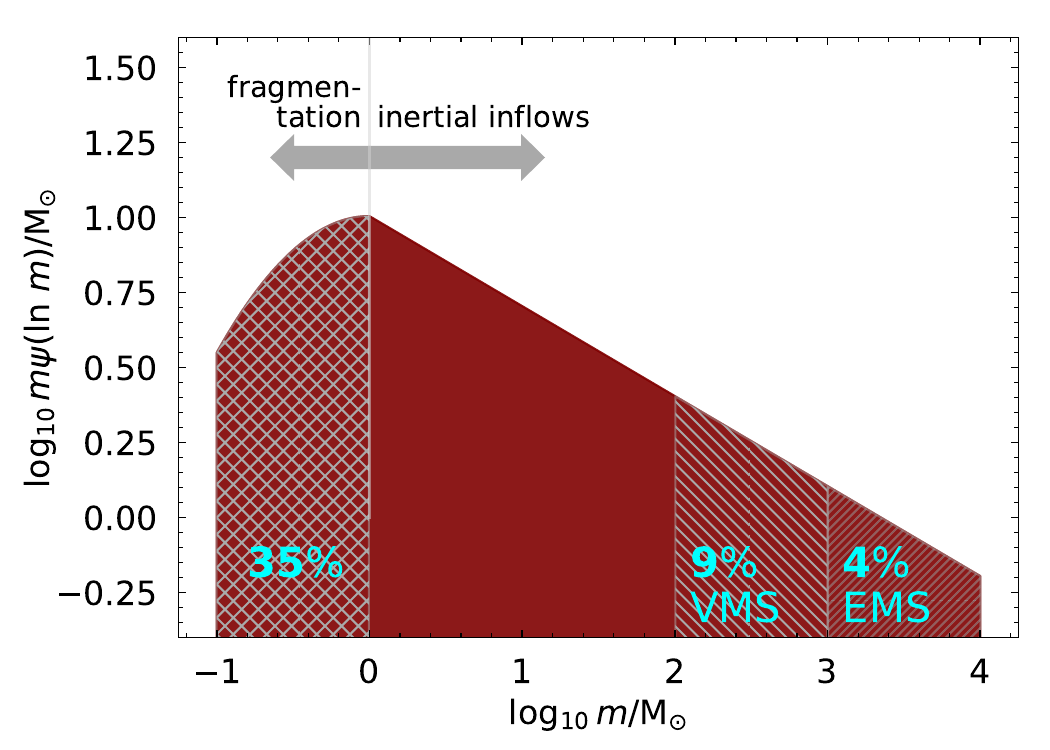}
\vspace{-5mm}
\caption{Schematic picture of the (normalised) IMF, where $m\psi(\ln m)=m^2\psi(m)$ and $\psi(m)$ is defined as the number of stars in the interval $[m, m+\dr m]$. Below $1\,\msun$ stars form  with a Chabrier-like IMF \citep{2003PASP..115..763C} via fragmentation. More massive stars grow in mass via accretion from  turbulence driven inertial inflows with a power-law  Salpeter/Kroupa slope. Allowing the IMF to extend to  $\sim10^4\,\msun$ enables a large fraction of the total cluster mass to be processed by \vemss\ that release hot-hydrogen burning yields via their strong winds.}
\label{fig:imf_schematic}
\end{figure}

An upper limit to the IMF that depends on GC mass is  a promising avenue to overcome the so-called `mass-budget problem', that is, the  (large) amount of polluted material required to explain observed abundances \citep{2006A&A...458..135P,2011MNRAS.413.2297S,2018ARA&A..56...83B}.
Assuming that the IMF slope above $100\,\msun$ is Salpeter-like up to $\sim10^4\,\msun$, then more than 10 per cent of the total cluster mass is in the form of \vemss\ (see Fig.~\ref{fig:imf_schematic}). \vemss\ have  high mass-loss rates \citep{2011A&A...531A.132V}, releasing a significant fraction of their initial mass during the hydrogen burning phase, ending their evolution $\lesssim 100\,\msun$ \citep[at least  for  solar metallicities,][]{2022MNRAS.514.3736S, 2023MNRAS.526..534H}. 

The theoretical abundances in the core of a main sequence EMS  resemble  observed GC abundances, after dilution with an order of magnitude larger amount of pristine material  \citep{2007A&A...470..179P, 2017A&A...608A..28P}.  If $\sim10$ per cent of the cluster mass is diluted by a factor of ten to create P2 stars, then the total mass in P2 stars is comparable to the mass in P1 stars.  This simple IMF argument is a further encouragement to consider GC pollution by EMSs.

In this work, we show first in Section~\ref{sec:inertial} that \vemss\ are a natural outcome of star formation on the mass scale of GCs in the inertial-inflow model. In Section~\ref{sec:model} we present a simple model for the growth of the stellar initial mass function (IMF) including inflows and accretion, wind mass loss and continuous low-mass star formation. In Section~\ref{sec:scaling} we show how results scale with GC mass and metallicity and compare  model predictions to observations. A discussion and our conclusions are presented in Sections~\ref{sec:discussion} and \ref{sec:conclusions}, respectively.

%%%%%%%%%%%%%%%%%%%%%%%%%%
\section{V/EMS formation in the inertial inflow model}
\label{sec:inertial}
\begin{table}
\caption{Description of symbols most used in this work. The two  parameters that control GC formation are the surface density of the gas, $\Sigmagas$, and the turbulence outer scale, $\Lgasout$. The five properties of a gas cloud on the outer scale are referred to with subscript $`\outer$' (that is, $\Lgasout, \Rgasout, \Mgasout, \tdynout, \vrmsgasout$). The maximum values of the three stellar parameters, corresponding to the properties of the most massive star in a cluster, are referred to with subscript `max', that is, $\mfmax, \tfmax, \mdotinmax, \mdotaccmax$.
\label{tab:syms}}
\begin{tabular}{ll p{3.2cm}}
\hline
& Symbol      & Description\\\hline
Model           & $\Sigmagas$           & surface density of the gas\\
                     & $\Lgasout$              & outer scale of turbulence\\\hline 
Gas clouds     & $\Lgas\le \Lgasout$   &   spatial scale/diameter   \\	   
                     & $\Rgas=\Lgas/2$     &   radius  \\
                     & $\Mgas = \pi  \Rgas^2\Sigmagas $ &  mass   \\
                     & $\tdyn$                     &  dynamical time  \\
                     & $\vrmsgas$               &  velocity dispersion \\\hline
Stars             & $\mf$                        & final  mass\\ 
                    & $\tf$                           & formation time\\
                   & $\mdotin$                &  inflow rate on star\\
                   & $\mdotacc=0.5\mdotin$   &  accretion rate on star\\
                   & $\mdwh(Z)$                &  Metallicity dependent (positive) wind mass-loss rate of a $100\,\msun$ star\\\hline
                   & $\mdwhz$                &  (positive) wind mass-loss rate of a $100\,\msun$ star of solar metallicity\\\hline
GCs              & $\sfe $           & star formation efficiency \\ 
                 & $\MGCi=\sfe\times\Mgas$                  & initial GC mass \\ 
                  & $\MGC$    & present-day GC mass \\\hline
 \end{tabular}
\end{table}

%___________
\subsection{The inertial-inflow model in short}
In the inertial-inflow model,  stars are fed by mass inflows that are not driven primarily by gravity, as they are an intrinsic feature of supersonic turbulence. The scenario is inspired by the IMF model of \citet{Padoan+Nordlund02imf}, where prestellar cores are formed by shocks in converging flows. The characteristic time of the compression is the turnover time of the turbulence on a given scale, which is generally larger than the free-fall time in the post-shock gas, so a pre-stellar core may collapse into a protostar well before the full stellar mass reservoir has reached the core \citep{Padoan+Nordlund11imf}. After the initial collapse, the star can continue to grow, as the converging flows that were feeding the prestellar core continue to feed the star, through the mediation of a disk. Thus, the stellar mass reservoir can extend over a turbulent and \emph{unbound} region much larger than the prestellar core. Because converging flows occur spontaneously in supersonic turbulence and can assemble the stellar mass without relying on a global collapse or on the gravity of the growing star (\citealt*{Padoan+14acc}; \citealt{Padoan+20massive}), the inertial-inflow model is fundamentally different from competitive-accretion \citep[for example,][]{Zinnecker82,Bonnell+2001a,Bonnell+2001b,Bonnell+Bate2006}. It is also different from the core-collapse model \citep[for example,][]{McKee+Tan02,McKee+Tan03}, because only low-mass stars (essentially below the IMF's turnover mass) are fully formed after the collapse of their prestellar core \citep{Pelkonen+21}.  
\begin{figure}
\includegraphics[width=\columnwidth]{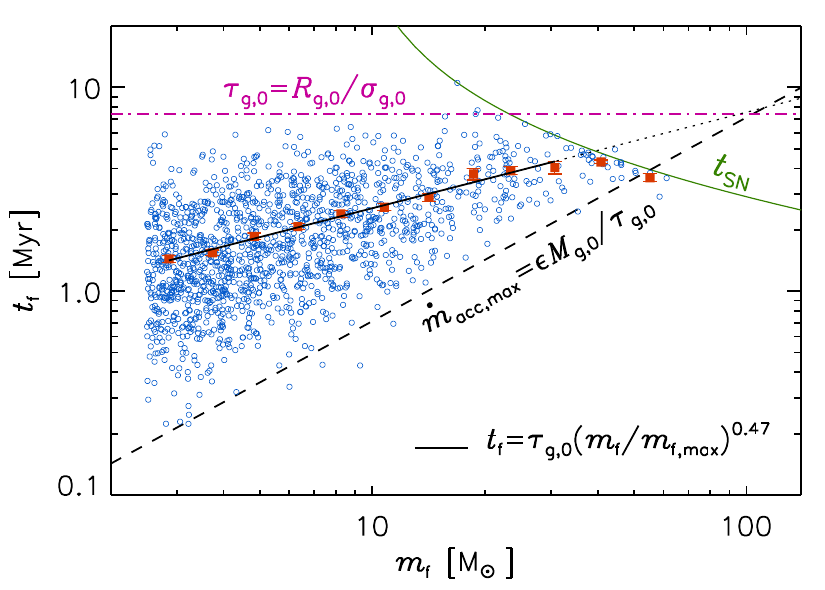}
\vspace{-5mm}
\caption[]{The time to reach 95 per cent of the final mass, $\tf$, versus the final mass, $\mf$, for 1,503 stars with mass $\ge 2.5\,\msun$ that have stopped accreting at the time $t=30\,\myr$ in the simulation from \citet{Padoan+20massive}. The curved solid line shows an analytic fit to the relation between stellar lifetime, $t_{\rm SN}$, and mass from \citet{Schaller+92}. The dashed line is the lower envelope of the scatter plot corresponding to the maximum accretion rate, while the dashed-dotted line corresponds to the dynamical time of the outer scale, $\tdynout$. The red square symbols with error bars show the mean values of $\tf$ in logarithmic intervals of $\mf$. The straight solid line is a least-squares fit to those mean values up to $\sim 30\,\msun$ ($\tf$ is limited by $t_{\rm SN}$ at larger masses), giving $t_{\rm f}=0.86\, (m_{\rm f}/{M}_{\odot})^{0.47\pm 0.02}$\,Myr. The extrapolation of the power law to larger masses is shown by the dotted line, which intersects the lower and upper envelopes of the scatter plot at approximately the same values of $m_{\rm f}=m_{\rm f,max}$ and $t_{\rm f}=\tdynout$. 
}
\label{fig:t95_v1}
\end{figure}
%
%___________
\subsection{Timescale for  the formation of massive stars}
\label{ssec:timescale}
The relation between the timescale to form a star, $\tf$, and the final mass of the star, $\mf$, is a scatter plot with a rather uniform distribution within well-defined upper and lower envelopes, as first shown in \citet{Padoan+14acc}, and later discussed in detail in \citet{Padoan+20massive}. The $\tf-\mf$ plot in Fig.~\ref{fig:t95_v1} uses the same data points as in \citet{Padoan+20massive}, a magneto-hydrodynamic (MHD) turbulence simulation representing a region of size 250\,pc, total mass $1.9\times 10^6\,\msun$, and a mean magnetic field strength of 4.6\,$\mu$G\footnote{The resulting rms value is 7.2 $\mu$G and the average of the absolute value is 6.0\,$\mu$G, consistent with the value of $6.0 \pm 1.8$\,$\mu$G from the `Millennium Arecibo 21-cm Absorption-Line Survey' by \citet{Heiles+Troland05}.}, where the only driving force is from supernova (SN) feedback. After a first period of 55.5\,Myr without self-gravity \citep{Padoan+16SN_I} and with artificially generated SNe randomly distributed in space and time, the simulation is then continued for over 30\,Myr with self-gravity and higher resolution (0.0076\,pc), resolving the formation of individual massive stars, whose positions and lifetimes determine self-consistently the SN explosions. The simulation has been shown to produce realistic molecular-cloud properties \citep{Padoan+16SN_III} and realistic star-formation rate values both in individual molecular clouds and globally \citep{Padoan+17SN_IV}.

The upper envelope of the scatter plot in Fig.~\ref{fig:t95_v1} corresponds to a constant maximum formation time, while the lower envelope rises linearly with increasing final stellar mass. As a result, the two envelopes intersect at a large mass, which defines the maximum stellar mass, $\mfmax$. This 
result shows that understanding the maximum stellar mass only requires a physical interpretation of the upper and lower envelopes, which is discussed at length in \citet{Padoan+20massive}, and only briefly summarized here. The maximum formation time is interpreted as the largest turnover time in the turbulent flow, because beyond that time the flow cannot remain coherent, hence the mass inflow towards a star cannot persist. Thus, we estimate this maximum time for a given cloud as 
\begin{equation}
\tfmax= \tdyn=\frac{\Rgas}{\vrmsgas},
\label{eq:tfmax}
\end{equation}
where $\Rgas=0.5\Lgas$ is the radius of the star forming gas region, $\Lgas$ is the spatial scale and $\vrmsgas$ the velocity dispersion on that scale (in Section\,\ref{ssec:scaling_to_GC} we will introduce the subscript `0' in reference to the turbulence outer scale, hence to the most massive cloud).
The lower envelope of the $\tf-\mf$ plot is fit by a linear relation, $\tfmin\propto m_{\rm f}$, shown by the dashed line in Fig.~\ref{fig:t95_v1}, corresponding to the maximum accretion rate, $\mdotaccmax$. We interpret this maximum accretion rate as a fraction, $\epsilon$, of the inflow rate corresponding to the whole turbulent region being compressed into a single star, $\Mgas/\tdyn$, where 
$\Mgas$ is the total gas mass within a radius $\Rgas$: 
\begin{equation}
\mdotaccmax = \epsilon\, \frac{\Mgas}{\tdyn}.
\label{Mdot_max}
\end{equation}
We assume that $\epsilon$ is a universal parameter, because it depends only on geometrical properties of supersonic turbulence, which determine the maximum efficiency with which the largest turbulent motions can be channeled into a single star.\footnote{For simplicity of notation, this maximum accretion rate already accounts for mass loss due to protostellar jets and radiative feedback, assumed to be 50 per cent of the inflow rate in the simulations (it does not account for the mass loss by stellar wind discussed in Section~\ref{ssec:wind}). Thus, with respect to the notation in \citet{Padoan+20massive}, $\epsilon=\epsout \epsilon_{\rm in}$, with $\epsout=0.5$.\label{fn:acceff}} Though related to the star-formation efficiency (after one turnover time of the outer scale), $\epsilon$ should be much smaller than that, because the stars that form with the largest accretion rate are only a small fraction of all stars formed within a region with radius $\Rgas$. Based on our simulations, we have established an approximate value of\footnote{Using the maximum accretion rate from the lower envelope of the relation between $\tf$ and $\mf$ (Fig.\,\ref{fig:t95_v1}), \citet{Padoan+20massive} derived a value $\epsilon=0.9\times 10^{-3}$ (including a factor of 0.5 to account for mass loss by protostellar jets and radiative feedback). Here we adopt a slightly larger value derived from the lower envelope of the same relation from \citet[][their figure\,11]{2018ApJ...854...35H}, given the much higher numerical resolution of those simulations.} $\epsilon\approx 2.5\times 10^{-3}$. This value is also supported by observations of star-forming clouds in the central molecular zone of the Galaxy. In Table\,4 of \cite{Barnes+17}, the average ratio between the estimated maximum stellar mass and the corresponding cloud mass is $\epsilon=1.8\pm 1.0\times 10^{-3}$, based on 10 clouds. The theory value we adopt  here is consistent with this empirically inferred value.

The lower envelope of the $\tf-\mf$ plot is then given by 
\begin{equation}
t_{\rm f,min}=\frac{\mf}{\epsilon\,\mdotaccmax}= \tdyn\frac{\mf}{\epsilon \Mgas}.
\label{eq:t95_min_2}
\end{equation}
Given the empirical value of $\epsilon$, the lower envelope is fully determined by the properties of the outer scale of the turbulence in the star-forming region, namely $\Mgas$ and $\tdyn$, according to equation~(\ref{eq:t95_min_2}). 
\begin{figure}
\includegraphics[width=\columnwidth]{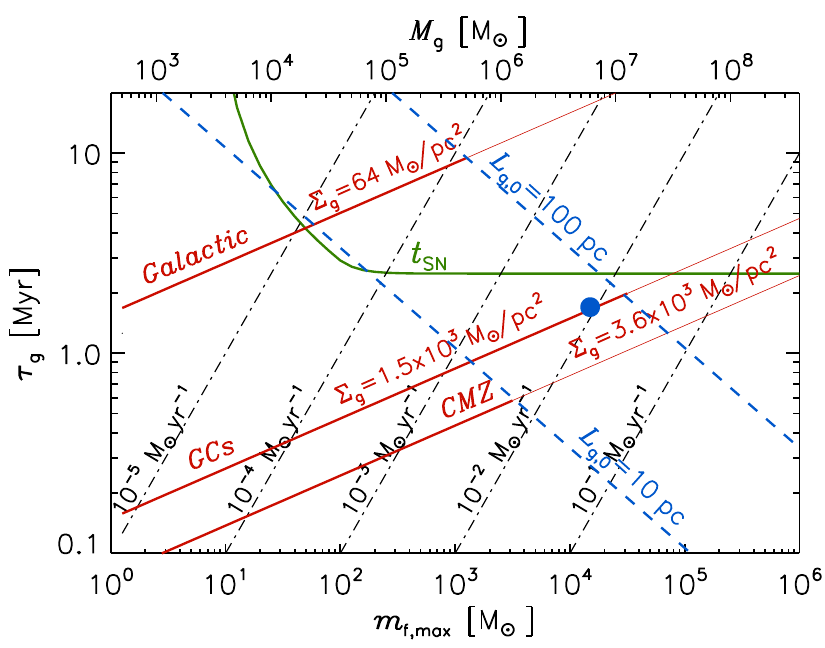}
\vspace{-5mm}
\caption[]{Cloud dynamical time, $\tdyn$, versus maximum stellar mass, $\mfmax$, for three different values of the gas column density, $\Sigmagas$, characterizing molecular clouds in the Galaxy, GC environments, and the Central Molecular Zone (red lines from top to bottom respectively). The filled blue circle corresponds to our prototypical GC progenitor with $\Mgas=6\times10^6\,\msun$. Lines of constant value of the maximum accretion rate, $\mdotaccmax=\mfmax/\tdyn$, are also shown (dashed-dotted lines). The blue dashed lines show the maximum dynamical time, $\tdynout$, corresponding to the turbulence outer scales $\Lgasout=100$\,pc (for Galactic and GC clouds) and $\Lgasout=10$\,pc (for CMZ clouds). Star forming clouds along the red lines cannot exist above the corresponding blue dashed line (hence the thinner red lines). Finally, the green line is an approximate model for the star lifetime (prior to SN explosion).}
\label{fig:t0_mmax}
\end{figure}
%
%___________
\subsection{The maximum stellar mass}
\label{ssec:mfmax}
The intersection of upper and lower envelopes determines the maximum stellar mass:
\begin{equation}
\mfmax=\epsilon  \Mgas. 
\label{eq:mfmax_Mgas}
\end{equation}
Interestingly, $\mfmax$ depends only on the total gas mass, and is independent of the turbulent rms velocity. In fact, an increase of the turbulent velocity lowers both the upper and lower envelopes of the $\tf-\mf$ plot, decreasing both by the same factor. Thus, their intersection would occur at the same value of $\mf$. A correlation between the maximum stellar mass and the mass of its host cluster/cloud has been found in other numerical simulations \citep{2023OJAp....6E..48G} and has been inferred from observations of nearby star clusters  \citep{2006MNRAS.365.1333W}.

The maximum accretion rate can also be expressed in terms of the virial parameter (for a uniform sphere), $\alpha_{\rm vir} = 5 \vrmsgas^2 \Rgas /(3  G \Mgas)$. Equation~(\ref{Mdot_max}) then becomes 
\begin{equation}
\mdotaccmax = \frac{5 \epsilon}{3\alpha_{\rm vir}}\frac{\vrmsgas^3}{G} 
\label{eq:Mdot_max_vir}
\end{equation}
Thus, in star-forming clouds with $\alpha_{\rm vir}\sim 1$, the maximum accretion rate is only a function of the rms velocity, $\mdotaccmax \propto \vrmsgas^3$.

Because of the rather uniform distribution of stars in the $\tf-\mf$ plot, the mean relation is approximately $\tf\propto \mf^{1/2}$ (black solid line in Fig.~\ref{fig:t95_v1}). This mean relation intersects the lower and upper envelopes of the scatter plot at approximately the same values of $\mf=\mfmax$ and $\tf=\tfmax=\tdyn$, so it can be expressed as
\begin{align}
\tf&\simeq\tfmax \left(\frac{\mf}{\mfmax}\right)^{1/2}.
\label{eq:tf_mf_mean}
\end{align}
We can use this mean formation time (an average over stars of the same mass) to estimate the mean accretion rate of an individual star (an average over time for that star), $\mdotacc$, as a function of the final stellar mass,
\begin{align}
\mdotacc \simeq \mdotaccmax\left(\frac{\mf}{\mfmax}\right)^{1/2}.
\label{eq:mdotacc}
\end{align}
%
%___________
\subsection{Scaling to GC formation environments} \label{ssec:scaling_to_GC}
To scale the model to different cloud masses, $\Mgas$, and different star-forming environments, we make the assumption that the condition $\alpha_{\rm vir}=1$ is approximately satisfied by star-forming regions at all scales. Combined with the virial relation we can then write $\vrmsgasout$ in terms of the two interstellar-medium (ISM) quantities that define the model: $\vrmsgasout \propto (\Sigmagas \Rgasout)^{1/2}$, where $\Sigmagas = M_0/(\pi R_0^2)=\Mgas/(
\pi \Rgas^2)$ is the (constant) surface density of clouds \citep{Larson81,Solomon+87}. 
Scaling to environments where we expect GCs to form \citep[$\Sigma\simeq10^3\,\msun\,\pc^{-2}$, ][]{2010Natur.463..781T,2023MNRAS.520.2180C,Messa+24} we can write for the maximum cloud mass
\begin{equation}
\Mgasout \simeq 7.9\times10^6\,\msun\,\frac{\Sigmagas}{10^3\,\msun\,\pc^{-2}} \left(\frac{\Rgasout}{50\,\pc}\right)^{2},
\end{equation}
which has a dynamical time
\begin{equation}
\tdynout \simeq 2.4\,\myr\,\left(\frac{\Sigmagas}{10^3\,\msun\,\pc^{-2}}\right)^{-1/2} \left(\frac{\Rgasout}{50\,\pc}\right)^{1/2}.
\end{equation}
Here we scaled to an outer radius of $\Rgasout=50\,\pc$, appropriate for turbulence driven by SNe ($\Rgasout\simeq35\,\pc$ in \citealt{Padoan+20massive}, 28 to 69\,pc in \citealt{Joung+MacLow09}, 100\,pc in \citealt{deAvillez+Breitschwerdt07}), as these are the primary energy sources of the interstellar medium in most star-forming galaxies \citep[for example,][]{Hopkins13,2023ApJ...948..107F}\footnote{Thanks to self-gravity and spiral-arm dynamics, the turbulence may extend even above the disk scale height to kpc scales, though with a change of slope of the velocity scaling, as found for example in the Large Magellanic Cloud \citep{Elmegreen+2001LMC,Padoan+2001lmc}. Turbulence at kpc scales is also generated in the virialization process of dark matter haloes of $\sim 10^8$\,$\msun$, due to hot and cold gas accretion \citep{Wise+Abel07,Greif+08}, or in the merging of even smaller haloes \citep{Prieto+12}.}. This outer scale is also consistent with the scale height of the youngest stellar disks in cosmological simulations of early galaxies \citep{Meng+Gnedin21}, where the gas scale height is controlled by the turbulence and thus should be of the order of the turbulence outer scale \citep[for example,][]{Ostriker+Shetty11,Krumholz+18}. It is also comparable to the sizes of dense star-forming clumps in cosmological simulations of early galaxies \citep{Tamburello+15,Andalman+25} and in observations of high-redshift lensed galaxies \citep{2023MNRAS.520.2180C,Messa+24}. For non-lensed galaxies, where the clump size is limited by spatial resolution, a size of order 100\,pc can still be derived by extrapolating the observed clumps' mass-size relation \citep[for example][]{Kalita+25} to masses of order $10^6\,\msun$ assumed here. In addition, for the physical parameters of our typical GC listed in Table~\ref{tab:typical}, an outer scale of 100\,pc corresponds to a velocity dispersion of $\sim 30$\,km\,s$^{-1}$, which is consistent with vertical velocity dispersion values  of the cold gas in simulations of early galaxies \citep{Jimenez+23}, and of molecular gas in real high-redshift galaxies \citep{Wisnioski+25}.

For a star formation efficiency of $\sim10$ per cent this results in a maximum initial GC mass of $\sim10^6\,\msun$ ($\sim5\times10^5\,\msun$ at present), which is approximately  what is found for the cut-off mass of the GC mass functions in different galaxies \citep{2007ApJS..171..101J}. We note that it is likely that  $\Rgasout\simeq50\,\pc$ in most environments, because SNe are the predominant drivers of turbulence. This implies a (near-)linear scaling between  the maximum initial GC mass and $\Sigmagas$, as is approximately found for young massive clusters in different galaxies \citep[][]{2017ApJ...839...78J,2020SSRv..216...69A,2020MNRAS.499.3267A,2023MNRAS.520.2180C}.
For a given $\Sigmagas$, lower-mass clusters form from lower-mass clouds ($\Mgas\le \Mgasout$) on smaller scales ($\Rgas=(\Mgas/\Mgasout)^{1/2}\Rgasout$) on  
 timescales
\begin{align}
\tdyn &= \tdynout\left(\frac{\Rgas}{\Rgasout}\right)^{1/2} = \tdynout\left(\frac{\Mgas}{\Mgasout}\right)^{1/4},\\
 &\simeq 2.2\,\myr\,\left(\frac{\Sigmagas}{10^3\,\msun\,\pc^{-2}}\right)^{-3/4} \left(\frac{\Mgas}{5\times10^6\,\msun}\right)^{1/4}.
\label{eq:tdyn}
\end{align}
We will assume that the cluster formation process stops after a time $\tdyn$, because inertial inflows in the turbulent flow cannot maintain coherence (that is, converging to the approximate same location) for longer than a dynamical time at that scale, as for the formation of the most massive stars (see Section\,\ref{ssec:timescale}). In fact, the inertial inflows forming the cluster are the same that form the most massive stars, except that only a fraction $\epsilon$ of their mass converges to those stars.

The maximum accretion rate on a star in a cloud with mass $\Mgas$ is then
\begin{equation}
\mdotaccmax\simeq 5.8\times10^{3}\,\msun\,\myr^{-1}\,\left(\frac{\Sigmagas}{10^3\,\msun\,\pc^{-2}}
\frac{\Mgas}{5\times10^6\,\msun}\right)^{3/4}.
\label{eq:mdamax}
\end{equation}
Fig.~\ref{fig:t0_mmax} shows this relation (red solid lines) for three different environments: Galactic molecular clouds (MCs), GC progenitors, and Galactic central molecular zone (CMZ) clouds. Each environment is fully determined by the constant surface density value, $\Sigmagas$. A specific star-forming region in a given environment is determined by its total gas mass, $\Mgas$, hence by the maximum stellar mass, $\mfmax=\epsilon \Mgas$, which further determines $\tdyn$, as per equation~(\ref{eq:tdyn}). The blue circle in Fig.~\ref{fig:t0_mmax} corresponds to our typical GC progenitor (described in Section~\ref{ssec:typical}), with $\Mgas=6\times10^6\,\msun$ in an environment with  $\Sigmagas=1.5\times10^3\,\msun\,\pc^{-2}$. For Galactic MCs we have adopted the mean surface density from \citet{Solomon+87}, while for CMZ clouds the median surface density of the clouds in table\,1 of \citet{Kauffmann+17}. The maximum cloud mass is set by the outer scale, $\Lgasout$ (blue dashed lines in Fig.~\ref{fig:t0_mmax}), assumed to be of order 100\,pc for Galactic MCs (as mentioned above), and 10\,pc for CMZ clouds \citep[that  is, the assumed scale height of the CMZ 100\,pc ring in][]{Kruijssen+14}. We note that for Galactic MCs, $\mfmax$ is limited by the star lifetime (see the green line in Fig.~\ref{fig:t0_mmax}), requiring a cloud surface density a few times larger than the average one to achieve values of $\mfmax\simeq 100\,\msun$. Additionally, most CMZ clouds are smaller than 10\,pc \citep[the median diameter is 5.6 in table\,1 of][]{Kauffmann+17}, so the typical value of $\mfmax$ in the CMZ is probably smaller than $10^3\,\msun$. Additionally, this maximum mass estimate does not account for mass loss by stellar winds, which should be very high in the Galactic centre because of the super-solar metallicity, further reducing the maximum mass estimate. The effect of wind mass loss will be considered in detail in Section~\ref{ssec:wind}.

Having provided arguments for why EMSs are expected in star clusters with  masses $\gtrsim10^5\,\msun$, we now proceed with the development of a parameterised model for the evolution of the stellar mass function and the self-enrichment by \vemss\ during GC formation. 

%%%%%%%%%%%%%%%%%%%%%%%%%%%%%%%%%%%%%%%%%%
\section{Globular cluster formation with \vemss}
\label{sec:model}
\begin{figure*}
\includegraphics[width=2\columnwidth]{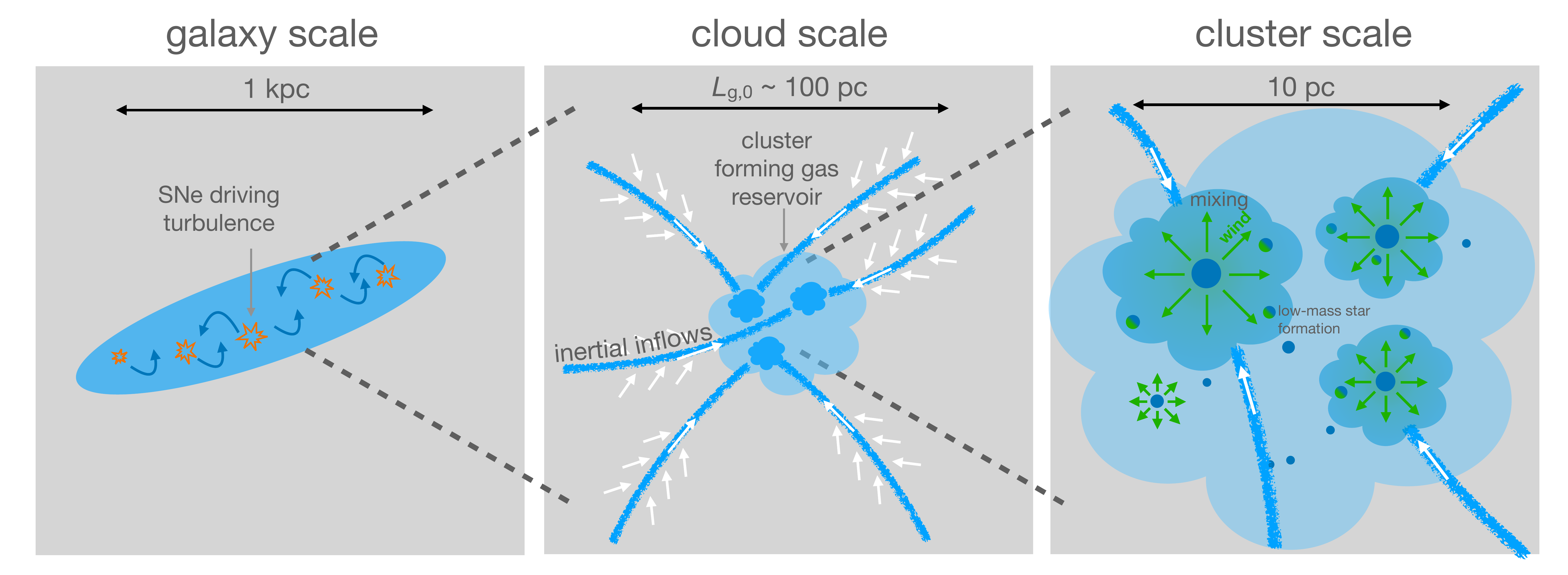}
\vspace{-1mm}
\caption{Schematic view of GC formation. Turbulence is driven by primarily by SNe on a scale of $\sim100\,\pc$, leading to inertial flows on that scale. In the centre, a cluster of stars accumulate, together with a reservoir of gas that did not make it onto stars. After some time, the most massive stars have reached the EMS regime ($\gtrsim10^3\,\msun$) and the pristine g as of the gas reservoir mixes with $\vems$ winds that contain hot-hydrogen burning products. Low-mass stars ($\lesssim1\,\msun$) continuously form from this mixture, which in approximately the second half of the GC formation process has approximately the abundances found in GCs (see Section~\ref{ssec:abundances} for details). }  
\label{fig:gcform}
\end{figure*}

In our picture of GC formation we step away from the classical idea of having a fully formed proto-cluster cloud at the start of star formation. Instead, the gas  forming the stellar cluster is brought in in unison with the star formation process \citep[an assumption also made in the cosmological zoom-in simulations of][]{2017ApJ...834...69L, 2018ApJ...861..107L}. 
{This is analogous to the formation of a massive star in the inertial-inflow scenario, where the stellar mass is gradually brought in from a large scale, rather than fully preassembled in a prestellar core.}
In Fig.~\ref{fig:gcform} we schematically illustrate this idea. On the galactic scale, turbulence is driven by SNe, gas accretion, galaxy mergers, etc. This sets an outer radius for the turbulent regions of $\Rgasout\simeq50\,\pc$,  containing a gas mass $\Mgasout$, which has a dynamical time $\tdynout$. We assume that an entire GC forms on this timescale, or shorter for less massive GCs. 

For sufficiently high $\Sigmagas\gtrsim10^3\,\msun\,\pc^{-2}$,  $\tdynout$ is shorter than the  timescale of the first supernova ($\sim2-3\,\myr$, see equation~\ref{eq:tdyn}) such that all GCs form without in-situ enrichment from supernovae (SNe), as required by the small/absent inferred iron spread in most GCs. Here we present a  model for the growth of the IMF, considering gas inflows and wind mass loss. We will show that the majority of GCs require stars in excess of $10^3\,\msun$ and that the ongoing accretion while they pollute the intra-cluster medium is essential. We will, therefore, from hereon refer to this GC pollution scenario as the `accreting EMS' (aEMS) model. 

%________________
\subsection{IMF and star formation efficiency}
During GC formation, we assume that stars\footnote{We use the term stars for all masses, but at low masses  ($\lesssim 1 \,\msun$) these are proto-stars or pre-main sequence (PMS) stars, because the duration of GC formation is short compared to the PMS phase.} form at a constant rate for a timescale $\tdyn$. At $t=\tdyn$ the accretion ends, and the IMF is fully populated. Stars with final masses $\le1\,\msun$ are assumed to fragment from the cold gas instantaneously, while  stars with final masses $>1\,\msun$ are first created with a mass of $1\,\msun$ and then grow with a constant rate corresponding to the predicted average inertial-inflow rate for their final masses. This is consistent with the fact that the peak of the prestellar core mass distribution is $\sim1\,\msun$ \citep{Padoan+20massive}. Inertial inflows may in principle contribute to the final mass of stars of all masses, as shown by Fig.~\ref{fig:t95_v1}\footnote{See also Fig.\,11 in \citet{2018ApJ...854...35H}, where the plot extends to brown-dwarf masses.}, though their contribution decreases towards lower-mass stars \citep{Pelkonen+21}. Thus, one could generate stars randomly with a distribution of formation times that covers the predicted scatter for all stellar masses. However, because low-mass stars do not affect the cluster chemical evolution, it is not necessary to follow their formation time in detail. As for the EMSs, the predicted scatter in their formation time is too small to affect our results, and adopting the average (mass-dependent) value leads to simple analytical predictions. 

Specifically, we assume that stars in the range  $0.1-1\,\msun$ follow a  \citet{2003PASP..115..763C} IMF. We define the mass function $\psi(m)$ as the number of stars in a mass interval $[m, m+\dr m]$, such that for the low-mass stars  
\begin{equation}
\psil(m) = \frac{A_1}{m}\exp\left[-\frac{\log_{10}^2 (m/0.08\,\msun)}{2\times0.69^2}\right], \hspace{0.2cm}0.1< m\le 1\,\msun.
\label{eq:psil}
\end{equation}
The final mass function for $\mf>1\,\msun$  is assumed to have the form
\begin{equation}
\psif(m) = A_2\mf^{\alpha},\hspace{0.5cm}1\,\msun< m\le\mfmax,
\label{eq:psif}
\end{equation}
with $\alpha=-2.3$ \citep{2001MNRAS.321..699K} and $A_2=A_1\exp[-\log_{10}^2(0.08)/(2\times0.69^2)]$, such that the IMF is continuous in the range  $[0.1\,\msun,\mfmax]$.

We now need to assume something about the star formation efficiency (\sfe). We define \sfe\ as the fraction of $\Mgas$ that ends up in stars at $\tdyn$, which defines the initial GC mass ($\MGCi$), ignoring winds: $\sfe=\MGCi/\Mgas$. 
We can either assume a constant \sfe\ for all $\Mgas$, or a fully populated IMF for different $\MGCi$, which gives rise to an increasing $\sfe$ with $\Mgas$. For simplicity, and for lack of guidance from numerical simulations, we adopt a constant $\sfe=0.1$. This value is of the order of the ratio between the mass of the most massive clusters and the total gas mass reservoir in simulations of star formation \citep[for example,][]{Padoan+17SN_IV} and in observations of hub-filament systems \citep[for example,][]{Rawat+24}, as discussed in Section~\ref{sec:hub_filaments}. Given our model assumption of $\alpha_{\rm vir}\sim 1$, it is also consistent with the star formation efficiency per free-fall time in models and simulations of supersonic turbulence \citep{Krumholz+McKee05sfr,Padoan+Nordlund11sfr,Federrath+Klessen12,Padoan+12sfr,Padoan+17SN_IV}. These models, where the SFE (after a free-fall time) is primarily controlled by the virial parameter of the turbulence, reproduce realistic galaxy properties when applied to cosmological simulations of galaxy formation \citep[for example,][]{Otero+25}, while giving local SFE values that peak around 0.1-0.3 \citep{Andalman+25}. 
Although we assume that a GC forms through the conversion of 10 per cent of the gas mass within a region with radius $\Rgasout$, and Fig. 4 schematically illustrates the formation of a single such cluster, star formation likely proceeds throughout the entire region. In addition to the main GC, other (typically lower-mass) clusters and isolated stars are expected to form in the surrounding environment, perhaps even another GC. As a result, the total star formation efficiency across a region comparable in size to the outer scale of the turbulence, over a timescale on the order of its dynamical time, is generally higher than the 10 per cent value adopted in our GC formation model. In this context, the GC represents the most massive cluster formed in the region and, in our approximation, the only one sufficiently massive to produce a stellar population significant enough to drive the observed enrichment in light elements.

The resulting $\mfmax(\MGCi)$ relation corresponding to equation~(\ref{eq:mfmax_Mgas}) is
\begin{equation}
\mfmax = 250\,\msun\,\frac{\MGCi}{10^4\,\msun}.
\label{eq:mfmax}
\end{equation}
We note that $\mfmax$ is a hypothetical maximum stellar mass that follows from the highest accretion rate ($\mdotaccmax$). For $\mfmax\gg100\,\msun$, this final mass is never reached because of wind mass-loss rates \citep{2018A&A...615A.119V}, which we discuss in Section~\ref{ssec:wind}. For the same reason, $\MGCi$ is never reached, because it also does not account for wind mass loss. The maximum stellar mass is also not reached  in clusters forming on timescales longer than the stellar lifetime at $\mfmax$, because the most massive stars would explode as a SN while accreting (see the green line in Fig.\,\ref{fig:t0_mmax}).

%________________
\subsection{Accretion}
The  time it takes to form a massive star depends on the final mass as $\tf\propto \mf^{1/2}$ and the  mass accretion rate, $\mdotacc\propto\mf^{1/2}$, is approximately constant during the star formation process \citep{Padoan+20massive}. If we assume that $\mf$ is much larger than the initial mass of the seed star ($\sim1\,\msun$), then $\mf \simeq\mdotacc\tf$, and $\mdotacc$ is given by equation~(\ref{eq:mdotacc}).
We assume that all stars above $1\,\msun$ grow in mass in this way. 
In Appendix~\ref{app:exact} we present  results for $\mdotacc(\mf)$ and $m(t)$ that are exact for all masses and that we use in the model. For readability,  we continue the discussion with the approximate proportionalities that are valid in the \vems\ mass range. 
 
%______________
\subsection{Wind mass-loss rate}
\label{ssec:wind}
The winds of stars $\gtrsim75\,\msun$ are optically thick and their fractional mass-loss rate is higher than for lower-mass stars, which have optically thin winds \citep{2011A&A...531A.132V,2011A&A...535A..56G}. For simplicity, we consider only winds of stars above $100\,\msun$ and assume that their winds are optically thick. This wind mass-loss rate, $\mdotw$,  depends on several parameters, and we here approximate it by the scaling
 \citep[][]{2018A&A...615A.119V}
\begin{equation}
\mdotw(Z, m) = -\mdwhz\left(\frac{m}{100\,\msun}\right)^2,
\label{eq:mdw}
\end{equation}
with
\begin{equation}
\mdwhz = \mdwh\left(\frac{Z}{\zsun}\right)^{\eta},
\label{eq:mdwh}
\end{equation}
where $\zsun = 0.014$ \citep{2021A&A...653A.141A},  $\mdwh$ is a (positive) constant of proportionality representing the wind-mass loss rate of a $100\,\msun$ star with $Z=\zsun$ and $\eta$ sets the strength of the $Z$ dependence.
The results for low-metallicity VMSs by \citet{2023MNRAS.524.1529S} suggest  $\eta\simeq0.6$ and $\mdwh\simeq10\,\msun\myr^{-1}$
for non-rotating stars at the zero-age main sequence (ZAMS), and approximately $\sim3$ times higher towards the end of the hydrogen burning phase as the result of their increased luminosity. These authors also show that rotation  enhances the wind mass-loss rate because stars close to the radiative Eddington limit reach the $\Omega\Gamma$-limit \citep{2000A&A...361..159M} already for rotation rates  below the critical rotation rate ($\Omega_{\rm crit}$). Specifically, they show that a model with $\Omega/\Omega_{\rm crit} = 0.6$ loses approximately 2.5 times more mass than a non-rotating model with the same metallicity ($Z=0.004$). 

In our model, \vemss\ lose most of their mass when  $\Delta Y\simeq0.2$ (Section~\ref{sssec:helium}) because of continuous rejuvenation by accretion and efficient mixing (Section~\ref{ssec:mixing}), such that their mass loss rates are expected to be higher than those at  the start of the evolution ($\Delta Y=0$) mentioned above. 
The accretion is expected to lead to high rotation rate \citep{Rosen+12} and therefore an enhanced mass-loss rate
compared to the rates of the non-rotating models of \citet{2023MNRAS.524.1529S}. Finally, accreting stars have a higher luminosity, also increasing the mass loss rate. We therefore settle on $\mdwh= 30\,\msun\,\myr^{-1}$. For $Z = 0.01\,\zsun(0.1\,\zsun)$ this results in $\mdwhz \simeq 1.9\,\msunmyr(7.5\,\msunmyr)$.

%___________________________________________
\subsection{Mass evolution due to accretion and mass loss}
\label{ssec:massevol}
We  assume that the effects of mass accretion and stellar winds can be added linearly and we discuss the validity of this assumption in more detail in Section~\ref{ssec:windacc}.
In this simple picture, the mass evolution is found from integrating $\dot{m}=\mdotacc+\mdotw$ (equations~\ref{eq:mdotacc} and \ref{eq:mdw}) starting from $1\,\msun$. For $m(t)\gg1\,\msun$ we find
\begin{equation}
m(t) \simeq \minf\tanh\left(\frac{t}{\tinf}\right).
\label{eq:mt}
\end{equation}
The stellar mass increases approximately linearly in time $m(t)\simeq\minf t/\tinf= \mdotacc t$ for $t\lesssim\tinf$  and asymptotically approaches $m(t)\simeq\minf$ for $t\gtrsim\tinf$. 
This asymptotic mass $\minf$ is set by the balance between wind mass loss and accretion from inflows \citep[$\mdotacc + \mdotw = 0$,][]{2018A&A...615A.119V} and is given by
\begin{equation}
\minf = 100\,\msun\sqrt{\frac{\mdotacc}{\mdwhz}}.
\label{eq:minf}
\end{equation}
The associated timescale to reach this mass, $\tinf$, is 
\begin{equation}
\tinf = \frac{\minf}{\mdotacc} = \frac{100\,\msun}{\sqrt{\mdotacc\mdwhz}}.
\label{eq:tinf}
\end{equation}
For some representative values of $\mdwhz=10\,\msunmyr$ and $\mdotacc=10^3\,\msunmyr$ we find $\minf=10^3\,\msun$ and $\tinf = 1\,\myr$, showing that a balance between accretion and wind mass loss can occur within the formation time of an EMS. 

Because $\tf\propto \mf^{1/2}$ and $\tinf\propto\mf^{-1/4}$, the moment when wind mass loss and accretion balance, relative to the formation time of the star, depends on  $\mf$ as $\tinf/\tf \propto \mf^{-3/4}$. For $\mf=10^3\,\msun$ and  $\mdotacc=10^3\,\msunmyr$ as before, we thus have $\tinf/\tf \simeq 1$. More massive  stars reach a mass close to $\minf$ before the accretion stops, while lower mass stars stop accreting before $\tinf$ has been reached. This means that the total wind mass loss is dominated by the most massive, accreting stars. The ratio $\tinf/\tf$ also depends on $\Sigmagas$, which we discuss in Section~\ref{sec:scaling}.

As a sanity check, we compare the resulting maximum $\minf$ to maximum stellar masses in known resolved YMCs. 
Galactic YMCs, such as Arches, NGC\,3603 and Westerlund~1, have masses of $\MGCi \simeq 10^4\,\msun$ \citep*[for example,][]{2010ARA&A..48..431P} and contain stars with masses up to $\sim100-150\,\msun$ \citep[for example,][]{2010MNRAS.408..731C}. Assuming that they formed from gas with $\Sigmagas\simeq10^2\,\msun\,\pc^{-2}$, with a SFE of 10 per cent (such that $\Mgas\simeq10^5\,\msun$), the maximum accretion rate is $\mdotaccmax\simeq55\,\msun\,\myr^{-1}$ (equation~\ref{eq:mdamax}), and combined with $\mdwhz=30
\,\msun\,\myr^{-1}$ and equation~(\ref{eq:minf}) we find that the maximum $\minf\simeq130\,\msun$, which is reached in a time $\tinf\simeq2.5\,\myr$, in perfect agreement with observational findings. 
The YMC R136 in the LMC has a mass of $\MGCi\simeq10^5\,\msun$ \citep{2009ApJ...707.1347A,2012A&A...546A..73H}, such that for the same $\Sigmagas$ and SFE we find $\mdotaccmax\simeq308\,\msun\,\myr^{-1}$. For half solar metallicity, $\mdwhz\simeq20\,\msun\,\myr$, such that $\minf\simeq395\,\msun$, in good agreement with the finding of several stars in excess of $150\,\msun$ and an estimated maximum initial mass of $\sim320\,\msun$ \citep{2010MNRAS.408..731C}.

\begin{figure}
\includegraphics[width=\columnwidth]{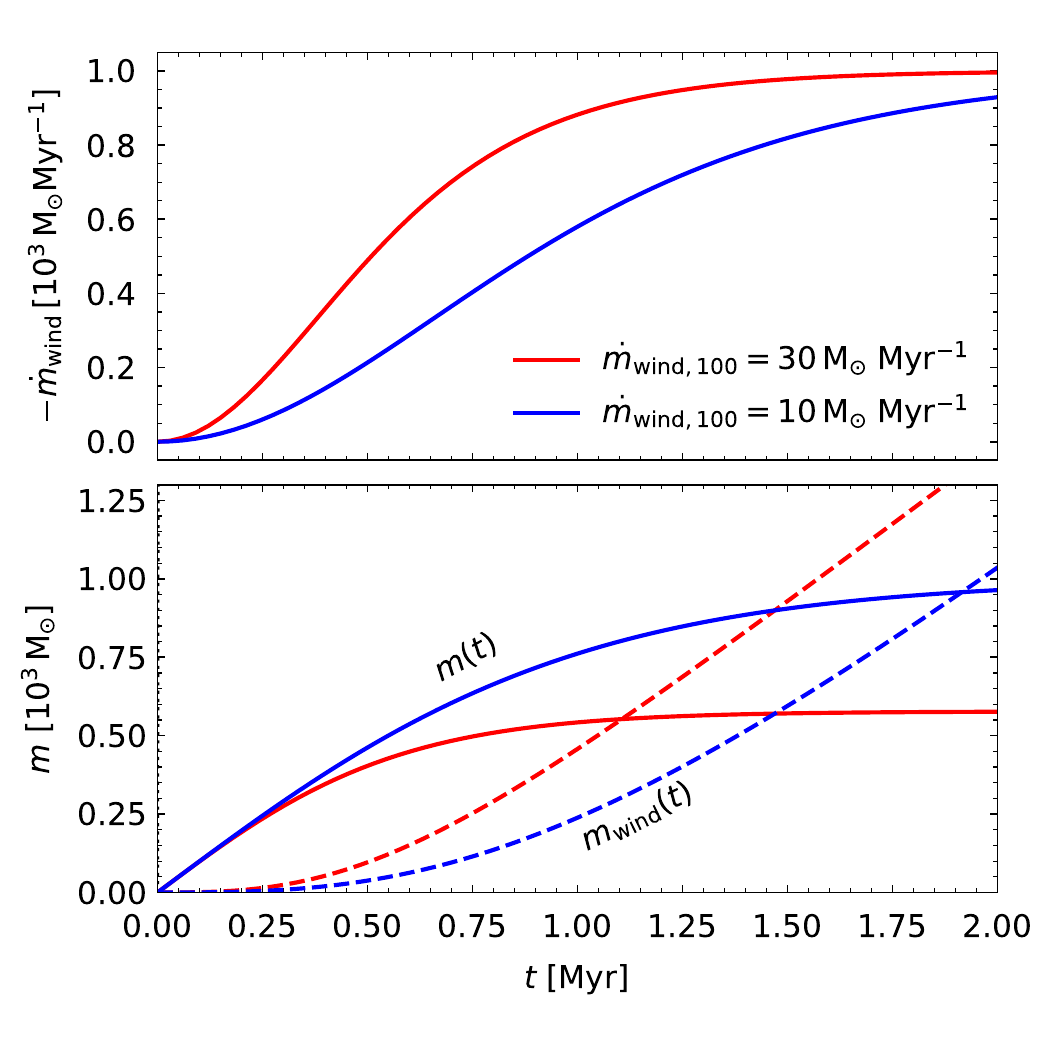}
\vspace{-5mm}
\caption{Wind mass loss rates (equation~\ref{eq:mdw}) (top panel) and masses (bottom panel) as a function of time for  two stars with different $\mdwhz$.  Both stars have $\mdotacc=10^3\,\msunmyr$, and the star with the higher(lower) $\mdwhz$ approaches lower(higher) $\minf$, and similar $\mdotw$.    }
\label{fig:mt}
\end{figure}

Because $\minf$ is set by a balance between $\mdotacc$ and $\mdotw$, the  wind mass-loss rate of a star with mass $\minf$ equals $-\mdotacc$, and is independent of $\mdwhz$. For a given $\mdotacc$, a lower(higher) $\mdwhz$ leads to a higher(lower) $\minf$, such that $\mdotw(\minf)$ is the same. We illustrate this in Fig.~\ref{fig:mt}. The top panel  shows   $\mdotw(t)$ for two values of $\mdwhz$  and the same $\mdotacc=10^3\,\msunmyr$. For both stars the wind mass-loss rate increases up to $\mdotw=-10^3\,\msunmyr$, the value for which $\mdotw=-\mdotacc$. 
The bottom panel shows how the corresponding $m(t)$ and the cumulative mass released in winds, $\mwind(t)\simeq \mdotacc t -m(t)$, evolve. Despite the factor of three difference in $\mdwhz$, the total amount of material released at $2\,\myr$ only differs by approximately $30$ per cent. This illustrates two important aspects of this pollution model, namely that for a given $\mdotacc$
\begin{enumerate}
\item metal-poor clusters have stars that are more massive with higher central temperatures;
\item metal-poor clusters lose a similar amount of mass as metal-rich clusters.
\end{enumerate}
We discuss this $Z$ dependence and its importance in reproducing observed scaling relations in more detail in Section~\ref{sec:scaling}.
%__________________________________
\subsection{Building the initial mass function} \label{ssec:3.5}
\begin{figure}
\includegraphics[width=\columnwidth]{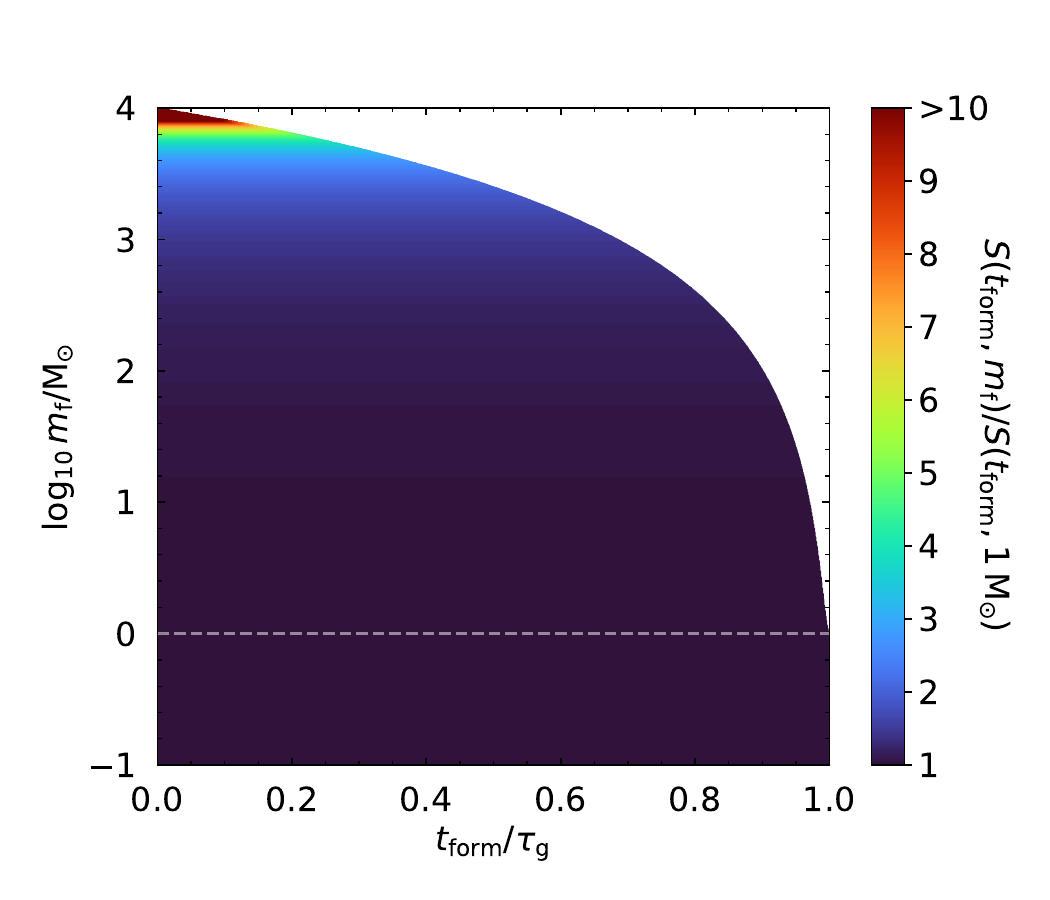}
\vspace{-5mm}
\caption{Source function $S(\tform, \mf)$ as a function of $\mf$ following from equation~(\ref{eq:Stform}). The formation of seeds of the most massive stars is skewed to earlier $\tform$, to ensure that all stars are done accreting at $\tdyn$. }
\label{fig:tform}
\end{figure}

Here we combine the scaling relations discussed up to here, and present a tractable, parameterised model of the evolution of the stellar mass function under the combined effects of accretion and wind mass loss. 
The mass function above $1\,\msun$ (where accretion is a dominant contribution to the final stellar mass) as a function  of time, $\psi(m,t)$, is found from
\begin{align}
\psi(m,t) &= \int_0^{t}S(\tform,\mf)\psif(m)\left|\frac{\partial \mf}{\partial m}\right| \dr \tform,
\end{align}
where $S(\tform, \mf)$ is a source function that describes the formation rate of seeds as a function of their final mass. We define this as
\begin{equation}
S(\tform,\mf) = \begin{cases}
({\tdyn-\tf})^{-1} ,&\tform< \tdyn-\tf,\\
0,&\tform>\tdyn-\tf.
\end{cases}
\label{eq:Stform}
\end{equation}
This results in a constant formation rate of low-mass stars ($\sim1\,\msun$) for the full duration of cluster formation. The formation rate of seeds for the more massive stars is also constant in time, but higher (relative to what is expected from the final IMF) and for a shorter duration, such that all stars are done accreting at $t=\tdyn$. We illustrate this  in Fig.~\ref{fig:tform}.
To model the mass function evolution, we sample values for $\mf$ from $\psif$ (equation~\ref{eq:psif}) and values for $\tform$ from $S(\tform, \mf)$. To reduce the noise, we sample more high-mass stars and weigh them according to $\psif$.  We then evolve the individual masses in time and create the mass functions at different times. During the accretion phase the time-dependent masses are given by equation~(\ref{eq:mt_exact}) and after the accretion ends (which for stars with $\mf<\mfmax$ happens before $\tdyn$ if $\tdyn-\tform > \tf$) by equation~(\ref{eq:mtpost}).

\begin{table*}
\caption{Summery of fixed aEMS model parameters and their origin.}
\label{tab:params}
\begin{tabular}{ll p{12cm}}
\hline
Parameter       &  Value & Comments\\\hline
$\alpha_{\rm vir}$ & 1 & Ratio of (turbulence) kinetic energy and gravitational energy, assuming equipartition in analogy with Galactic star-forming regions \citep{Larson81}.\\
$\epsilon$      &  $2.5\times10^{-3}$    &Relates $\mfmax$ to $\Mgas$ (equation~\ref{eq:mfmax_Mgas}). Derived from the star formation models of \citet{2018ApJ...854...35H,Padoan+20massive}, see Section~\ref{ssec:mfmax} for details.\\
$\sfe$             &  $0.1$                         & Star formation efficiency defined as the fraction of the total mass reservoir in the outer scale that ends up in the GC stars (neglecting the wind mass loss) \citep{Padoan+17SN_IV}\\
  $\mdwh$       & $30\,\msunmyr$ & Constant of proportionality in equation~(\ref{eq:mdw}). Three times the value of \citet{2023MNRAS.524.1529S} for non-rotating ZAMS stars, to include the effect of internal evolution and rotation. \\
  $\eta$            & $0.6$ & Power-law index in the expression for $\mdotw$ (equation~\ref{eq:mdw}), from \citet{2023MNRAS.524.1529S}. \\
 \hline
\hline
 \end{tabular}
\end{table*}
%________________________
\subsection{A model for a typical GC}
\label{ssec:typical}
A typical GC today has a mass of $\sim2\times10^5\,\msun$ \citep[for example,][]{2020PASA...37...46B}. Assuming that it has lost  $\sim10^5\,\msun$ due to evaporation in the Galactic tidal field \citep[for example,][]{2001ApJ...561..751F,2023MNRAS.522.5340G} and  50 per cent\footnote{For a canonical IMF in the range $0.1-100\,\msun$ this fraction is closer to 40 per cent, but because we here include the wind mass loss from \vemss,  50 per cent is more appropriate.} of its initial mass due to stellar evolution   then the initial mass was $\MGCi\simeq6\times10^5\,\msun$. We adopt $\sfe=0.1$, such that the initial cloud mass is $\Mgas\simeq6\times10^6\,\msun$. 
Observations of high-redshift galaxies and  the constraint that GCs form before SNe occur imply that $\Sigmagas\gtrsim10^3\,\msun\,\pc^{2}$ (Section~\ref{sec:inertial}). We will later show that $\Delta Y$ is sensitive to $\Sigmagas$ and  we find that for $\Sigmagas=1.5\times10^3\,\msun\,\pc^{-2}$ we obtain satisfactory agreement between the model and observed helium spreads, hence we adopt this from hereon. In Section~\ref{ssec:fp2} we show the effect of varying $\Sigmagas$ on the  helium spread.
For the adopted $\Sigmagas$, 
the radius of the gas clouds is $\Rgas\simeq36\,\pc$ ($\Lgas\simeq71\,\pc$)  and the dynamical time is $\tdyn\simeq1.7\,\myr$. For the wind mass loss we adopt an intermediate metallicity of $Z=0.1\,\zsun$, resulting in $\mdwhz\simeq7.5\,\msunmyr$. Other derived properties are given in Table~\ref{tab:typical}.
\begin{table}
\caption{Adopted values for a typical GC. Note that the first three variables ($\Mgas, \Sigmagas,  Z$) define all the variables below the line with the use of the model parameters that are given in Table~\ref{tab:params}.} 
\label{tab:typical}
\begin{tabular}{ll p{3.5cm}}
\hline
 Variable       &  Value & Comment\\\hline
  $\Mgas$       &  $6\times10^6\,\msun$    & $=\MGCi/\sfe$\\
 $\Sigmagas$   & $ 1.5\times10^3\,\msun\,\pc^{-2}$  &Motivated by inferred gas densities at high redshift (see Section~\ref{ssec:scaling_to_GC}) and the resulting $\Delta Y$ (see Section~\ref{ssec:helium}). \\
  $Z$       & $0.1\,\zsun$ & equation~(\ref{eq:mdw}) \\\hline
$\MGCi$         & $6\times10^5\,\msun$      &  present day: $\MGC\simeq2\times10^5\,\msun$   \\
 $\mfmax$        &  $1.5\times10^4\,\msun$    &  $= \epsilon\Mgas $\\
 $\tdyn$            & $ 1.7\,\myr$  & equation~(\ref{eq:tdyn}) \\
 $\mdotaccmax$  & $9.0\times10^3\,\msunmyr$  & $=\mfmax/\tdyn$ \\
 $\minf$       & $3.5\times10^3\,\msun$ & equation~(\ref{eq:minf}) \\ 
 $\tinf$       & $0.39\,\myr$ & equation~(\ref{eq:tinf}) \\
\hline
 \end{tabular}
\end{table}

\begin{figure}
\includegraphics[width=\columnwidth]{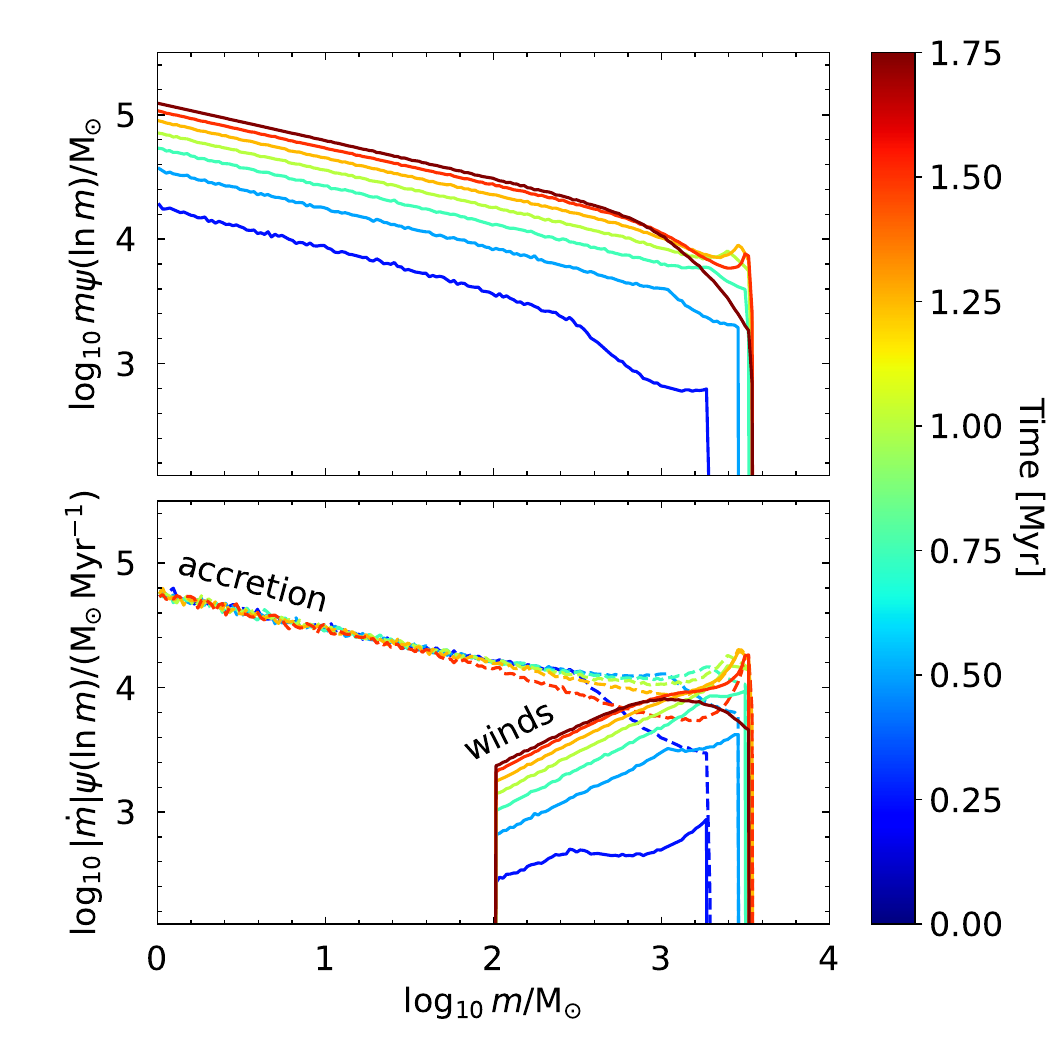}
\vspace{-5mm}
\caption{Evolution of the IMF (top) and the corresponding accretion and wind mass-loss rates  (bottom) for typical GC parameters (see Table~\ref{tab:typical}). }
\label{fig:imf}
\end{figure}

Fig.~\ref{fig:imf} shows the time evolution of the IMF for this cluster (top) as well as the total accretion rates and \vems\ wind mass-loss rates as a function of mass (bottom). We plot the mass function as $m\psi(\ln m)$, where $\psi(\ln m)\equiv\dr N/\dr\ln(m)=m\psi(m)$, such that a flat $m\psi(\ln m)$ corresponds to $\psi(m)\propto m^{-2}$. Equivalently, we plot the rates as $\dot{m}\psi(\ln m)$, such that a horizontal line in the bottom panel implies that the mass accreted/lost per unit of time is the same in each decade of $m$, and declining(rising) lines correspond to bottom(top) heavy contributions.
The shape of the IMF in the early accretion phase resembles what is found in MHD simulations: the high-mass end starts steeper than Salpeter/Kroupa and gradually approaches the Salpeter/Kroupa power-law slope \citep*{2018ApJ...854...35H}.
 The bottom panel of Fig.~\ref{fig:imf} shows that the most massive stars ($\gtrsim10^3\,\msun$) achieve an equilibrium between accretion and wind mass loss (dashed and full lines of the same colour coincide). This figure also shows that most mass is released by EMSs. These two aspects have important consequences for GC abundances:
\begin{enumerate}
\item The yields of the `pile up' of stars near $\minf$ of the most massive stars are similar, leading to a discrete abundance pattern separated from the abundance of the pristine stars (we discuss this further in Section~\ref{ssec:discreteness}); 
\item A `conveyer belt' is established at $t\simeq\tinf\simeq0.4\,\myr$, which leads to an efficient production of hot-hydrogen burning yields, at relatively low He.
\end{enumerate}
We discuss the implications for abundances in more detail next.

%________________________
\subsection{Abundances}
\label{ssec:abundances}
\subsubsection{Helium}
\label{sssec:helium}
An important observational constraint is that the typical He spreads of GCs, expressed as the difference between He mass fraction ($Y$) of P2 and P1 stars,  is  low \citep[$\DYobs \simeq 0.01$,][]{2018MNRAS.481.5098M}.
One of the promising aspects of the aEMS model is that aEMSs accrete pristine gas while they blow strong winds of processed material. Assuming that these stars are fully convective (see Section~\ref{ssec:mixing}), this gas accretion rejuvenates the stars and keeps $Y$ low. To quantify the typical $\Delta Y$  we can expect, we propose a simple model for the evolution of the surface $Y$ of fully convective stars. We assume that the pristine $Y$ is $Y=0.25$ and that it increases by H-burning on a nuclear timescale, $\tnuc\simeq3\,\myr$, and that $Y$ decreases by accretion of pristine gas. We can then write
\begin{equation}
\Ydot =  \frac{0.75}{\tnuc} + (0.25-Y)\frac{\mdotacc}{m},
\label{eq:Ydot}
\end{equation}
for $0.25< Y\lesssim 1$ and $0<t<\min(\tnuc, \tdyn)$.  
The wind mass-loss rate does not enter in equation~(\ref{eq:Ydot}), because we assume a fully mixed star, such that the abundances of the winds are the same as those in the stellar interior.
In the absence of accretion, the first term causes $Y$ to increase to $Y=1$ on the nuclear timescale, independent of the wind mass loss. This is what is found for the VMS models of \citet{2023MNRAS.526..534H}, which have hydrogen burning phases that last $2.5-3\,\myr$ for stars with ZAMS masses in the range $500-100\,\msun$ \citep[see also][for other metallicities in this mass range]{ 2022A&A...659A.163M,2024A&A...690A..91S}. Because $Y>0.25$, the second term in equation~(\ref{eq:Ydot}) is negative, implying a slower increase of $Y$ because of rejuvenation. 

The competing effects of nuclear burning (increasing $Y$) and rejuvenation by gas accretion (reducing $Y$)  eventually balance  ($\dot{Y} = 0$), such that $Y$ approaches a constant $\Yinf$.
The corresponding $\Delta \Yinf$,  which is the $\Delta Y$ of a star with mass $\minf$, is
\begin{equation}
\Delta \Yinf =  \frac{0.75}{\tnuc}\frac{\minf}{\mdotacc} = 0.75\frac{\tinf}{\tnuc},
\end{equation}
which for our typical GC parameters gives $\Delta \Yinf\simeq 0.12$ (that is, $ \Yinf \simeq 0.37$). Note that  $\Delta \Yinf$ can not exceed $\sim0.75$, even if $\tinf\gtrsim\tnuc$, because the positive contribution to $\Ydot$ in equation~(\ref{eq:Ydot}) stops at $t=\tnuc$.

\begin{figure}
\includegraphics[width=\columnwidth]{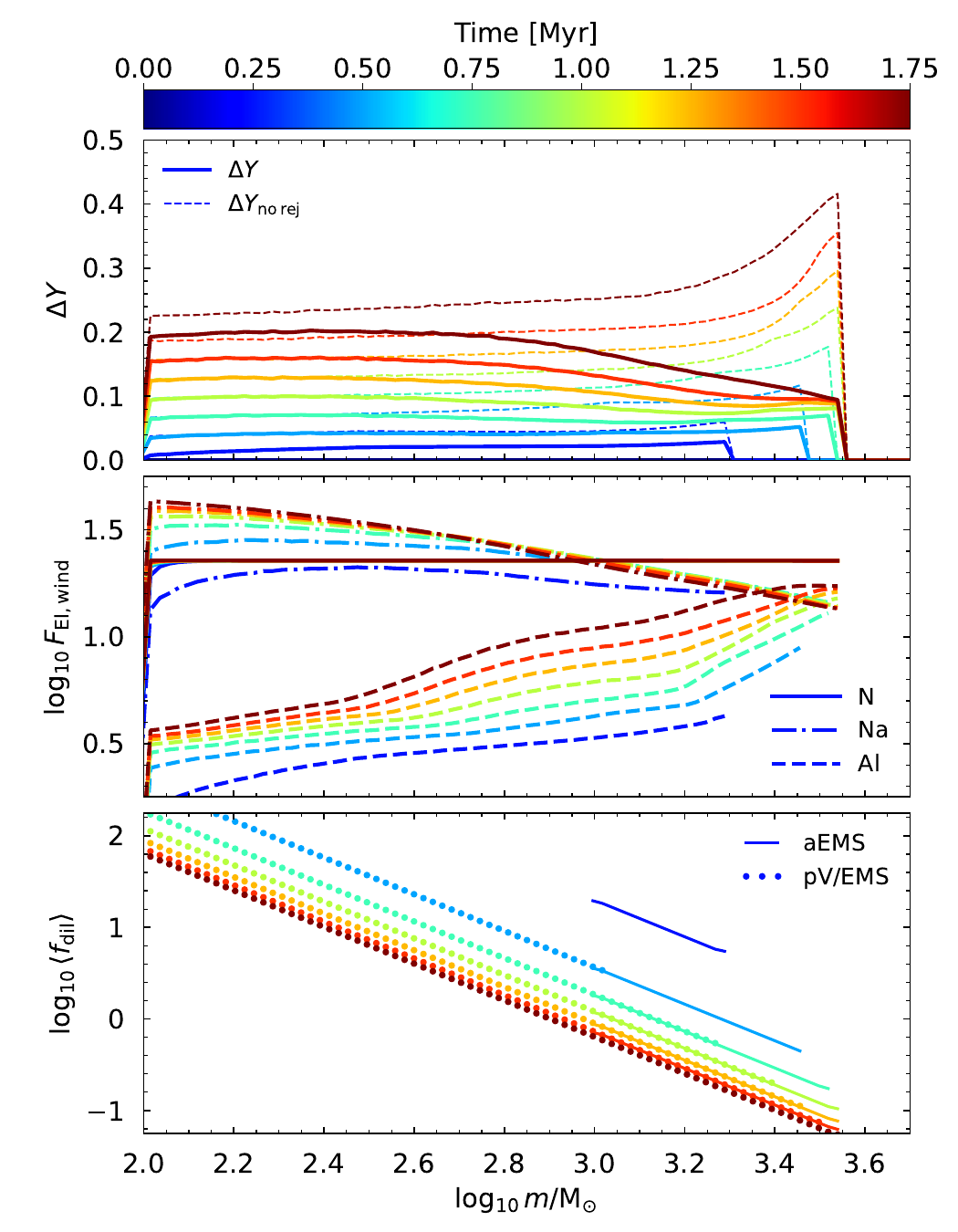}
\vspace{-5mm}
\caption{Evolution of $\Delta Y$ (top), $\Flin{N}{wind}$, $\Flin{Na}{wind}$, $\Flin{Al}{wind}$ (middle) and $\fdil$ (bottom) as a function of V/EMS mass for the time evolving stellar mass function of Fig.~\ref{fig:imf}. For initial $\ElFe=0$, $\log_{10}\!\Flin{El}{wind}$ can be interpreted as $\Delta\ElFe$.
The  $\Delta Y$ values in the absence of rejuvenation ($\DYnor$) are shown as dashed lines. The dilution factors ($\fdil$) are shown for the two types of polluters we distinguish: accreting EMS (aEMS) and V/EMS that are  no longer accreting (pV/EMS). The mass dependence is the same (equation~\ref{eq:fdil}), but they contribute at slightly different times and mass ranges.
}
\label{fig:abundances_mass}
\end{figure}

We now discuss the results of numerically integrating equation~(\ref{eq:Ydot}) to obtain $Y(t)$ for individual stars. 
In Fig.~\ref{fig:abundances_mass} (top) we show $\Delta Y$ as a function of stellar mass for the model  shown in Fig.~\ref{fig:imf}. The He abundances of the most massive stars -- which dominate the wind mass loss -- increase approximately linearly in time and $\Delta Y$ of a $5\times10^3\,\msun$ star at $t=\tdyn$ has reached the asymptotic $\Delta \Yinf=0.12$. 
The average He increase of all ejecta in this model is $\DYw\simeq0.09$.
\citet{2018MNRAS.481.5098M} finds that the maximum difference between P1 and P2 in a typical GC is  $\max(\DYobs ) \simeq 0.05$,  suggesting that the most extreme stars are made of 50 per cent material (1:1 dilution).
Also, \citet{2018MNRAS.481.5098M} finds a typical difference in helium between P1 and P2 of $\DYobs \simeq 0.01$. This requires a factor of $\sim10$ dilution with pristine material. We discuss the expected dilution with pristine gas in Section~\ref{ssec:dilution} and a more detailed comparison with observations in Section~\ref{sec:scaling}. 

%________________________
\subsubsection{Other light elements}
\label{ssec:other}
\begin{figure}
\includegraphics[width=\columnwidth]{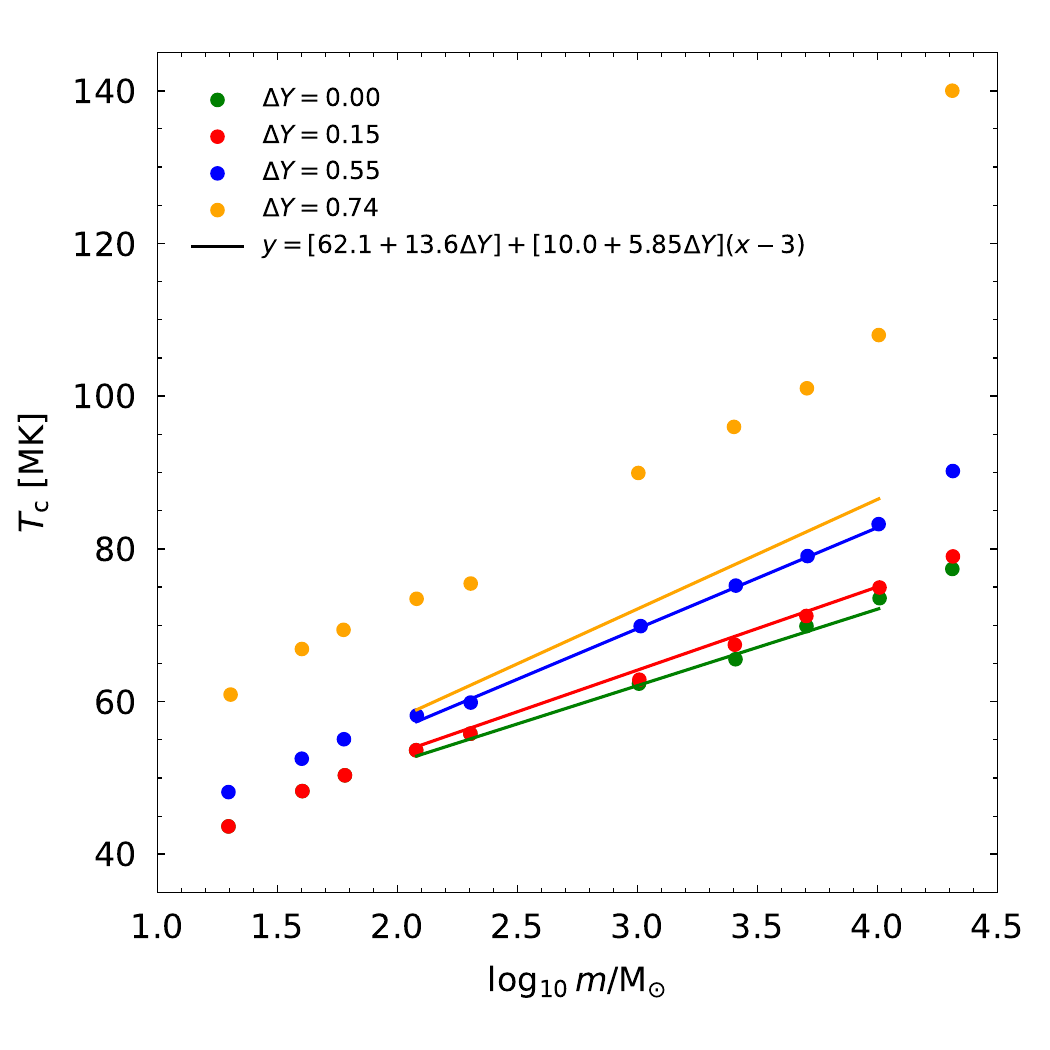}
\vspace{-5mm}
\caption{Central temperatures, $\Tc$, of stars of different masses, at different evolutionary stages($\Delta Y$) for $\feh=-1.5$ from \citet{2017A&A...608A..28P}. A simple functional fit over the relevant mass range is shown, that reproduces $\Tc$ well for $\Delta Y\le 0.55$. }
\label{fig:tcmod}
\end{figure}

For the abundances of other light elements we need to consider how the central temperature of a star, $\Tc$, depends on its mass and helium abundance. In Fig.~\ref{fig:tcmod} we show $\Tc$ from stellar models with $\feh=-1.5$ \citep[see][for details]{2017A&A...608A..28P} as a function of mass, for different $\Delta Y$.  For $\Delta Y\le0.55$ and  in the mass range $10^2\lesssim m/\msun\lesssim 10^4$, $\Tc(m, \Delta Y)$ is well approximated by a linear dependence on $\log m$, with both the slope and the intercept  linear functions of $\Delta Y$:
\begin{equation}
\frac{\Tc}{\MK} = (62.1 + 13.6\Delta Y) + (10.0 + 5.85\Delta Y)\left(\log_{10}\!\frac{m}{\msun}-3\right).
\end{equation}
The relation was found from a least-square fit to the stellar models with $\Delta Y\le0.55$ and reproduces the model results with a precision of one per cent. The central temperature at the end of the hydrogen burning phase ($\Delta Y = 0.74$) are not reproduced by this relation, but because the vast majority of wind mass is released at $\Delta Y\lesssim0.5$, this is not an issue. 
From this we see that for stars with masses $(1-5)\times10^3\,\msun$ and $\Delta Y = 0.1$, hydrogen is burnt at $\sim63-71\,\MK$. This temperature range is where the relative abundances of ONaMgAl are similar to the observations \citep{2017A&A...608A..28P}. Next we discuss how  we compute the abundances of these elements in the model.

\begin{figure}
\includegraphics[width=\columnwidth]{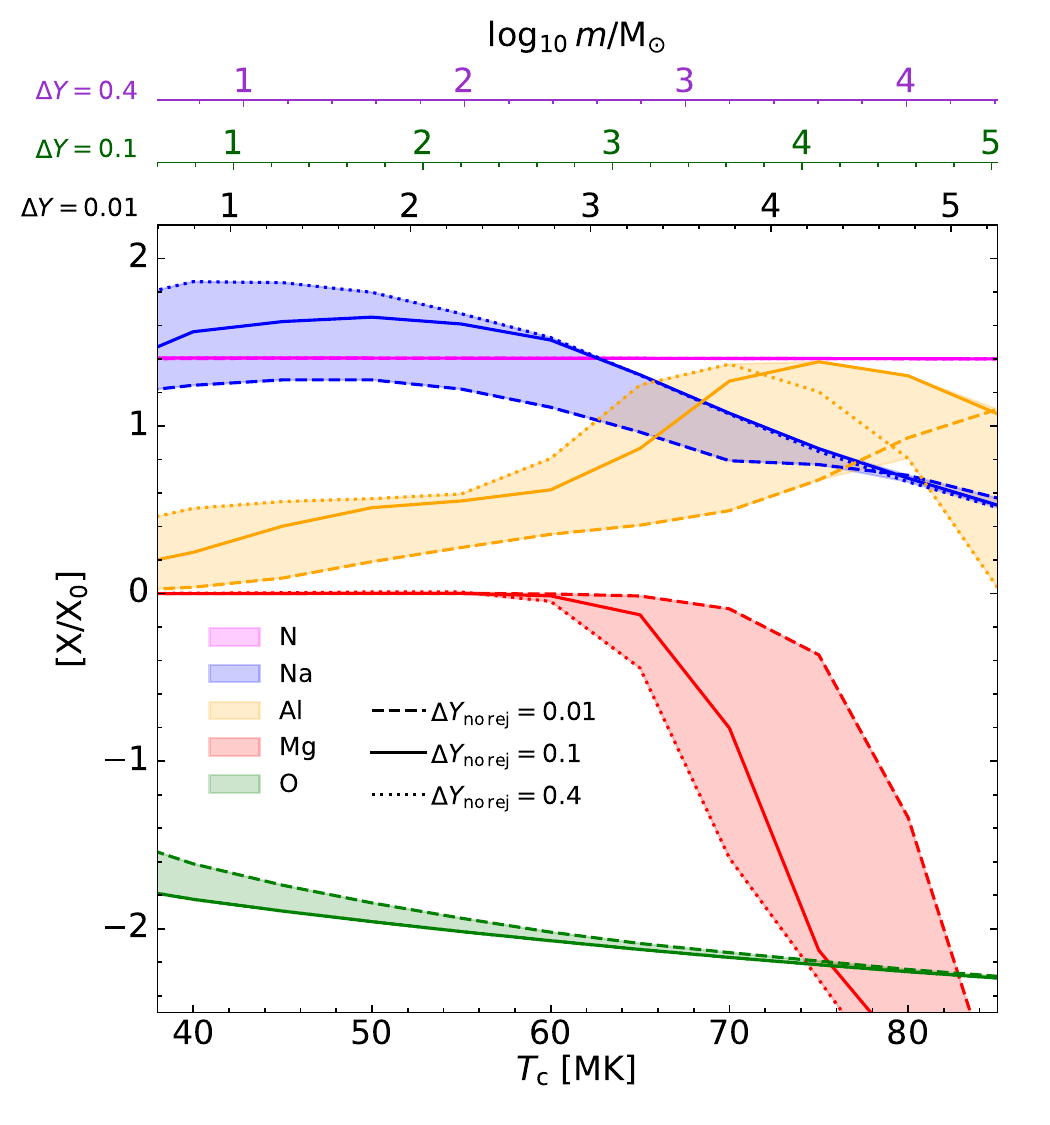}
\vspace{-5mm}
\caption{Relation between abundance variations and central temperature, $\Tc$, for different moments in the evolution quantified by $\DYnor$ from the nucleosynthesis results of \citet{2017A&A...608A..28P}. The top axes show corresponding stellar masses, $m$, for different $\Delta Y$, based on the relation shown in Fig.~\ref{fig:tcmod}. 
 All stars $\gtrsim10\,\msun$ are able to give rise to the O-Na anticorrelation, and masses $\gtrsim3\times10^3\,\msun$ are needed to get a significant Mg depletion and Al increase. }
\label{fig:tc_m}
\end{figure}

We use the nucleosynthesis results at fixed temperature and density of \citet{2017A&A...608A..28P}. Because these models did not consider rejuvenation, we interpret their $\Delta Y$ as our $\DYnor$ as a proxy of time.
Fig.~\ref{fig:tc_m}  shows the logarithmic changes in abundances of N, O, Na, Mg and Al, for three values of $\DYnor$. From this we see that the stars that dominate the wind mass loss ($\sim3\times10^3\,\msun$, $\Delta Y\simeq0.1$), have $\Tc\simeq69\,\MK$ and combined with their $\DYnor\simeq0.4$, we see that the Na and Al enhancement are a factor of $\Flin{Na}{wind}\simeq\Flin{Al}{wind}\simeq15$\footnote{We define linear abundances increase of some element El as $\Flin{El}{env}\equiv 10^{{\rm [El/El_0]}_{\rm env}}$, where `env' denotes the environment, such as `wind' or `GC'.}. From this we can make a rough estimate of how much dilution is required. The total Na enhancement of a typical GC is a factor of $\sim2-3$ \citep{2017A&A...601A..96L, 2009A&A...505..117C}, meaning that EMS yields require dilution with pristine material by a factor of $\sim10$. This required dilution factor is similar as what was needed for helium, which is an encouraging step towards addressing the relative abundance problem (\citealt*{2015MNRAS.449.3333B}; \citealt{2024A&A...690A.199V}).  

In Fig.~\ref{fig:abundances_mass} (middle panels) we show $\Flin{N}{wind}$, $\Flin{Na}{wind}$ and $\Flin{Al}{wind}$ as a function of stellar mass and time for the typical GC model. This shows in an alternative way that N is insensitive to $\Delta Y$, while Na is relatively insensitive to $\Delta Y$ and Al continues to increase with increasing $\Delta Y$. 

From Fig.~\ref{fig:tc_m} we also see that the most massive stars of $\sim5\times10^3\,\msun$ in our typical GC  give rise to a  Mg depletion of $\sim-1$\,dex (undiluted), that is, an order of magnitude higher than O depletion (undiluted).
Significant Mg depletion is only seen in the most massive and metal-poor GC and significant Al spreads are common \citep[][]{2009A&A...505..139C,2017A&A...601A.112P,2020MNRAS.492.1641M}. 
In the aEMS  model we expect a similar mass and metallicity dependence for the Mg spread, because
\begin{enumerate}
\item more massive GCs hosted more massive stars. The expected maximum stellar mass scales linearly as $\mfmax\propto\MGCi$ and if we include the effect of winds  we find that the actual maximum scales as $\minf\propto \MGCi^{3/8}$ (equations~\ref{eq:mdamax}, \ref{eq:mfmax} and \ref{eq:minf});
\item lower metallicity GCs have weaker winds leading to a higher $\minf$ for a given $\mdotacc$ (see Section~\ref{ssec:massevol}). For our adopted scaling $\mdwhz\propto Z^{0.6}$ we find $\minf\propto Z^{-0.3}$ (equation~\ref{eq:minf}).
\end{enumerate}
Both effects result in higher $\Tc$ in more massive and metal-poor GCs. We note that we do not include the effect that more metal-poor stars have a higher $\Tc$, for a given mass. In the metallicity range covered by Galactic GCs, the central temperature on the early main sequence for a star of a given mass would typically vary by a couple of million K, due to opacity effects. This would enhance the metallicity dependence. In Section~\ref{ssec:anticorrs} we present a detailed comparison between our model predictions and observed O-Na and Mg-Al anticorrelations of Milky Way GCs. 

%________________________
\subsection{Dilution and mass budget}
\label{ssec:dilution}
VMSs are not expected to be fully convective, so we assume they contribute hot-hydrogen burning yields only when they lose their radiative layer, after their accretion has stopped. Thus, we consider the wind mass loss  of accreting EMSs (aEMSs), and, for both VMSs and EMSs after accretion has stopped (hereafter pV/EMSs). 
The  wind material, containing hot-hydrogen burning yields, is released while the cluster is still deeply embedded in high-density gas. \citet{2018MNRAS.478.2461G} showed that high-velocity winds ($\gtrsim10^3\,\kms$) shock with dense, cold gas and subsequently cools to become available for star formation. The winds are expected to stall at distances of $\ll\pc$ \citep[section~4.1 of][]{2018MNRAS.478.2461G} and we can  therefore assume that the mixing of wind material occurs near each star. 
We made the assumption that 50 per cent of the gas that flows towards massive stars is not accreted, as the accretion requires angular momentum transport and thus mass loss through jets and accretion-disk winds (Section~\ref{ssec:timescale}, footnote \ref{fn:acceff}). Near each aEMS, we therefore have an outflow of gas with pristine abundances at an approximate rate $\mdotacc$ (protostellar disk winds and jets) mixing with an outflow of processed gas at a rate $|\mdotw|$. New (low-mass) P2 stars are assumed to form from this mixture of pristine and processed material, while the much denser gas accreting onto the EMSs (through large-scale inflows and eventually the circumstellar disk) is assumed to remain unmixed, at pristine composition.

Defining  dilution as one part of wind material is diluted by $\fdil$ parts of pristine material, then the dilution of aEMS winds with the outflowing gas with pristine abundances leads to $\fdil \simeq \mdotacc/|\mdotw|$, because the outflow rate equals the accretion rate.
This type of `local dilution' would lead to a population of low-mass stars with abundance patterns originating from a single aEMS. This is a promising way to create more than two discrete MPs in GCs (which we discuss further in Section~\ref{ssec:discreteness}). Because the physics of mixing of high-velocity winds with outflows is complex, and its outcome uncertain, we here use a simple dilution model, in which we only account for $\mdotw$:
\begin{equation}
\fdil(m) =  \frac{\langle\dot{m}_{\rm prist}\rangle}{|\mdotw|},
\label{eq:fdil}
\end{equation}
where $\langle\dot{m}_{\rm prist}\rangle$ is the average rate at which pristine gas is added to  the formation rate of low-mass stars. We compute this as
\begin{equation}
\langle\dot{m}_{\rm prist}\rangle = \frac{\Mdotlow - |\Mdwindpoll| }{N_{\rm poll}}.
\end{equation}
Here $\Mdotlow$ is the total star formation rate of low-mass stars, $\Mdwindpoll$ the total wind mass-loss rate of polluting stars (aEMS and pV/EMSs) and $N_{\rm poll}$ is the total number of polluting stars. Because $\Mdotlow$ is assumed to be constant during GC formation, the resulting dilution depends on stellar mass approximately as $\fdil\propto m^{-2}$ (from the stellar mass dependence of the wind, equation~\ref{eq:mdw}). The constant of proportionality reduces in time, because of the increase of $|\Mdwindpoll|$.
This physically motivated form for $\fdil$ results in less dilution of the winds of the most massive stars and at later times, such that the most extreme abundances originate from the most massive stars towards the end of GC formation. 

We have made the optimistic assumption that all wind material released before $\tdyn$ is used in low-mass  star formation and that all the material ends up in low-mass stars only. Diluting the wind material over the total IMF would reduce the wind abundances by a factor of $\sim3$. Distributing the winds only among low-mass stars is justified by the fact that massive stars are fed by pristine gas inflows from larger scales, while low-mass stars form locally, hence from gas inside the star forming cluster that is mostly mixed with the wind material. On the other hand, we neglect all the wind mass loss that occurs after $\tdyn$, which is a factor of $\sim2-3$ larger than the mass released before $\tdyn$, comparable to the abundance reduction factor from the dilution of the wind material with the full IMF. In our current model we do not use this mass, but it could contribute if star formation ceased gradually after $\tdyn$, rather than stopping abruptly as we assume now.

We illustrate the mass-dependent $\fdil$ at different times in the bottom panel of Fig.~\ref{fig:abundances_mass}. Lower mass \vemss\ (few $100\,\msun$) have high $\fdil\gtrsim10$, while the most massive EMSs have a much lower $\fdil\lesssim1$ which decreases with time. This mass-dependent dilution has important consequences for the shape of the O-Na and Mg-Al anticorrelations, which we  discuss in Section~\ref{ssec:anticorrs}.

\begin{figure}
\includegraphics[width=\columnwidth]{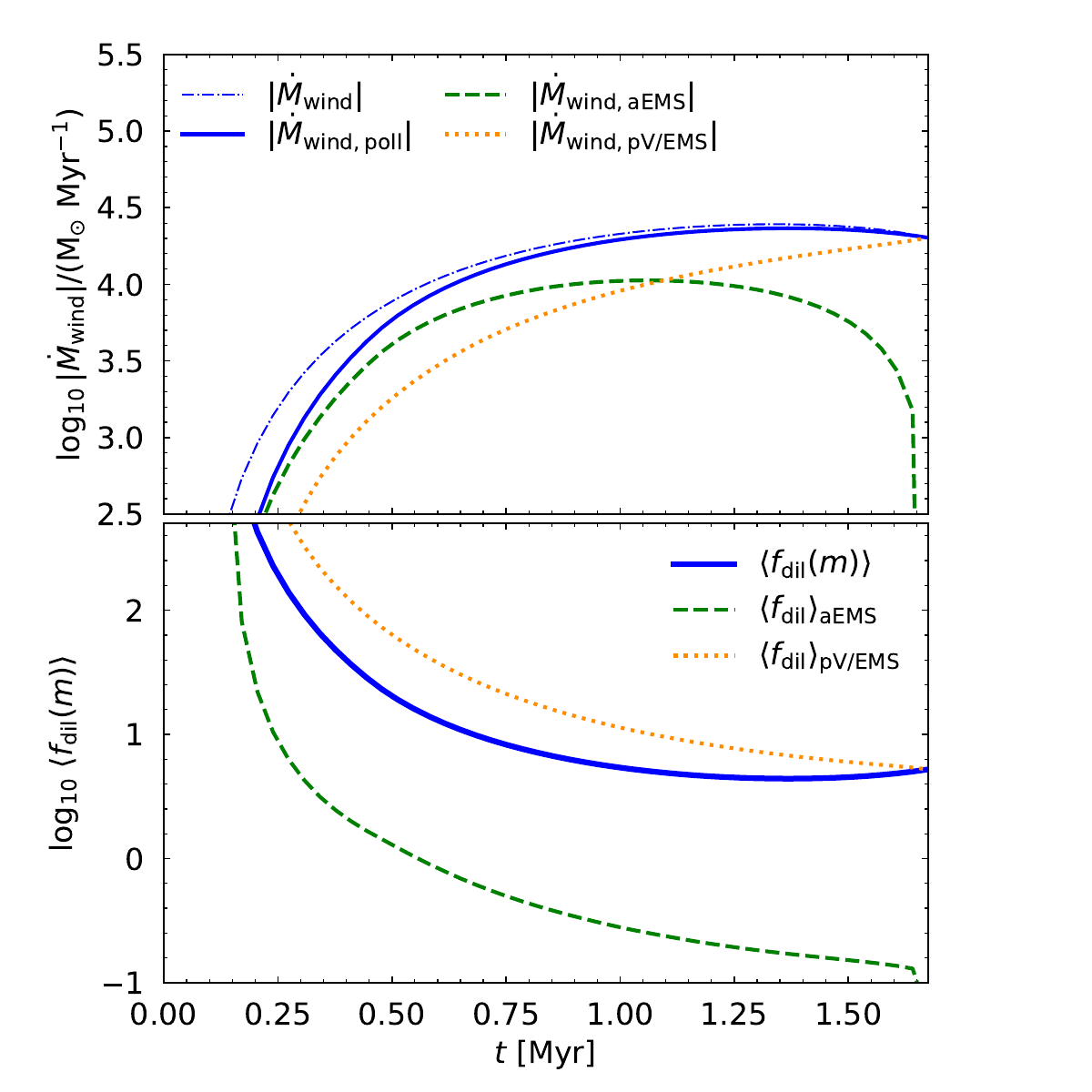}
\vspace{-5mm}
\caption{Top: Different contributions to the total wind mass-loss rate $|\Mdwind|$, and the wind mass loss rate of star contributing to the pollution, $\Mdwindpoll=\dot{M}_{\rm wind,aEMS}+ \dot{M}_{\rm wind,pV/EMS}$. Bottom: Average dilution factor (blue),  and those of the contributing components. See Section~\ref{ssec:dilution} for details.}
\label{fig:fdil}
\end{figure}

In Fig.~\ref{fig:fdil} (top) we show the wind mass-loss rates of the different contributions from aEMSs and pV/EMSs. We also compare $\Mdwindpoll=\dot{M}_{\rm wind, aEMS} + \dot{M}_{\rm wind, pV/EMS}$  to the total wind mass-loss rate $\Mdwind$.
From this we see that  accreting VMSs only  contribute a small fraction to  $\Mdwind$. 
We can also see that in approximately the first half of GC formation the dilution with pristine gas is high, and in the second half the average dilution factor settles at a value of $\mfdil\simeq5$ and the aEMSs have a typical $\fdil\simeq1$, from which we can conclude:
\begin{enumerate}
\item Approximately half of the low-mass stars (forming early on) have pristine composition and the other half (forming later) are polluted by winds, satisfying the observational mass budget constraint that $\fpoll\simeq0.5$ \citep{2017MNRAS.464.3636M};
\item The predicted $\Delta Y$ of a typical GC is $\DYobs\simeq 0.12/(1+\mfdil)\simeq0.02$, and the maximum $\DYmax\simeq0.12/(1+\fdilinf)\simeq0.06$,  in excellent agreement with what was found by  \citet{2018MNRAS.481.5098M} for Milky Way GCs;
\item The undiluted $\Flin{Na}{wind}\simeq\Flin{Al}{wind}\simeq15$ ($\Delta\NaFe\simeq\Delta\AlFe\simeq1.2$), when diluted 1:1, results in a  maximum $\Delta\NaFemax\simeq\Delta\AlFemax\simeq1$\,dex, in good agreement with the spreads found by \citet{2009A&A...505..117C,2009A&A...505..139C}; 
\item The undiluted $-1\,$dex Mg depletion, when diluted 1:1,  results in a (modest)  depletion of $\sim-0.26\,$dex, consistent with the fact that significant Mg depletion is only seen in the most massive, metal-poor GCs;
\item The average Na and Al abundance increase for the whole GC is $\Flin{Al}{GC}\simeq\Flin{Na}{GC} = (\mfdil + \Flin{Na}{wind})/(1+\mfdil)\simeq3.3$, which agrees well with the findings of  Na enhancement of MilkWay GCs of a factor of $\sim2-3$ \citep[$\Delta\NaFe\simeq0.3-0.5$,][]{2017A&A...601A..96L, 2009A&A...505..117C}.
\end{enumerate}
We next consider how the results of our typical GC change when varying GC mass and metallicity. 
 
%%%%%%%%%%%%%%%%%%%%%%%%%%%%%%%%%%%%%%%%%%%
\section{Abundance trends with GC mass and metallicity }
\label{sec:scaling}

%_________________________________________
\subsection{Model parameters}
The three main parameters that determine the abundance pattern in our model are the surface density of the gas, $\Sigmagas$, the initial GC mass, $\MGCi$, and the GC metallicity, $\feh$, through the wind mass-loss rates parameter $\mdwhz$. 
We note that the turbulence outer scale $\Lgasout$ only sets the mass of the most massive GC, and it does not affect the properties of lower mass GCs. We can then consider how GC formation depends on  $\Sigmagas$, $\MGCi$  and $\feh$ to  understand the origin of  observed abundance trends with $\MGC$ and $\feh$. We model GCs with $\feh = [-2.5, -2, -1.5, -1, -0.5, 0]$ and initial  masses $\log_{10}(\MGCi/\msun) = [4.5, 5, 5.5, 6, 6.5, 7.0]$. For these models we adopt $\Sigmagas=1.5\times10^3\,\msun\,\pc^{-2}$.
%____________
\subsection{Helium}
The first scaling we look at is the dependence of the undiluted helium spread, $\Delta Y$, on $\MGCi$ and $\feh$. 
In the top panel of Fig.~\ref{fig:scaling1} we show the average $\Delta Y$ in the mass released in winds by \vemss, $\DYw$, as a function of $\MGCi$ (left) and $\feh$ (right). The subscript `wind' indicates that this is the abundance of the material released in winds, that is, before dilution. We also show $\DYnor$ (dashed lines) which is what we  use for deriving other light element abundances (Section~\ref{ssec:other}). The increase of $\DYw$ with $\MGCi$ is because the time to form a GC, relative to $\tnuc$, scales as $\tdyn/\tnuc \propto (\Rgas^3/\Mgas)^{1/2}\propto \Sigmagas^{-3/4}\MGCi^{1/4}$, that is, $\tdyn/\tnuc\propto \MGCi^{1/4}$ because $\Sigmagas$ is assumed to be constant. For almost all models $\tdyn<\tnuc$,  such that  helium steadily increases during GC formation, and \vemss\ in more massive GCs (with longer $\tdyn$), therefore, have higher helium abundances.
As such,  in higher gas-density environments, GCs form faster, allowing less time for helium to increase. This is an important aspect of the aEMS model: at higher $\Sigmagas$ we expect a lower $\DYw$, but other light-element abundances are not sensitive to $\Sigmagas$, because the nuclear reaction chains reach equilibrium on a timescale much shorter than the nuclear timescale. The ability to vary the helium spread, independently of other light-elements abundances (that is, before dilution that affects all elements in the same way), allows our model to address the `relative abundance' problem pointed out by \citet{2015MNRAS.449.3333B,2024A&A...690A.199V}. 

The modest (factor of $\sim3$) increase of $\DYw$ over two decades of $\MGCi$ seen in Fig.~\ref{fig:scaling1} is the result of dependence of $\tdyn$ on $\MGC^{1/4}$. We also find a modest decrease of $\DYw$ with $\feh$. As far as we are aware, no dependence of the He spread on $\feh$ has been reported. But we recall that this is the He spread in the wind material.
To translate $\DYw$ to the observed He spread, $\DYobs$, we need to consider how dilution depends on $\MGC$ and $\feh$.

%%%%%%%%%%%%%
\subsubsection{Dilution}
\label{sssec:dilution}
\begin{figure*}
\includegraphics[width=2\columnwidth]{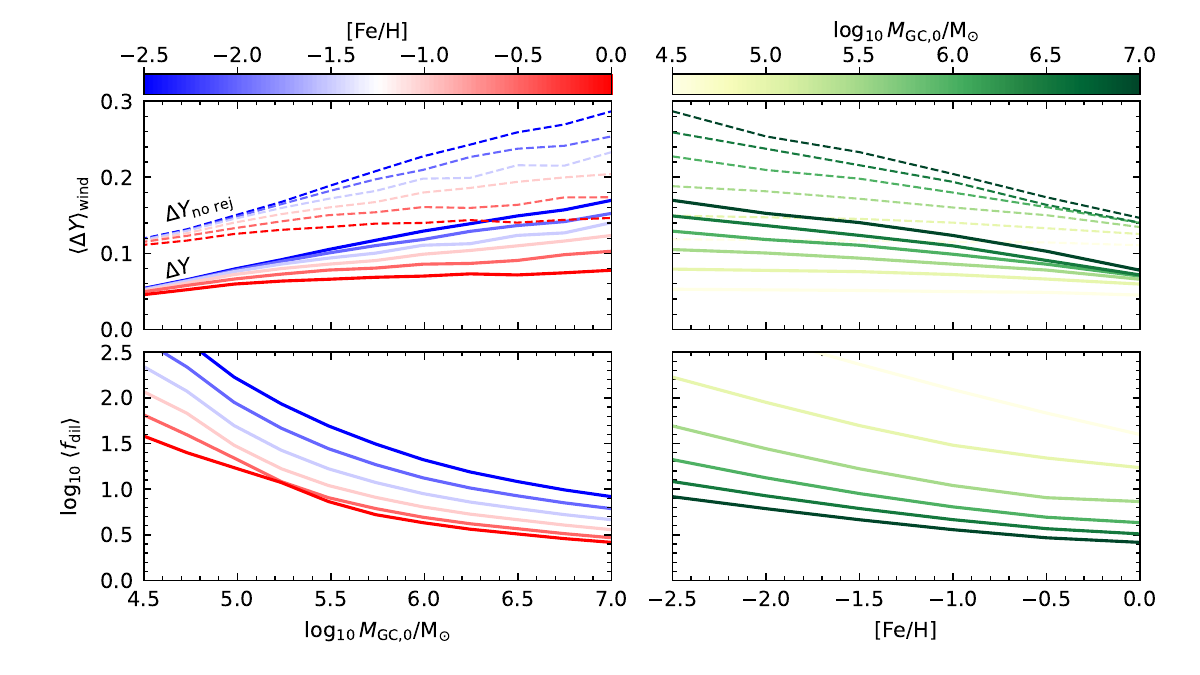}
\vspace{0mm}
\caption{Wind mass-loss averaged $\Delta Y$ ($\DYw$, top) and average $\fdil$ (bottom) for different $\MGC$ and $\feh$. Note that $\DYw$ is before dilution, so it is the maximum $\Delta Y$ a P2 star can have, if it formed for 100\% of wind material.}
\label{fig:scaling1}
\end{figure*}

In the bottom panel of Fig.~\ref{fig:scaling1}  we show the dependence of $\fdil$ on $\MGCi$ and $\feh$. There is a clear decrease of $\fdil$ with increasing $\MGCi$ and $\feh$ because of their higher mass loss rates. 

The observed helium spread, $\DYobs=\DYw/(1+\fdil)$, which enhances the $\DYw(\MGCi)$ correlation and reduces/cancels the $\DYw(\feh)$ anticorrelation. We  discuss a direct comparison with observations in Section~\ref{ssec:helium}.

%___________________
\subsection{Maximum  mass and central temperatures}
As explained in Section~\ref{ssec:other}, the masses of the most massive EMSs in the model depend on both $\MGCi$ and $\feh$. Here we quantify how  their central temperatures, $\Tc$, depend on these GC properties.  

\begin{figure*}
\includegraphics[width=2\columnwidth]{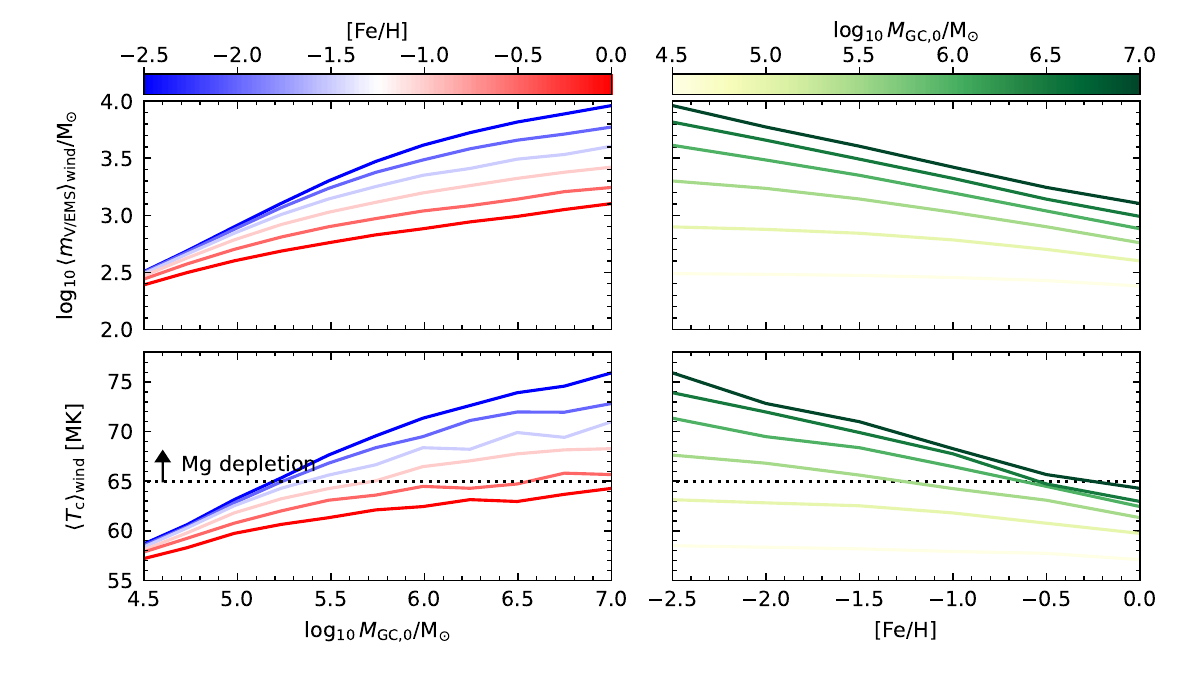}
\vspace{0mm}
\caption{Wind mass-loss averaged \vems\ mass ($\mvmsw$, top) and central temperature ($\Tcw$, bottom) for different $\MGC$ and $\feh$.  Critical $\Tc$ for Mg depletion to become important is shown as a horizontal dashed line. Massive and metal-poor GCs had EMSs with $\Tc$ above this critical temperature and should therefore be Mg depleted, as is observed \citep[for example,][]{2009A&A...505..139C,2017A&A...601A.112P}.}
\label{fig:scaling3}
\end{figure*}

In the top panel of Fig.~\ref{fig:scaling3} we show wind mass  averaged \vems\ mass, $\mvmsw$, as a function of $\MGCi$ and $\feh$. In the bottom panel we show the corresponding wind mass averaged $\Tc$, $\Tcw$. Note that this is not the temperature of the wind, but the central temperature at which the material released in winds was processed. This shows that for more massive and more metal-poor GCs, the material in  winds was processed at higher temperatures. We note  that we do not include the effect that metal-poor stars (at fixed mass) are hotter. The $\Tc$ dependence is because metal-poor GCs had more massive stars, due to their lower $\mdwhz$, allowing EMSs to grow to larger masses (for a given accretion rate). 
The dependence of $\Tcw$ on $\MGCi$ and $\feh$ provides a qualitative  explanation for the observed increase of the Al spread with increasing(decreasing) $\MGC$($\feh$) found by  \citet{2009A&A...505..139C,2017A&A...601A.112P}. In Section~\ref{ssec:anticorrs} we further quantify this by directly comparing abundance distributions  from the aEMS model to observed anticorrelations. 

%_______________________________
\section{Comparison to observations}
\label{ssec:obs}
\begin{table}
\caption{Data of Milky Way GCs used in Fig.~\ref{fig:scaling2}. Initial GC masses are found from their present-day mass and their orbits by \citet{2019MNRAS.482.5138B} and $\DYobs$ and $\fpoll$ are taken from \citet{2018MNRAS.481.5098M} and \citet{2017MNRAS.464.3636M}, respectively. Details are given in the text of Section~\ref{sssec:data}. \label{tab:mwgc}}
\begin{tabular}{lcclrr}
\hline
     Name & $\log_{10}\MGC\!$ & $\log_{10}\MGCi\!\!$ &   [Fe/H]& $\DYobs$ & $\fpoll$ \\
          & $[{\rm M}_\odot]$ & $[{\rm M}_\odot]$ &         &          &         \\ \hline
    IC4499 & 5.18 & 5.53 & $ -1.53$ & $ 0.004$ & $     -$ \\ 
 NGC\,0104 & 5.95 & 6.24 & $ -0.72$ & $ 0.011$ & $  0.82$ \\ 
 NGC\,0288 & 4.94 & 5.60 & $ -1.32$ & $ 0.015$ & $  0.46$ \\ 
 NGC\,0362 & 5.44 & 6.05 & $ -1.26$ & $ 0.008$ & $  0.72$ \\ 
 NGC\,1261 & 5.26 & 5.75 & $ -1.27$ & $ 0.004$ & $  0.64$ \\ 
 NGC\,1851 & 5.51 & 6.02 & $ -1.18$ & $ 0.007$ & $  0.74$ \\ 
 NGC\,1904 & 5.26 & 6.46 & $ -1.60$ & $     -$ & $     -$ \\ 
 NGC\,2298 & 4.70 & 5.78 & $ -1.92$ & $-0.003$ & $  0.63$ \\ 
 NGC\,2808 & 5.99 & 6.29 & $ -1.14$ & $ 0.048$ & $  0.77$ \\ 
 NGC\,3201 & 5.26 & 5.61 & $ -1.59$ & $-0.001$ & $  0.56$ \\ 
 NGC\,4590 & 5.11 & 5.46 & $ -2.23$ & $ 0.007$ & $  0.62$ \\ 
 NGC\,4833 & 5.27 & 6.01 & $ -1.85$ & $ 0.016$ & $  0.64$ \\ 
 NGC\,5024 & 5.72 & 6.00 & $ -2.10$ & $ 0.013$ & $  0.67$ \\ 
 NGC\,5053 & 4.80 & 5.21 & $ -2.27$ & $-0.002$ & $  0.46$ \\ 
 NGC\,5139 & 6.51 & 6.90 & $ -1.53$ & $ 0.033$ & $  0.91$ \\ 
 NGC\,5272 & 5.69 & 5.93 & $ -1.50$ & $ 0.016$ & $  0.70$ \\ 
 NGC\,5286 & 5.63 & 6.12 & $ -1.69$ & $ 0.007$ & $  0.66$ \\ 
 NGC\,5466 & 4.75 & 5.20 & $ -1.98$ & $ 0.002$ & $  0.53$ \\ 
 NGC\,5897 & 5.22 & 5.72 & $ -1.90$ & $     -$ & $  0.45$ \\ 
 NGC\,5904 & 5.59 & 5.94 & $ -1.29$ & $ 0.012$ & $  0.76$ \\ 
 NGC\,5927 & 5.47 & 5.86 & $ -0.49$ & $ 0.011$ & $     -$ \\ 
 NGC\,5986 & 5.47 & 6.25 & $ -1.59$ & $ 0.005$ & $  0.75$ \\ 
 NGC\,6093 & 5.48 & 6.14 & $ -1.75$ & $ 0.011$ & $  0.65$ \\ 
 NGC\,6101 & 5.24 & 5.56 & $ -1.98$ & $ 0.005$ & $  0.35$ \\ 
 NGC\,6121 & 4.95 & 6.01 & $ -1.16$ & $ 0.009$ & $  0.72$ \\ 
 NGC\,6144 & 4.93 & 5.73 & $ -1.76$ & $ 0.009$ & $  0.56$ \\ 
 NGC\,6171 & 4.80 & 5.81 & $ -1.02$ & $ 0.019$ & $  0.60$ \\ 
 NGC\,6205 & 5.63 & 6.11 & $ -1.53$ & $ 0.020$ & $  0.82$ \\ 
 NGC\,6218 & 5.00 & 5.69 & $ -1.37$ & $ 0.009$ & $  0.60$ \\ 
 NGC\,6254 & 5.32 & 5.83 & $ -1.56$ & $ 0.006$ & $  0.64$ \\ 
 NGC\,6304 & 5.01 & 5.79 & $ -0.45$ & $ 0.008$ & $     -$ \\ 
 NGC\,6341 & 5.48 & 5.98 & $ -2.31$ & $ 0.022$ & $  0.70$ \\ 
 NGC\,6352 & 4.99 & 5.49 & $ -0.64$ & $ 0.019$ & $  0.53$ \\ 
 NGC\,6362 & 5.05 & 5.65 & $ -0.99$ & $ 0.003$ & $  0.43$ \\ 
 NGC\,6366 & 4.51 & 5.47 & $ -0.59$ & $ 0.011$ & $  0.58$ \\ 
 NGC\,6388 & 6.12 & 6.48 & $ -0.55$ & $ 0.019$ & $  0.76$ \\ 
 NGC\,6397 & 5.03 & 5.57 & $ -2.02$ & $ 0.006$ & $  0.66$ \\ 
 NGC\,6441 & 6.14 & 6.48 & $ -0.46$ & $ 0.029$ & $     -$ \\ 
 NGC\,6496 & 4.87 & 5.54 & $ -0.46$ & $ 0.009$ & $  0.33$ \\ 
 NGC\,6535 & 4.30 & 5.77 & $ -1.79$ & $ 0.003$ & $  0.46$ \\ 
 NGC\,6541 & 5.34 & 5.97 & $ -1.81$ & $ 0.024$ & $  0.60$ \\ 
 NGC\,6584 & 5.06 & 5.60 & $ -1.50$ & $ 0.000$ & $  0.55$ \\ 
 NGC\,6624 & 5.01 & 6.41 & $ -0.44$ & $ 0.010$ & $  0.72$ \\ 
 NGC\,6637 & 5.14 & 6.30 & $ -0.64$ & $ 0.004$ & $  0.57$ \\ 
 NGC\,6652 & 4.61 & 6.33 & $ -0.81$ & $ 0.008$ & $  0.66$ \\ 
 NGC\,6656 & 5.67 & 6.04 & $ -1.70$ & $ 0.005$ & $  0.73$ \\ 
 NGC\,6681 & 4.98 & 5.97 & $ -1.62$ & $ 0.009$ & $  0.77$ \\ 
 NGC\,6715 & 6.20 & 6.47 & $ -1.49$ & $ 0.012$ & $  0.73$ \\ 
 NGC\,6717 & 4.42 & 5.97 & $ -1.26$ & $ 0.003$ & $  0.36$ \\ 
 NGC\,6723 & 5.25 & 5.88 & $ -1.10$ & $ 0.005$ & $  0.64$ \\ 
 NGC\,6752 & 5.32 & 5.84 & $ -1.54$ & $ 0.015$ & $  0.71$ \\ 
 NGC\,6779 & 5.23 & 5.85 & $ -1.98$ & $ 0.011$ & $  0.53$ \\ 
 NGC\,6809 & 5.26 & 5.88 & $ -1.94$ & $ 0.014$ & $  0.69$ \\ 
 NGC\,6838 & 4.58 & 5.24 & $ -0.78$ & $ 0.005$ & $  0.38$ \\ 
 NGC\,6934 & 5.18 & 5.60 & $ -1.47$ & $ 0.006$ & $  0.67$ \\ 
 NGC\,6981 & 4.91 & 5.89 & $ -1.42$ & $ 0.011$ & $  0.46$ \\ 
 NGC\,7078 & 5.79 & 6.05 & $ -2.37$ & $ 0.021$ & $  0.60$ \\ 
 NGC\,7089 & 5.80 & 6.26 & $ -1.65$ & $ 0.013$ & $  0.78$ \\ 
 NGC\,7099 & 5.15 & 5.76 & $ -2.27$ & $ 0.015$ & $  0.62$ \\
\hline
\end{tabular}
\end{table}

In this section we compare aEMS model predictions to observations.
%_____
\subsection{Description of the data}
\label{sssec:data}
\subsubsection{Photometry}
We use the  difference in the helium abundance between P1 and P2, $\DYobs$, for 57 Milky Way GCs from \citet[][their table~4]{2018MNRAS.481.5098M} and $\fpoll$
for 53 Milky Way GCs in \citet[][$\fpoll = 1-N_1/N_{\rm tot}$, with $N_1/N_{\rm tot}$ from their table 2]{2017MNRAS.464.3636M}. Combining these results yields 58 unique GCs and for 53 of those  both $\DYobs$ and $\fpoll$ are available. Iron abundances for all GCs are taken from the Harris catalogue \citep[][\citeyear{2010arXiv1012.3224H} edition]{1996AJ....112.1487H}. For present-day GC masses we use the results of \citet[][33 GCs]{2023MNRAS.522.5320D}, complimented by values from the catalogue of \cite{2018MNRAS.478.1520B}. We use initial GC mass estimates from  \citet{2019MNRAS.482.5138B}. 
The data is summarised in Table~\ref{tab:mwgc}.

\subsubsection{Spectroscopy}
We focus on  GCs with O, Na, Mg and Al abundance available from high-resolution spectroscopy (UVES@VLT) from \citet{2009A&A...505..139C}, complemented with lower resolution O-Na data from GIRAFFE  \citep{2009A&A...505..117C} and Mg-Al data from APOGEE     \citep{2020MNRAS.492.1641M}. From the 19 GCs with UVES data, we focus on 16 GCs that have detections for O, Na, Mg and Al in both high and lower resolution data\footnote{There are no Al data for NGC\,6397, and no APOGEE data for NGC\,7099 and we drop NGC\,6441 because there are only 5 stars for this cluster and it has a very similar mass and $\feh$ as NGC\,6388.}. 

%__________________-
\subsection{Helium}
\label{ssec:helium}
\begin{figure*}
\includegraphics[width=2\columnwidth]{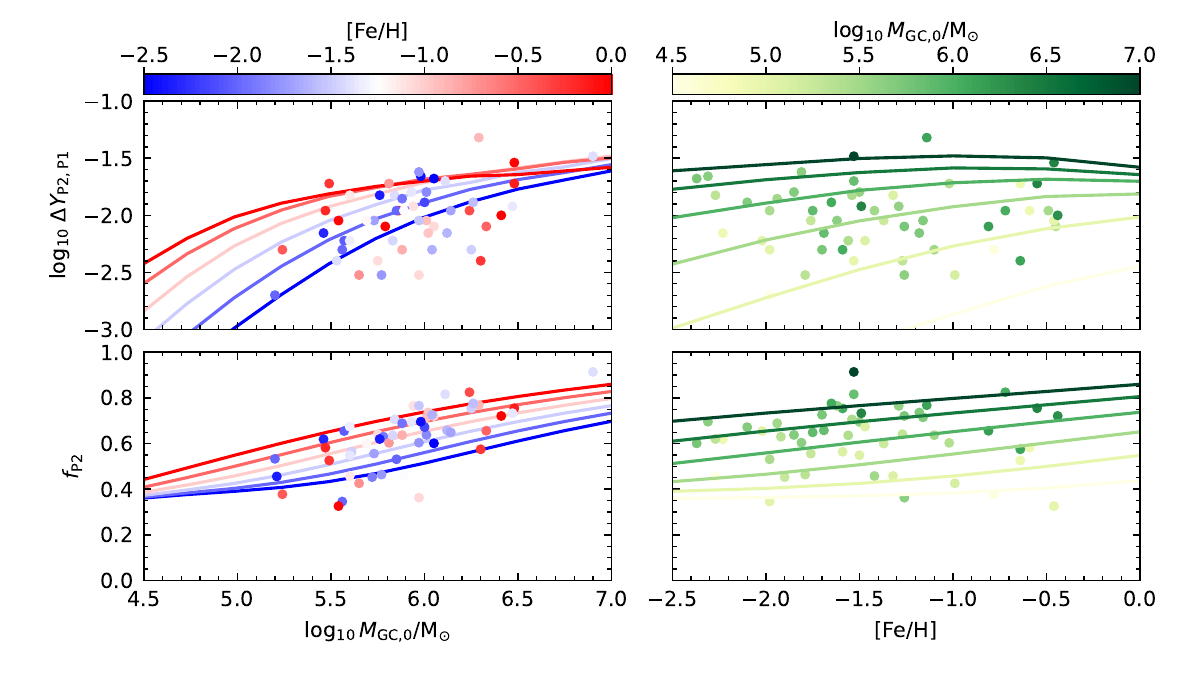}
\vspace{0mm}
\caption{Helium spread ($\DYobs$, top) and fraction of polluted stars ($\fpoll$, bottom) for different $\MGCi$ and $\feh$ in the aEMS model. 
Data points are Milky Way GCs from \citet{2018MNRAS.481.5098M} (top) and \citet{2017MNRAS.464.3636M} (bottom). The aEMS model reproduces observed typical $\DYobs\simeq0.01$ and $\fpoll\simeq0.5$ as well as the increase of both $\DYobs$ and $\fpoll$ with GC mass. }
\label{fig:scaling2}
\end{figure*}

\begin{figure}
\includegraphics[width=\columnwidth]{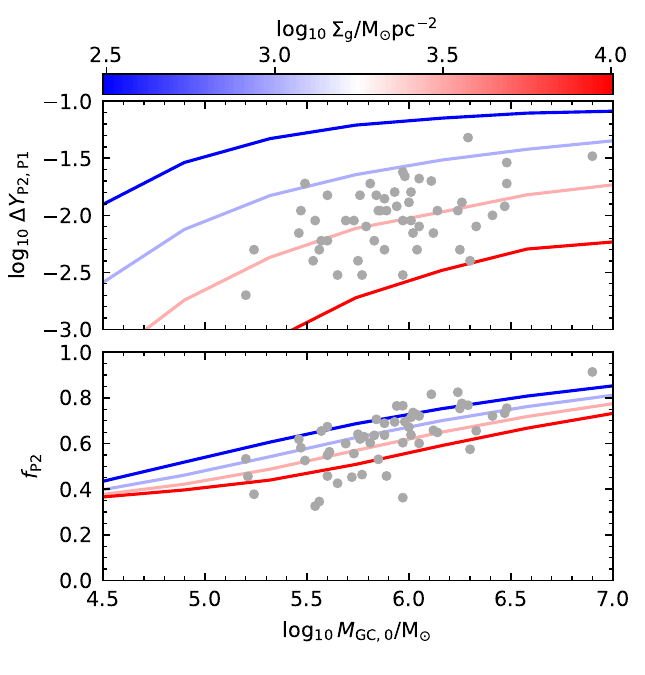}
\vspace{0mm}
\caption{Variation of $\DYobs$ (top) and $\fpoll$ (bottom) for different $\Sigmagas$. All models adopted $\feh=-1$. Grey dots are Milky Way GCs, also shown in Fig.~\ref{fig:scaling2}. This shows that $\DYobs$ is more sensitive to the gas density than $\fpoll$. }
\label{fig:scaling_Sigma}
\end{figure}

In Fig.~\ref{fig:scaling2} (top panel) we compare $\DYobs$ of the  model to the observations. We define the time when P1 star formation stops and P2 star formation starts as the moment when $|\Mdwind|$ reaches half of its maximum value. This definition is somewhat arbitrary, and using lower(higher) fraction leads to higher(lower) $\fpoll$\footnote{We note that the empirical $\fpoll$ is also uncertain, with the values determined from APOGEE data of the outskirts of GCs by \citet[][their figure 4]{2023MNRAS.525.4456B} being approximately half the values derived  by \citet{2017MNRAS.464.3636M} from Hubble Space Telescope ({\it HST}) observations of the central regions of GCs. }. It nevertheless serves as a practical separation between P1 and P2.
The observed increase of $\DYobs$ with $\MGCi$ is stronger then the scaling of the undiluted $\DYw$ (Fig.~\ref{fig:scaling1}). As discussed above,  this is because of the decreasing $\fdil$ with increasing $\MGCi$. The resulting slope matches remarkably well with the observed scaling $\DYobs(\MGCi)$. 
The fact that the vertical offset is reproduced so well is because it is sensitive to $\Sigmagas$, whose value we picked to match these observations (as mentioned in Section~\ref{ssec:typical}).
There is more spread in the observations than in our model, which might be due to our assumption of a constant $\Sigmagas$ for all GCs. In reality, there is a distribution of $\Sigmagas$ from which GCs form, and $\Sigmagas$ may  depend on $\feh$ and/or $\MGCi$. We note that a factor of three higher(lower) $\Sigmagas$  captures the lower(upper) envelope of the data points (Fig.~\ref{fig:scaling_Sigma}).

%_____________________________________
\subsection{Fraction of polluted stars ($\fpoll$)}
\label{ssec:fp2}
In the bottom panel of Fig.~\ref{fig:scaling2} we show the fraction of P2 stars in GCs ($\fpoll$).
In the model we find $\fpoll$ by using the same separation between P1 and P2 as  in Section~\ref{ssec:helium}. 
We  find an increase of $\fpoll$ with $\MGCi$ in the model.
This follows from the timescale on which \vems\ winds and accretion  reach an equilibrium ($\tinf$, equation~\ref{eq:tinf}) relative to $\tdyn$. For larger ratio $\tdyn/\tinf$, a larger fraction of low-mass stars  form after $\tinf$, and this ratio  therefore correlates with $\fpoll$.
From equation~(\ref{eq:tinf}) we see that for the most massive stars $\tinf\propto \mdotaccmax^{-1/2}$. Because $\mdotaccmax \propto \Mgas/\tdyn \propto  (\Sigmagas\MGCi)^{3/4}$, we then find $\tinf \propto (\Sigmagas\MGCi)^{-3/8}$.  Then the ratio $\tdyn/\tinf \propto \Sigmagas^{-3/8} \MGCi^{5/8}$. Thus, at higher $\Sigmagas$, winds become important later in the GC formation process, but the dependence on $\Sigmagas$ is not very strong. Increasing $\Sigmagas$ by an order of magnitude reduces $\fpoll$ by $\sim0.05$ and still reasonably reproduces the observations (see Fig.~\ref{fig:scaling_Sigma}). Hence, $\Sigmagas$ needs to be high enough to keep $\Delta Y$ low (Section~\ref{ssec:helium}), but not too high such that $\fpoll$ reduces by too  much. The helium spread combined with the vertical scaling of the observed $\fpoll(\MGCi)$ are therefore a measure of the surface density of the gas that GCs formed from. 
We note that $\Sigmagas$ is the average surface density within $\Rgas$, and central gas densities can therefore be higher. The dependence of both $\DYobs$ and $\fpoll$ on $\Sigmagas$ means that the adopted value of $\Sigmagas=1.5\times10^3\,\msun\,\pc^{-2}$ has now  become empirically constrained by the properties of MPs in MW GCs. The fact that this value is similar to what is found in star forming galaxies at high redshift mentioned in Section~\ref{ssec:scaling_to_GC} \citep{2010Natur.463..781T,2023MNRAS.520.2180C}, lends support to the aEMS model presented here.

The increase of $\tdyn/\tinf$ with $\MGCi^{5/8}$ explains why $\fpoll$ increases with $\MGCi$. 
Also for the dependence of $\fpoll$ on $\MGCi$ and $\feh$ we find good agreement between our model and observations of Milky Way GCs by \citet{2017MNRAS.464.3636M}.
The absolute minimum (initial) GC mass to posses MPs in the aEMS model is $\MGCi\gtrsim10^4\,\msun$, because these clusters barely have VMSs (equation~\ref{eq:mfmax}).
A minimum mass of $\sim10^5\,\msun$ is often mentioned as a minimum mass for MPs to occur \citep[for example,][]{2016A&A...587A..53K}, in
 practise it is hard to find this minimum GC mass, because it is comparable or lower than the mass loss by evaporation of a typical GC \citep[for example,][]{2023MNRAS.522.5340G}. Also, the dilution factors are $\fdil\gtrsim100$ for GCs with $\MGCi\simeq10^5\,\msun$ (Fig.~\ref{fig:scaling1}), making it very hard to prove the absence of MPs in low-mass GCs. There is therefore no strict mass limit to the existence of MPs, and a detection depends on the precision of the observations. 

We note a mild increase of $\fpoll$ with $\feh$ both in the model as well as in the observations. We are not aware that this observed $\feh$ dependence of $\fpoll$ has been noted in literature, but it is well reproduced by the model. 
In the model the increase is the result of the lower $\mdwhz$ at lower $\feh$. Because $\tinf\propto\mdwhz^{-1/2}$ (equation~\ref{eq:tinf}) and $\mdwhz\propto Z^{0.6}$, such that $\tinf\propto Z^{-0.3}$ and hence $\tdyn/\tinf\propto Z^{0.3}$. This correlation is the cause for the increase of $\fpoll$ with $\feh$. We note, however, that the correlation in the data is not significant: the Pearson correlation coefficient is $\sim0$ for all GCs and $\sim0.2$ for  GCs with $\MGCi>10^6\,\msun$. We note that \citet{2010A&A...516A..55C} found that the fraction of `extreme' stars correlates with $\feh$, but no correlation was found for the fraction `intermediate' population.
%

%________________________________
\subsection{O-Na and Mg-Al anticorrelations}
\label{ssec:anticorrs}
\begin{figure*}
\includegraphics[width=2\columnwidth]{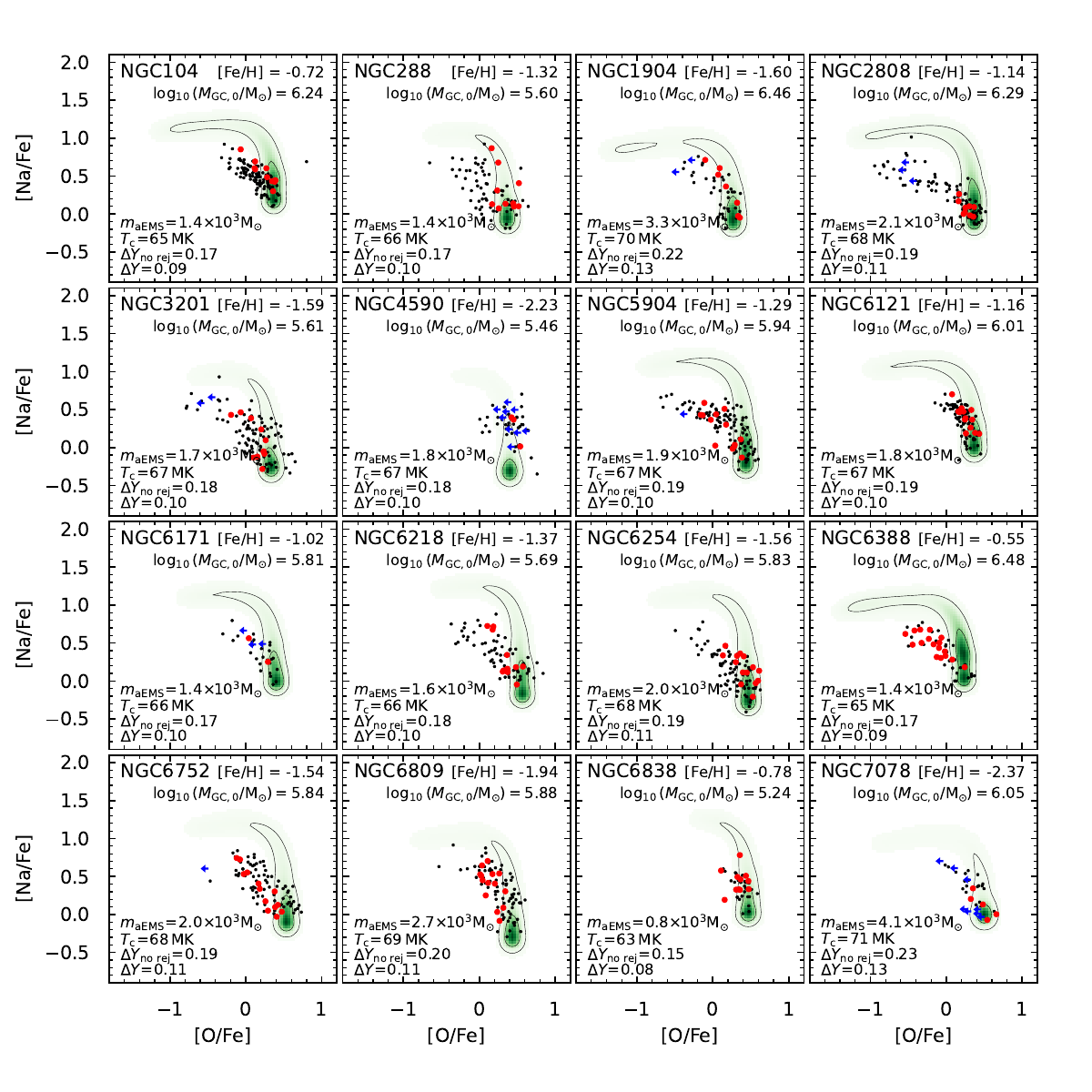}
\vspace{-5mm}
\caption{O-Na abundances for stars in 16 GCs derives from high-resolution UVES spectra by  \citet{2009A&A...505..139C}. Red dots show detections for both Na and O, and blue arrows indicate stars for which only an upper limit for O was available. Black dots  show results from lower resolution GIRAFFE data \citep{2009A&A...505..117C}. Overplotted are  histograms of the model abundances, with the intensity proportional to the  mass in each bin. Contours indicate regions containing 50\% and 90\% of the mass. See Section~\ref{ssec:anticorrs} for details. } \label{fig:anticorrs1}
\end{figure*}

\begin{figure*}
\includegraphics[width=2\columnwidth]{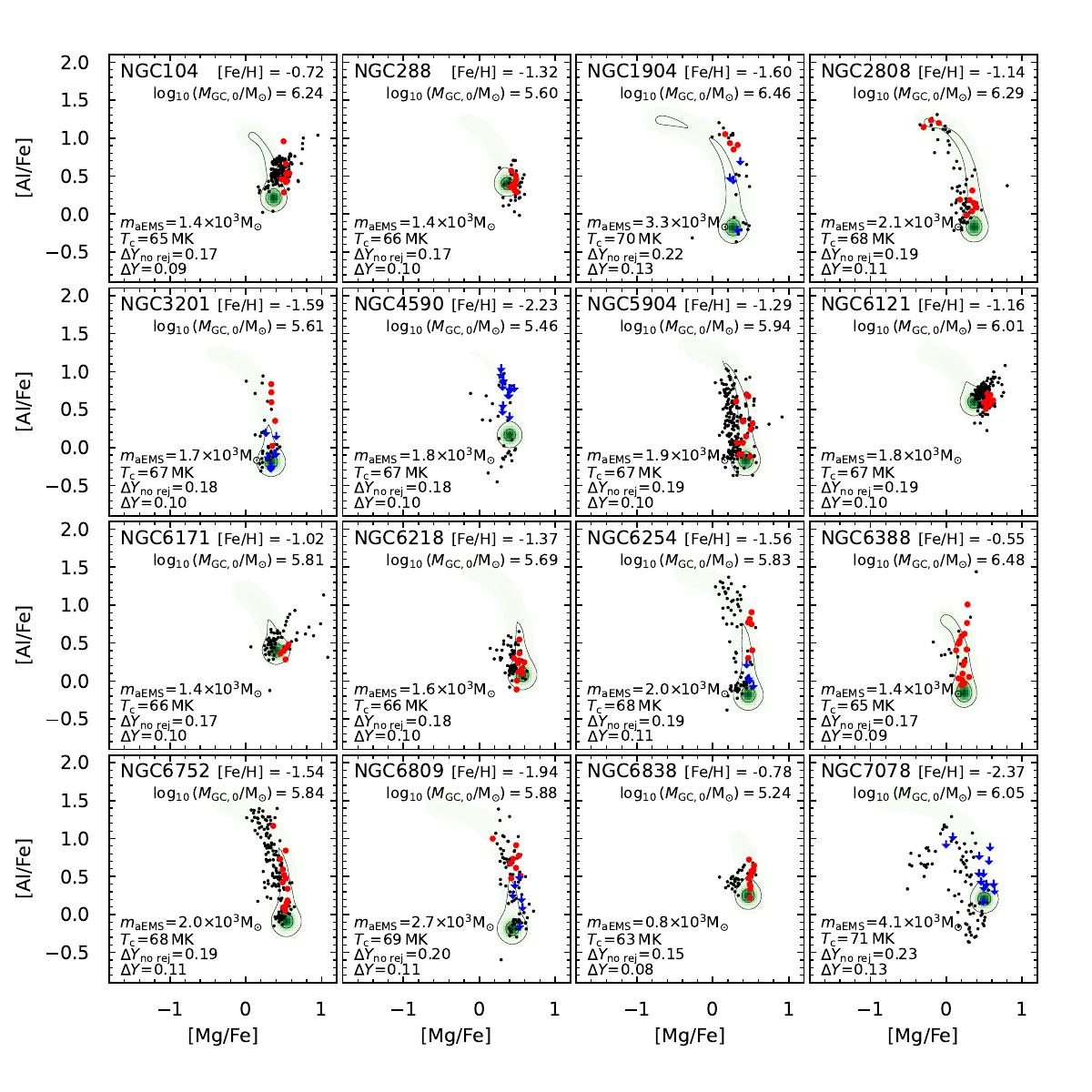}
\vspace{-5mm}
\caption{Mg-Al abundances for stars in 16 GCs derives from high-resolution UVES spectra by  \citet{2009A&A...505..139C}. Red dots show detections for both Mg and Al, and blue arrows indicate stars for which only an upper limit for Al was available. Black dots  show results from lower resolution APOGEE data \citep{2020MNRAS.492.1641M}. Overplotted are  histograms of the model abundances, with the intensity proportional to the  mass in each bin.  Contours indicate regions containing 68\% and 95\% of the mass. See  Section~\ref{ssec:anticorrs} for details. }
\label{fig:anticorrs2}
\end{figure*}

In this section we compare abundance distributions from the aEMS model to observed O-Na and Mg-Al abundances of stars in Milky Way GCs. 
We show with red dots the O-Na and Mg-Al abundances from UVES in Figs.~\ref{fig:anticorrs1} and \ref{fig:anticorrs2}, respectively. Upper limits for $\OFe$ and $\AlFe$ are shown with blue arrows.  Small black dots  in Fig.~\ref{fig:anticorrs1}(Fig.~\ref{fig:anticorrs2}) show the GIRAFFE(APOGEE) data.

To model the anticorrelations, we use for each GC $\MGCi$ and $\feh$ from Table~\ref{tab:mwgc} as computed in Section~\ref{sssec:data} (NGC\,1904 is the only GC of the Carretta sample that does not appear in the Milone data).
We create a model for the MPs for each of the 16 GCs based on their $\MGCi$, $\feh$ and $\Sigmagas=1.5\times10^3\,\msun\,\pc^{-2}$. For each model we obtain $\Tc$ and $\DYnor$ for each stellar mass and 30 equally spaced times, which we use to interpolate the nucleosynthesis results from \citet{2017A&A...608A..28P}. We use $\DYnor$, rather than $\Delta Y$, because in \citet{2017A&A...608A..28P} $\Delta Y$ serves as a proxy for time, and in the model $\Delta Y$ remains low because of rejuvenation and using this would give us abundances too early in the evolution of the star. 
These nucleosynthesis calculations are done at constant temperature and for $\feh=-1.5$\footnote{Prantzos (2024, private communication) made us aware that $\feh$ and initial mass fraction abundances in \citet{2017A&A...608A..28P} are the same as in \citet{2007A&A...470..179P}.}. This gives us O, Na, Mg and Al abundances for the  yields as a function of mass and time,  which we then dilute with pristine material using the dilution function described in Section~\ref{ssec:dilution}. For the pristine abundances, we adopt a constant $\NaFe_0$ and $\AlFe_0$ from \citep{2009A&A...505..139C} and $\OFe_0$ from \citet{2009A&A...505..117C}, which were derived from fitting dilution curves to the anticorrelations. After visual inspection, we updated  $\OFe_0$ for NGC\,4590  to $\OFe_0=0.4$ and $
\AlFe_0$ for NGC\,1904, NGC\,2808, NGC\,3201, NGC\,6254, NGC\,6809 and NGC\,6388 to $
\AlFe_0=-0.2$. The Prantzos models are based on $\OFe_0=\MgFe_0 =0$ and $\NaFe_0=\AlFe_0=0$. To model clusters with different initial abundances, we assume that the $\alpha$-elements scale similarly compare to Fe, that is, $\NeFe_0=\OFe_0=\MgFe_0$ and that
\begin{itemize}
\item  the final Na abundance is proportional to the initial Ne abundance;
\item the final Al abundance is proportional to the initial Mg abundance.
\end{itemize}
We then bin the resulting abundances in bins of 0.033 dex and blur the resulting 2-dimensional histograms with a Gaussian with two pixel width, so that the model `resolution' is $\sim0.05$ dex, comparable to the typical uncertainty in the Carretta et al. observations. 
The resulting O-Na and Mg-Al distributions are shown in Figs.~\ref{fig:anticorrs1} and \ref{fig:anticorrs2}, respectively. We note that these are not dilution tracks, because stars with different masses and helium abundances contribute to different parts of the distribution. Because of the mass-dependent dilution, the most massive stars, with the highest $Y$ (and hence $\Tc$) contribute to extreme end of the distributions (highest Na and Al). A large fraction of the wind material is diluted to nearly pristine values, such that there is no clear distinction between P1 and P2 stars, in almost all GCs, $\sim90-95\%$ of the mass contains wind material. We note that the extent of the anticorrelations in the model is the combined results of
\begin{itemize}
\item $\Tc$, in the sense that spreads in Na decreases with increasing $\Tc$ (more massive, metal-poor GCs), while the spreads in Mg and Al increase;
\item $\Mwind$, in the sense that spreads of all elements increase when there is more wind mass (less dilution), which happens in more massive, metal-rich GCs.
\end{itemize}

\begin{figure}
\includegraphics[width=\columnwidth]{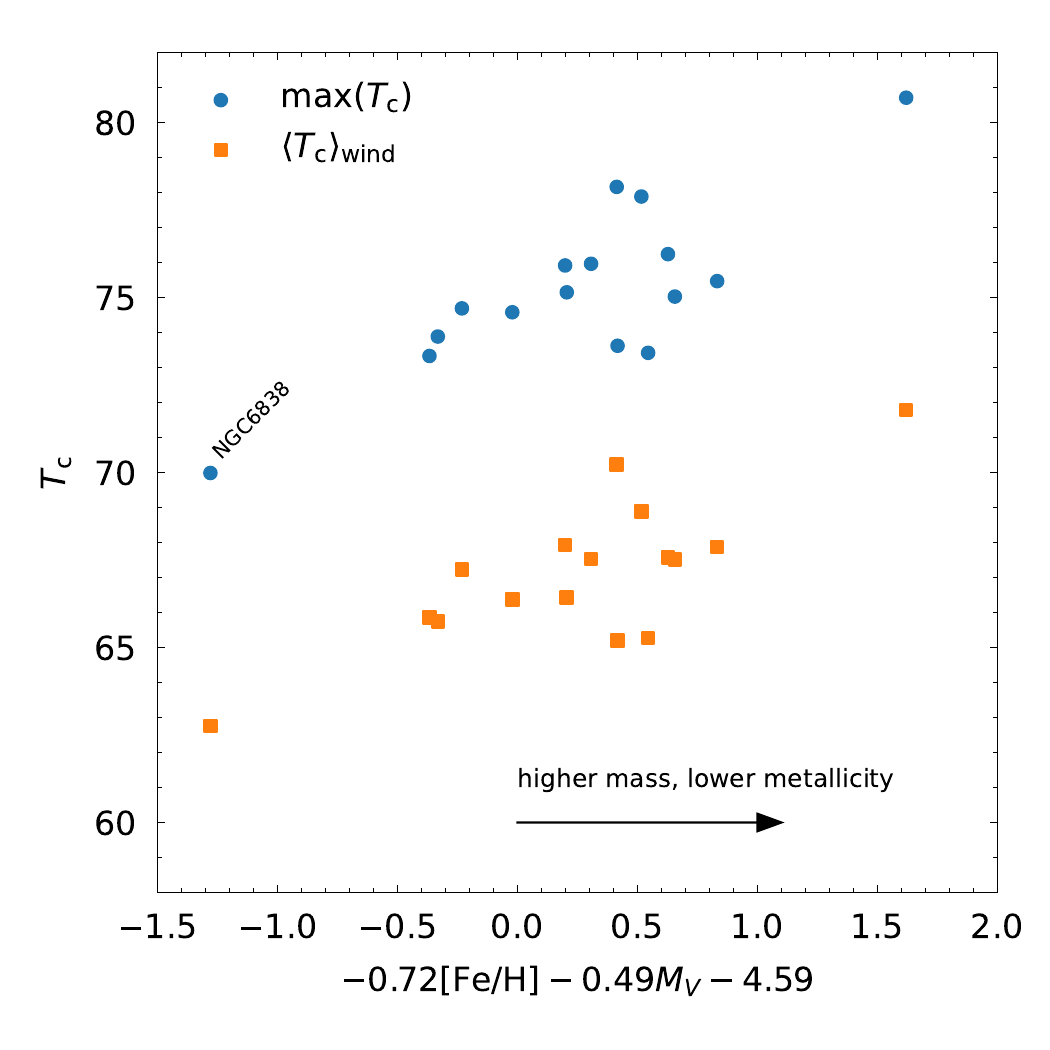}
\vspace{-5mm}
\caption{Comparison of $\Tcw$ with Carretta's bivariate relation for all 16 GCs with UVES data available. The increase shows that Carretta's empirical quantity is a good indicator of temperature.} 
\label{fig:tc_carretta}
\end{figure}

To  illustrate the dependence of the extreme abundances on GC mass and metallicity, we compare in Fig.~\ref{fig:tc_carretta} the maximum and average temperatures in the aEMS models to the empirical proxy for temperature suggested by \citet{2009A&A...505..139C}, which  correlates with the Al spread: $-0.72\feh-0.49M_V-4.59$. We confirm that Carretta's bivariate relation of luminosity and $\feh$ is indeed a good proxy for temperature.
Carretta et al. note that NGC\,6838 (left-most data point) is an outlier in their figure 13, in the sense that its Al spread is too high. This GC follows the general trend very well. This is likely because the cluster has a low present-day mass of $\sim3\times10^4\,\msun$, but a $\sim5$ times large initial mass (Table~\ref{tab:mwgc}). Hence, its Al spread is large for its present-day mass, but expected based on its initial mass. A similar argument applies to NGC\,6535, the `lowest mass Milky Way globular cluster with a O-Na anti-correlation' \citep{2017A&A...607A..44B}.

For most GCs, the model appears to overproduce Na at a given O, or not deplete O enough, for a given Na. A naive solution is to increase the temperatures to reduce Na, but we note that this would create tension with Mg, because if we would simply multiply the masses by a factor of $\sim3$ to reduce Na, there would be too much Mg depletion. We note that we use highly idealised nucleosynthesis results, and therefore caution from over-interpreting this mismatch between the model and the observations. More detailed modelling of accreting stars and their nuclear burning is required before we can make any firm conclusions. The difference between the O-Na model and the data could be interpreted as a mismatch in the O distribution. Because O is well understood from a nucleosynthesis perspective, this could only be explained  by a systematic underestimation of the measured $\OFe$. 

The O abundances reported by Carretta et al. are derived from equivalent width (EW) measurements of the [O I] lines at 630 and 636 nm, which are blended with a Ni I line and weak CN lines, respectively. Oxygen abundances based on EW measurements appear to yield lower $\OFe$ values compared to those obtained from spectral synthesis analyses, with the differences possibly being more pronounced for weaker lines (that is, at lower $\OFe$; Lardo, private communication). This could suggest that spectral synthesis results may produce a steeper O-Na anticorrelation, in better agreement with the nucleosynthesis predictions

The model correctly reproduces the shape of the Mg-Al anticorrelation for most GCs, but the Al abundances in the model appear to be more peaked towards pristine values than in the observations. This may in part be explained by the fact that for several GCs there are only Al upper limits for the lower values. The distribution of points in the model is extremely sensitive to the details of the dilution model, which is an uncertain aspect in the model.  This can be improved, with better understanding of how winds mix with pristine material. 

\citet{2009A&A...505..117C} note that the minimum $\OFe$, $\OFemin$,  is lower in massive, metal-poor GCs. This scaling is expected from the undiluted abundances of \vems\ (Fig.~\ref{fig:tc_m} and Fig.~\ref{fig:scaling3}). However, the decrease of $\OFemin$ with $\Tc$ in the regime of \vems\ is  minor (Fig.~\ref{fig:tc_m}), hence dilution likely plays a larger role in shaping the dependence of the observed $\OFemin$ and GC properties. Wind material is more diluted in low-mass, low-metallicity GCs, hence dilution will enhance the anti-correlation between $\OFemin$ and GC mass. However, dilution likely turns the $\OFemin-\feh$ correlation produced by nucleosynthesis into an anticorrelation. This is assuming that the maximum $\OFe$ is the same for all GCs.
However, from the data in table 7 of  \citet{2009A&A...505..117C} we find that $\OFemax$ correlates with $\OFemin$ (Pearson correlation coefficient of $\sim0.3$), which affects the comparison with model predictions. It is nevertheless interesting to keep this difference in mind.

%______________
\subsection{Lithium}
\label{ssec:lithium}

Lithium (Li-7) is a fragile element that is destroyed  at temperatures above $\sim2.5\,\MK$ in stellar interiors, such that the yields of  massive stars are expected to be devoid of Li. If EMSs are the source of pollution, an anticorrelation is expected between Li and elements that are enhanced in P2 stars (N, Na, Al) and a correlation between Li and the elements that P2 stars are depleted in  (C, O and sometimes Mg). In fact, an almost linear correlation between Li and O is expected because in undiluted yields,  O is depleted by two orders of magnitude \citep[for example,][]{2007A&A...470..179P}. 
There is no consensus, however,  regarding the existence of a correlation between Li and O (or an anticorrelation between Li and Na) in P1 and P2 GC stars, with different authors finding contradicting results \citep{2005A&A...441..549P,2007A&A...470..153B,2009A&A...503..545L,2010ApJ...713L...1D,2010A&A...524L...2S,2011MNRAS.412...81M,2012A&A...539A.157M,2014A&A...565A.121D,2024A&A...690A.245B}. We think that part of the inconsistency  is due to the fact that the O and Na abundance ranges covered by P2 stars in the above mentioned studies are not large enough to populate the Li-dilution curve, as shown below. 

Furthermore, we note that the Li abundances we determine today in old GC stars are not expected to reflect the initial abundances the stars were born with.  Surface  Li depletion should indeed occur in a similar way to halo stars (at least in P1 stars), first along the main sequence due to non-standard transport processes as required to explain the so-called cosmological lithium problem \citep[for example,][and references therein]{2024A&A...690A.245B}, and later on the red giant branch due to the first dredge-up \citep[for example,][]{2022A&A...661A.153M}. This is supported by the fact that P1 GC dwarf and turn-off stars have similar Li abundances along the so-called Li plateau \citep[for example,][]{1982A&A...115..357S,2007ApJ...671..402K} as their field counterparts with similar metallicity and Na and O content. Here we assume that P1 and P2 stars undergo similar Li depletion on the main sequence, and we focus on observations for dwarf and turnoff stars that were determined in several GCs; we do not discuss the case of red giants to avoid further uncertainties related to the first dredge-up.

\begin{figure}
\includegraphics[width=\columnwidth]{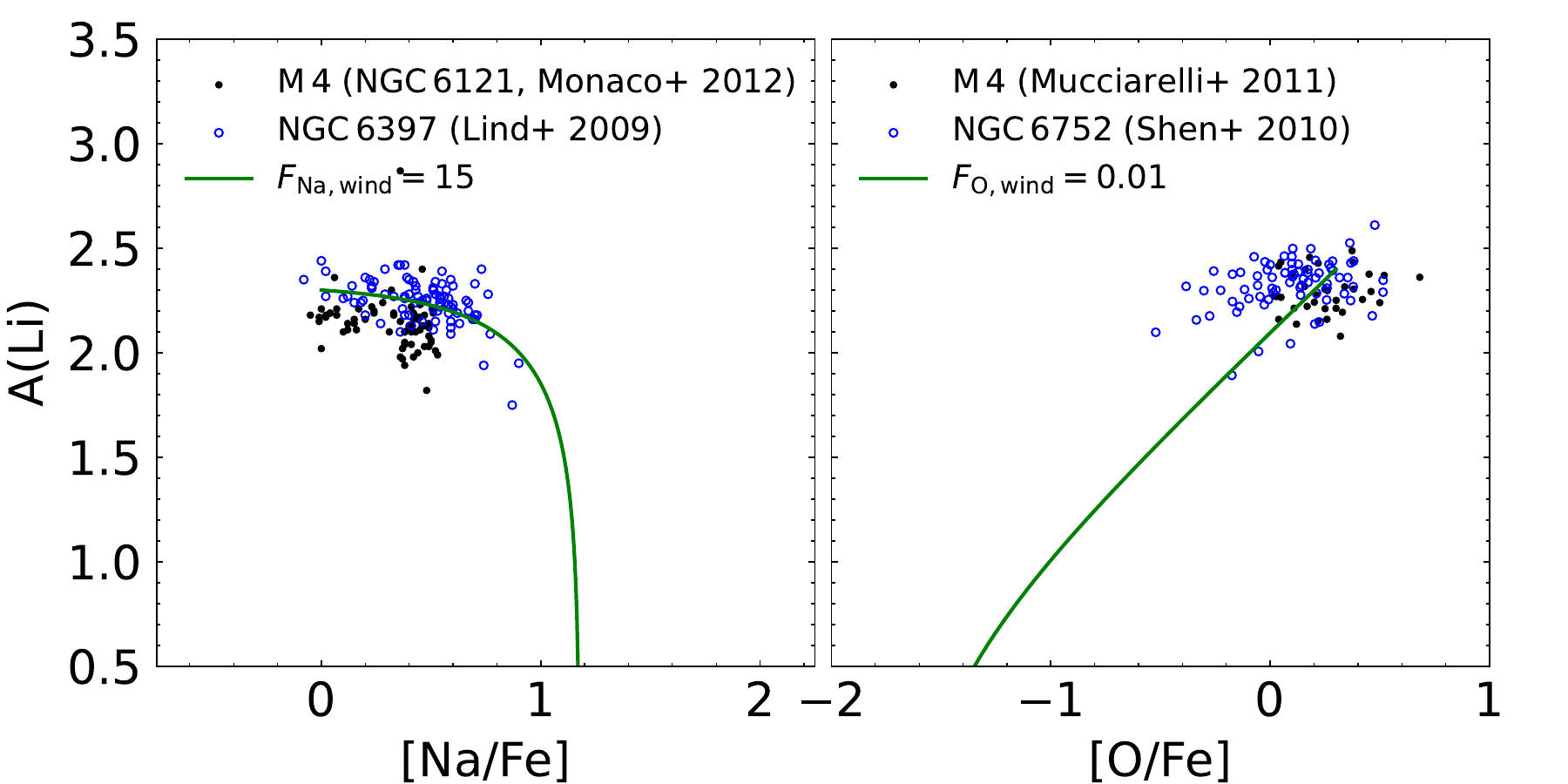}
\vspace{-5mm}
\caption{Left: Abundance of Li ($\ALi$) as a function of $\NaFe$ for for main sequence stars in M\,4 (NGC\,6121, $\Teff>5880\,$K) and NGC\,6397 ($V<3.3$ mag) combined with a dilution 
track for typical \vems\ yields, assuming \vems\ are devoid of Li. The model can reproduce  the 
`mild' observed anticorrelation. Right: $\ALi$ as a function of $\OFe$ for M\,4 and NGC\,6752. A near-linear relation is expected, but the narrow range $\OFe$ makes it hard to confidently confirm this or rule this out.}
\label{fig:li}
\end{figure}

We quantify what the expected slopes are of the Li-Na anticorrelation and the Li-O correlation based on \vems\ yields and 
assuming that \vems\ winds are devoid of Li ($10^\ALi=0$) and oxygen. For $\NaFe_0=0$, the dilution tracks can be written as 
\begin{align}
\ALi  & = \ALi_0 + \log_{10}(1-f),\label{eq:ali}\\
\NaFe  & = \log_{10}\left[1+ f\left(\Flin{Na}{wind}-1\right)\right]\label{eq:nafe},\\
\OFe  & = \OFe_0 + \log_{10}(1-f)
\end{align}
where $f$ goes from 0 for pristine stars to 1 for stars that are for 100 per cent made of \vems\ yields. We adopt $\ALi_0 = 2.2$  to account for a similar main sequence depletion in halo, P1 and P2 stars (\citealt{2024A&A...690A.245B} and Charbonnel, in prep.). 
The dilution predictions are shown in Fig.~\ref{fig:li} for  $\Flin{Na}{wind}=15$ (left panel) and  $\Flin{O}{wind}=0.01$ (right panel) that are appropriate for \vemss\ (see Fig.~\ref{fig:tc_m}).

We compare this to Li data for turnoff stars in M4 (NGC\,6121;  \citealt{2011MNRAS.412...81M,2012A&A...539A.157M}),  
 NGC\,6397 \citep{2009A&A...503..545L},  and  NGC~6752 \citep{2010A&A...524L...2S}\footnote{We  added 0.19 dex to the $\OFe$ to correct for NLTE effects discussed in \citet{2010A&A...524L...2S}.} as a function of $\NaFe$ (left) and $\OFe$ (right).
The Li-Na behaviour of the dilution track is in good agreement with the observed mild slope in the Na range
covered by the observations. To further quantify this, we use equations~(\ref{eq:ali}) and (\ref{eq:nafe}) to eliminate $f$ and write $\ALi$ as a function of $\NaFe$. Using a  Tayler expansion near $\NaFe=0$ we find that the slope in the $\ALi-\NaFe$ diagram is $-1/(\Flin{Na}{wind}-1)$, which is  $-0.07$ for the plotted scenario. Near $\NaFe=0.3$ the slope is $\sim - 2/(\Flin{Na}{wind}-2)$, which is $-0.15$ for the adopted $\Flin{Na}{wind}=15$.  
In M4,  \citet{2012A&A...539A.157M} find a slope of $-0.30\pm0.09$ when using LTE Na abundances and $-0.22\pm0.07$ when using NLTE Na abundances. 
A similar slope  ($-0.18$) is found in NGC~6397 among turnoff stars ($V<3.3$ mag). In addition, more Na-enriched stars in this GC drive a more significant Li-Na anticorrelation which is compatible with dilution between Li-free ejecta and pristine material, as already underlined by \citet{2009A&A...503..545L}. 

We show in the right panel of Fig.~\ref{fig:li} the theoretical O-Li correlations for  $\Flin{O}{wind}=0.01$  and we compare this to observations of turnoff stars from M\,4 and NGC~6752. In the small $\OFe$-range covered by stars in M4, \citet{2011MNRAS.412...81M} find a near constant $\ALi$. On the other hand, \citet{2010A&A...524L...2S} find in NGC\,6397 a slope of 0.4 for the O-Li correlation, which is lower than the theoretical one. The observational slope, however, is driven by Li and O upper limits. We note that there is also tension between the modelled and observed O-Na anticorrelation of NGC\,6752, with the modelled $\OFe$ being larger, at a given Na (Fig.~\ref{fig:anticorrs1}). We also recall that there may be systematic issues with $\OFe$ 
 that imply that the lowest $\OFe$ may be underestimated (Section~\ref{ssec:anticorrs}).
We are therefore cautious with drawing any strong conclusion from the slope of the O-Li correlation.

In the case of 47 Tucanae (47 Tuc), an Na-Li anti-correlation was initially found  by \citet{2007A&A...470..153B} for a very small sample of stars covering $\sim$0.6~dex in Na. In a larger sample covering a similar Na range,  \citet{2014A&A...565A.121D} found an anticorrelation that was not retrieved by \citet{2010ApJ...713L...1D}. Both studies revealed a large dispersion in Li among dwarf stars in this cluster. Following  \citet{2014A&A...565A.121D}, we propose that this is due to a change of Li depletion regime around $\feh\sim-1$~dex, as predicted by \citet{2024A&A...690A.245B}. 

The presence of $\ALi$ in P2 stars with similar abundances as in P1 stars is often used as an argument against massive stars as polluters, and in favour of AGB stars  \citep[for example,][]{2019A&ARv..27....8G}, because AGB stars are able to produce Li via the Cameron-Fowler mechanism \citep{1971ApJ...164..111C}.
We have shown, however, that the available data support  dilution between Li-free ejecta and pristine material, and that 
part of the  differences in findings for Li in different cluster studies comes from the range of Na and O covered by the stars where the Li is detected. We call for the observers to go and look for Li in the P2 stars with the highest Na enrichment and O depletion and predict that stars with lower $\ALi$  exist. 
%________________________________
\section{Discussion}
\label{sec:discussion}

\subsection{Discreteness}
\label{ssec:discreteness}
\begin{figure}
\includegraphics[width=\columnwidth]{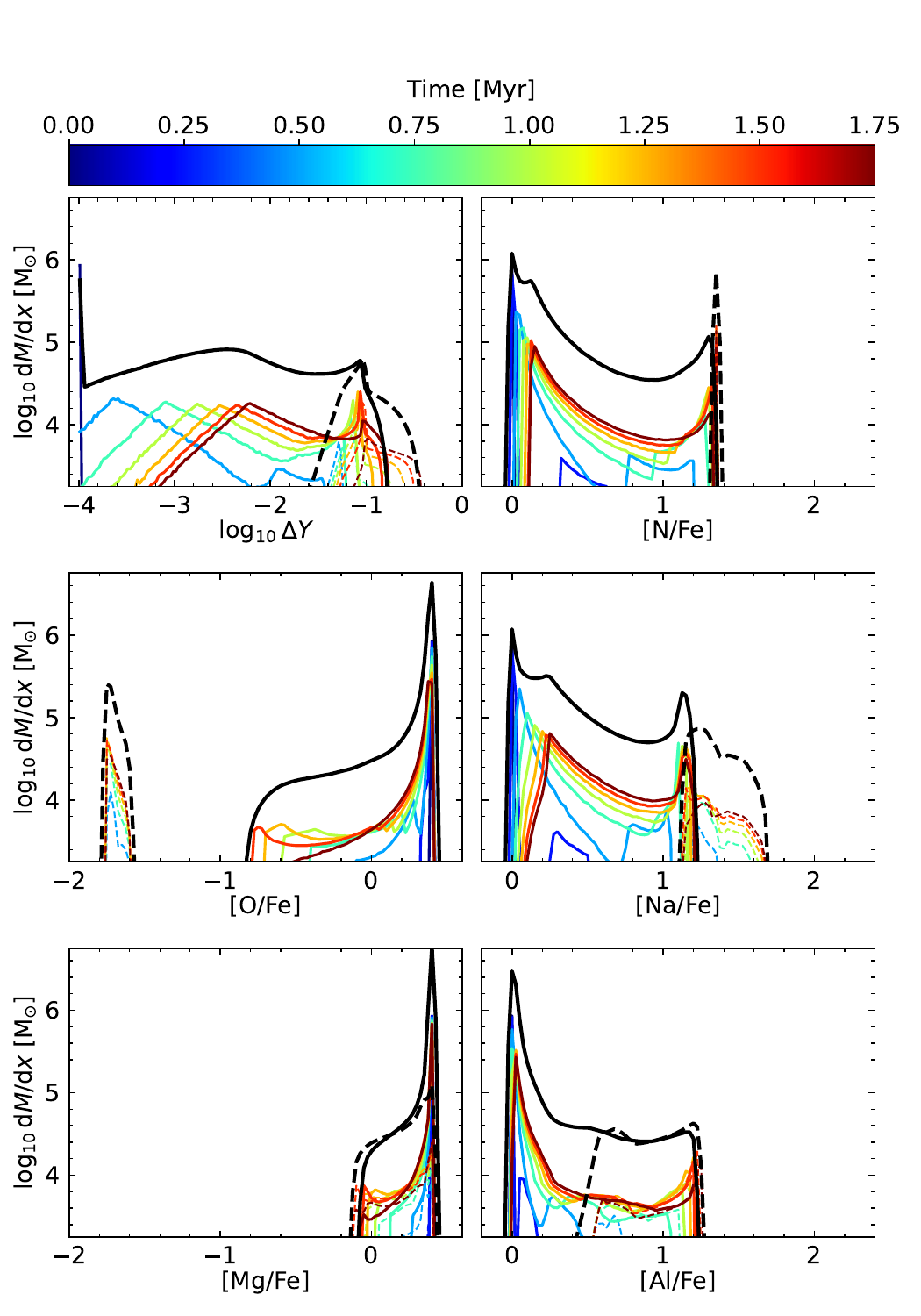}
\vspace{-5mm}
\caption{Distributions of the main abundances for the typical GC model of Section~\ref{ssec:typical}. The undiluted(diluted) yields are shown with dashed(full) lines, where the coloured lines show the distributions at different times, and the black lines show the totals. Clear bi-modality is seen in the diluted distributions of N, Na and a bit for He and Al. The distributions of O and Mg are mostly broadened. Note that the maximum Na and Al in the diluted distributions are similar, despite the fact that the undiluted maximum Na is a factor of three higher than Al. This is because of our mass-dependent dilution: the yields of the most massive stars are least diluted, and these stars have the lowest Na and the highest Al (see Fig.~\ref{fig:abundances_mass}).}
\label{fig:abundance_distributions}
\end{figure}

An important property of the MPs of GCs is the observed discreteness of the abundance pattern in most GCs, for some elements. In the {\it HST} pseudo-colour maps \citep[or, `chromosome maps', ][]{2017MNRAS.464.3636M}, the P1 and P2 stars appear for most GCs as distinct populations, suggesting a bi-modality in the distribution of N (and to lesser extent He). For NGC\,2808, five distinct populations were reported based on the photometry \citep{2015ApJ...808...51M}.   Spectroscopically derived Na distributions appear unimodal for most GCs, but this could be because of large measurement uncertainties. Some cases of clear bimodal O and Na distributions are known \citep{2008A&A...490..625M} and three distinct Al distributions have been reported for NGC\,2808 \citep[][]{2014ApJ...795L..28C}. In Section~\ref{ssec:typical} we note that the `pile up' in the mass function near the most massive stars as the results of winds, results in a discrete abundance pattern for the undiluted yields.  
In Fig.~\ref{fig:abundance_distributions} we show the distribution of undiluted and diluted yields for six elements for the typical GC. 
There is clear bimodality in the distributions of N and Na, and to lesser extent in He and Al. The small third peak seen in the distributions of He, N, Na, and Al are more likely a `feature' of the specific choice of dilution model, in which towards the end of GC formation there are no stars with $\fdil\gtrsim30$ (see bottom panel of Fig.~\ref{fig:abundances_mass}). We note that the high-N peak in the N distribution is the result of the pile up of stars near $\minf$ of the most massive stars (bottom panel of Fig.~\ref{fig:imf}), and not the result of choice of dilution. A power-law mass function would result in a unimodal distribution of N distribution, peaking near pristine abundance. The discreteness, is therefore an important aspect of the aEMS model. 

The model as presented is not able to create more than two peaks. We here discuss four aspects not considered in the model that could lead to more discrete populations:
\begin{itemize}
\item Discreteness in the (I)MF and mixing: we model the mass function as a smooth distribution. If we would sample individual stars from the IMF of the typical GC, we find that there are $10_{-3}^{+3}$ stars that reach masses $>10^3\,\msun$. The two most massive aEMSs are $\sim20$ per cent different in mass. If dilution happens near the aEMSs, this could lead to different abundances;
\item Sub-clustering: GCs probably form hierarchically, meaning that the final cluster is the result of the merger of several smaller sub-clusters \citep[for example][]{2025arXiv250418620L}. If each of these sub-clusters is massive enough to form EMSs, and the mass of their most massive EMS scales with the mass of the sub-cluster, more than two discrete abundances patters can be created;
\item Stellar collisions: a few of the most massive stars may undergo stellar collisions amplifying the differences in mass and abundance patterns;
\item Bondi-Hoyle accretion: the most massive stars may undergo a runaway growth via Bondi-Hoyle accretion of the gas that sits in the cluster centre, amplifying differences in mass and abundances at the upper end of the IMF, similar to collisions.
\end{itemize}
In conclusion, the aEMS model as presented naturally produces bi-modality/discreteness in the distributions of He, N, Na and Al and there are several additional ways to create more than two populations.

\subsection{Mixing}
\label{ssec:mixing}
An important aspect of the aEMS model presented here is that surface abundances are similar to the central abundances, even while stars are still accreting.
Stellar evolution models  of non-rotating, non accreting main sequence SMSs ($\gtrsim10^4\,\msun$) computed with MESA \citep[Modules for Experiments in Stellar Astrophysics,][]{2011ApJS..192....3P} with a typical GC metallicity ($\feh=-1.6$ dex) are fully convective \citep{2014MNRAS.437L..21D} hence in that  mass range this is a safe assumption for non-accreting stars. However, in the aEMS model most of the mass comes from stars of $\sim{\rm few}\times10^3\,\msun$, and accretion might build positive entropy gradients in the stellar envelope that prevent the extension of the convective core towards the surface \citep[for a discussion and references see][]{2025arXiv250612132R}. 
Thus, the minimum mass of fully convective accreting stars is uncertain. \citet{2025arXiv250612132R} find that non-rotating, compact aEMSs with high, and luminosity-dependent, mass accretion rates (following the Churchwell-Henning relation and rising up to $0.1\,\msun\,\yr^{-1}$) become fully convective between $\sim10^3$ and $\sim10^4\,\msun$  (for $\feh =-2$), with the minimum mass being strongly dependent on the treatment of superadiabatic convection. For comparison, we computed a model with the same evolution code (MESA, version 22.11.1, \citealt[][]{2011ApJS..192....3P,2013ApJS..208....4P, 2015ApJS..220...15P, 2018ApJS..234...34P, 2019ApJS..243...10P, 2023ApJS..265...15J}) and similar input physics except for the gas accretion rate, which we fix at a typical value we consider here (that is, $10^{-3}\,\msun\,\yr$). In this case, the minimum mass for the model to be fully convective is $\sim 1300 \,\msun$, but for different treatments of the transport of energy, the model is never fully convective until the end of the main sequence. Beyond the current uncertainties on superadiabatic convection in \vemss, various hydrodynamical mechanisms might lead to the transport of chemicals between the core and the very external layers.
During the accretion phase, \vemss\ are expected to have high/near critical rotation rates \citep{Rosen+12}, 
which {likely} leads to the transport of the burning products  
from the core to the external radiative layers towards the surface as in the case of lower mass stars   \citep{2000ARA&A..38..143M,2011A&A...530A.115B, 2023MNRAS.524.1529S,2023A&A...679A.137M}.  Last, but not least, an additional  source for the transport of entropy, angular momentum, and  chemicals in the  radiative layers of the aEMS could be turbulence induced by fast accretion shocks and/or collisions 
\citep[for a discussion and references see][]{2025arXiv250612132R}.
Therefore, while modeling is required to address these questions in the case of aEMS, we consider our assumption about the fully mixed state of these objects ($\gtrsim10^3\,\msun$) to be a realistic representation.

%_________________________________
\subsection{Nucleosynthesis uncertainties}
The theoretical abundance variations we describe are based on the  behavior of the  yields of the CNO cycle and NeNa, MgAl chains  from \citet[][see Sections \ref{ssec:other} and \ref{ssec:anticorrs}]{2007A&A...470..179P,2017A&A...608A..28P}. Their  nucleosynthesis calculations were run at constant temperature and density, over a broad range of temperatures between 25 and 200~MK, and for $\feh=-1.5$. As already mentioned, the evolution of both the central temperature and density along the main sequence depends on the actual metallicity, which is expected to slightly affect the extent of the anticorrelations compared to predictions shown in Figs.~\ref{fig:anticorrs1} and \ref{fig:anticorrs2}. Furthermore, another source of uncertainty comes from those on the nuclear reaction rates, with typical  factors between 2 and 20 in the 60-80~MK range for the reactions involved in the NeNa and MgAl chains
\citep[][see also \citealt{2007A&A...466..641I} and \citealt{2007A&A...464.1029D} who discuss these aspects in the cases of the yields from AGB and FRMS, respectively]{PhysRevC.107.035806,2007ApJ...654..835W,PhysRevC.95.015806}. The yields of the various isotopes being interconnected, it is difficult to estimate the impact of these  uncertainties for aEMS without performing extensive calculations that are out of the scope of our study. Finally, the initial abundances of the different isotopes involved in the nuclear reactions at a given metallicity are challenging to ascertain. In particular, no spectroscopic data exist to constrain the amount of $^{20,22}$Ne in the ISM at different metallicities. In their calculations,  \citet{2007A&A...470..179P,2017A&A...608A..28P} assumed the predictions of Galactic chemical evolution models for Ne which behaves as an $\alpha$ element, but which production factor along time and metallicity depends on stellar yields that have their own caveats \citep[see for example,][]{2018MNRAS.476.3432P}.  

\subsection{Wind mass-loss rate of accreting stars}
\label{ssec:windacc}
The interplay between (disc) accretion and outflows has long been studied in low-mass stars, where the paradigm is that of magnetospheric accretion \citep{2016ARA&A..54..135H}. The situation for (very) massive stars is far from clear. 
For decades it was thought that massive stars can only form via stellar
collisions \citep{2002MNRAS.336..659B} as feedback in the form of radiation pressure on dust opacity
would not produce stars over $\sim10\,\msun$ \citep{1971A&A....13..190L, 1987ApJ...319..850W}.
However, such feedback models were only available in one dimension, while 
more recent multi-dimensional massive star formation models 
\citep{2002ApJ...569..846Y, 2009Sci...323..754K, 2010ApJ...722.1556K, 2016MNRAS.463.2553R}, show that
the accretion occurs via an equatorial disc while radiative feedback
blows predominately over the polar regions \citep{2015ASSL..412...43K}.

Until 2018, feedback during (very) massive star formation did not include radiation pressure on atomic line opacity, while this source of opacity is what drives the stellar winds from (very) massive stars during stellar 
evolution. \citet{2018A&A...615A.119V} suggested that as the stellar surface is not in contact
with the stellar core, the onset of H burning on the zero-age main sequence 
(ZAMS) should not be considered as the key division point between the 
formation of stars with accretion rates, versus stellar evolution with mass-loss
rates, but that accretion and winds should be considered simultaneously, and
that the upper-mass limit of the initial mass function (IMF) should actually
be referred to as the `effective' upper stellar mass, and that this upper 
limit is likely $Z$-dependent due to winds, with lower $Z$ stars producing a 
higher upper mass limit. 

In these one-dimensional considerations, \citet{2018A&A...615A.119V} simply balanced
the mass-accretion rate of order $10^{-3}\,\msunyr$ \citep[for example,][]{2017ApJ...835...32T} versus
the radiation-driven mass-loss rate in the VMS regime \citep{2011A&A...531A.132V}, but it 
would probably be more realistic to consider this accretion-feedback 
process to be multi-dimensional, with the accretion occurring over the equator, and simultaneously 
the wind mass loss at high Eddington factor \citep{2011A&A...531A.132V} 
to be predominately over the polar regions \citep{1994ApJ...424..887O, 1999A&A...347..185M, 2000A&A...359..695P}.

%______________
\subsection{Type II GCs}
Based on pseudo-colour maps, \citet{2017MNRAS.464.3636M} find  10 (out of 57) GCs that display photometric features in addition to the light elements variations we discussed so far. These so-called Type II GCs (as opposed to the 47 Type I GCs), are characterised by multiple sub-giant branches in optical filters and a split red giant branch. This points at either variations in C+N+O which were not found in Type~I clusters, and which would result from He-burning, or age differences of $\sim1-2\,\gyr$. 
Through a comparison with spectroscopy results \citep{2009ApJ...695L..62Y, 2015MNRAS.450..815M}, Milone et al. conclude that the peculiar stars in Type II GCs are enhanced in C+N+O, $s$-process elements and $\feh$. Various authors have shown that the reported $\feh$ enhancements of the Type II stars (from hereon defined as stars enhanced in $s$-process elements and C+N+O) are spurious, or at least overestimated  \citep{2015ApJ...809..128M,2016MNRAS.457...51L,2018ApJ...859...75M}. It is also worth noting that for 6/10 Type II GCs, the fraction of Type II stars is only a few per cent, but for the extreme case of $\omega$ Centauri ($\omega$\,Cen, NGC\,5139) they constitute the majority of stars in the sample. We note that due to the close distance of $\omega$\,Cen, the {\it HST} data is dominated by the high-density central region of the cluster, such that the global fraction of Type II stars could be lower. 

We note that massive stars at low metallicity can produce $s$-process elements (with large discrepancies between the models) \citep[$20-150\,\msun$, $\feh=-2$,][]{2018A&A...618A.133C}, with rotation boosting $s$-process production for stars with masses between 20 and 60 $\msun$ \citep{2018A&A...618A.133C,2018ApJS..237...13L,2011Natur.472..454C,2016MNRAS.456.1803F}. 
This occurs during helium burning, so it requires cluster formation to last longer than $\sim3\,\myr$. This is not impossible, and it could also explain $\feh$ spreads (see Section~\ref{ssec:ironspread}).
The aEMS model presented here can therefore only explain C+N+O variations or $s$-process enhancements if $\tdyn\gtrsim3\,\myr$. The former feature can 
also result  from He burning and second dredge-up 
in rotating intermediate-mass AGB stars \citep{2009A&A...505..727D}, the second from neutron-captures in low- and/or intermediate-mass AGB stars \citep{2014PASA...31...30K,2021ApJ...908...55B,2022Univ....8..220V,2023ARNPS..73..315L}. In the model, GC formation finishes before ($\sim1-2\,\myr$) H burning ends ($\sim 2 -  3\,\myr$) and well before  the appearance of low-mass AGB stars  ($\gtrsim 80 - 100\,\myr$).  \citet{2015MNRAS.450..815M} notices that the ratio $\LaEu$ of the $s$-process enhanced stars is very similar to that of CEMP/s stars, of which many are known to be binaries \citep{2005ApJ...625..825L,2016A&A...586A.158J}. The peculiar stars in Type II GCs could therefore also be the result of binary interactions involving AGB stars, and have nothing to do with the GC formation process. Some support for this idea comes from the fact that the Type~II GCs are more massive ($\langle\log_{10}M/\msun\rangle= 5.7\pm0.3$, mean and standard deviation) then Type~I GCs ($\langle\log_{10}M/\msun\rangle= 5.2\pm0.4$), for the GCs in the Milone et al. sample. The dynamical interaction rate of binaries with other stars (and binaries) is higher in more massive GCs, and this leads to more triggered binary interactions and collisions. 
This is also why the binary fraction of GCs decreases with $\MGC$ \citep[for example][]{2012A&A...540A..16M} and why the dynamical production of blue stragglers is  more efficient (per unit mass) in more massive GCs \citep*{2004MNRAS.349..129D}. A binary origin is also appealing because Type~II stars have their own light element anticorrelations \citep[for example][]{2011A&A...533A..69C,2011A&A...532A...8M}, and this could be explained if the light elements anticorrelations and helium spreads are the result of GC formation, and the $s$-process and C+N+O enhancements are the result of binary physics that happens later. 

%________________________
\subsection{Iron spreads}
\label{ssec:ironspread}
Iron spreads of $\Delta\feh\sim0.1-0.2$ dex in P1 stars were reported for NGC\,2808 \citep{2023A&A...669A..19L} and 47\,Tuc \citep{2023ApJ...958...31M}, which have masses ($\sim10^6\,\msun$). Curiously, \citet{2023ApJ...958...31M} find tentative evidence for a smaller iron spread in P2 stars, but this is based on only three stars and there may issues with the abundance analysis  \citep{2025A&A...695A.150C}. \citet{2023A&A...669A..19L} did not have spectra of P2 stars. 
In the aEMS model an iron spread could result in the P2 stars, because  GCs ($\MGCi\gtrsim{\rm few}\times10^6$) have a formation time of $\sim3\,\myr$ for the adopted surface density of $1.5\times10^3\,\msun\,\pc^{-2}$ (equation~\ref{eq:tdyn}). This may result in some SNe pollution towards the very end of GC formation. If indeed the iron spread is only present in P1 stars, then this can only be taken as a characterisation of the pristine gas in the aEMS model, that is to say an input to the model rather than a prediction from self-enrichment. In fact, one can use simple arguments to show that mixing of SN ejecta is less efficient in dense gas, simply because the timescale between SN explosions from pre-existing field stars is longer than the gas mixing time within the dense cluster-forming cloud. Assuming that all stars above $10\,\msun$ go SN and adopting a canonical IMF ($\sim10^{-2}\,{\rm SNe}\,\msun^{-1}$), then the supernova rate in some volume $V$, $\rsn$, relates to the star formation per unit volume, $\sfrv$, as $\rsn \simeq 10^{-2}\,\msun^{-1}\sfrv V\simeq10^{-4}\,\msun^{-1}\Mgas(G\rhogas)^{1/2}$, where we assumed a Kennicutt-Schmidt star formation law: $\sfrv\simeq 10^{-2}\rhogas (G\rhogas)^{1/2}$, where $\rhogas$ is the average gas density in the volume and we used $\Mgas = \rhogas V$.
The mixing timescale of the gas within that volume is the dynamical  timescale, $\tdyn$ (equation~\ref{eq:tfmax}), which can be approximated by $\tdyn\simeq(G\rhogas)^{-1/2}$, such that the product $\tdyn\rsn\simeq \Mgas/(10^{4}\,\msun)$. This implies that the mixing time is approximately similar to the time in between SNe for $\Mgas\simeq10^{4}\,\msun$. For higher mass clouds, the mixing time is longer than the time in between SNe, 
so indeed, massive GCs should form from gas with relatively unmixed SN ejecta and hence appreciable $\feh$ spreads in P1 stars.

This is supported by numerical simulations of GC formation in dwarf galaxies by
\citet{2024MNRAS.530..645L}, where small iron spreads  occur because of unmixed gas clouds. This could explain why iron spreads occur in P1 only, because P2 stars form in higher gas densities and had more time to mix. Further confirmation with observations would be a useful test.  

There are even more massive GCs with convincing iron spreads, and these are accreted nuclear clusters, such as M\,54 (NGC\,6715), $\omega$\,Cen and NGC\,6273 \citep{2021MNRAS.500.2514P}. Sub-populations with different $\feh$ in $\omega$\,Cen each display a O-Na anticorrelation. These iron spreads could be the results of cluster mergers due to inspiral by dynamical friction in the progenitor galaxy, which provides a natural explanation for the O-Na anticorrelations in each sub-population. The iron spread could also be the result of self-enrichment, or multiple starbursts due to various episodes of gas inflow, in which case the O-Na are not trivially explained, because star formation in an existing cluster may lead to different results, but explaining these complexities is  beyond the scope of this paper.

%______________
\subsection{Structure and kinematics of P1 and P2 stars in GCs today}
Several studies have looked at the relative spatial distribution and kinematics of P1 and P2 stars in Milky Way GCs \citep{2015MNRAS.450.1164H,2018MNRAS.479.5005M,2020ApJ...889...18C}. The model presented here does not make specific predictions for the kinematics, but it is expected the P2 stars form more centrally concentrated because it is expected the EMS winds quickly stall when hitting the high density pristine gas in the centre of the cluster. Observations of Milky Way GCs with long relaxation times (so there was not enough time for the two populations to spatially mix) suggest that P2 stars are indeed more centrally concentrated \citep{2024A&A...691A..94D, 2025A&A...694A.184L}, possible exception being NGC\,6101 \citep{2023MNRAS.520.1456L}. More detailed numerical simulations of the MP formation process described in this paper are needed to be able to make reliable predictions for the spatial distribution and kinematics of P2 relative to P1.

%______________________________________________
\subsection{Multiple populations in young massive clusters}
\label{ssec:ymcs}
No MPs have been found in YMCs $\lesssim2\,\gyr$   \citep{2016MNRAS.460.1869C, 2018MNRAS.473.2688M}. In our aEMS model, there is no reason why MPs can only form at high redshift, but several conditions  need to be met for a cluster to show MPs that  favour conditions in the early Universe: the gas cloud from the cluster forms needs to be massive enough to make a sufficient number of EMSs ($\Mgas\sim{\rm few}\times10^6\,\msun$). This mass threshold increases with  $\feh$, because a higher accretion rate is needed to overcome the stronger winds at higher $\feh$; also, the gas density needs to be high enough for EMSs to appear well before the first SNe.  For a gas density that is an order of magnitude lower than used here, the formation time of a $10^6\,\msun$ cluster is $\sim10\,\myr$ (equation~\ref{eq:tdyn}), such that SNe feedback interrupts cluster formation. 
No significant Al spreads were found in YMCs in the Antennae galaxies \citep{2017MNRAS.468.2482L}. Although some of the molecular gas in the centre of the Antennae galaxies has densities in excess of $10^3\,\msun\,\pc^{-2}$, most gas is found at lower densities \citep{2024MNRAS.530..597B} and MPs are then not expected.  Directly comparing the locations of three YMCs for which \citet{2017MNRAS.468.2482L} determined Al abundances from integrated light \citep[figure 1 in ][]{2015ApJ...812..160L} to the gas surface density maps of \citet[][their figure 2]{2024MNRAS.530..597B}, the gas densities in the immediate surroundings of the YMCs is $\sim10^2\,\msun\,\pc^{-2}$, too low for MPs to form.

Also, the YMCs in the Antennae have  (slightly) super-solar metallicity, such that the most massive stars in such a YMC would  be $\sim10^3\,\msun$  (equation~\ref{eq:minf}). Such a star has a temperature of $\lesssim63\,\MK$ (Fig.~\ref{fig:tc_m}), where Al enhancement is only moderate. Accounting for the fact that stars with $\feh\simeq0$ are cooler in their centres, even the stars in the most massive YMCs in the Antennae galaxies may not be high enough to show Al enhancement. It is also not clear to what extent massive stars ($\sim10-100\,\msun$) that dominate the light share the abundance pattern of the lowest mass stars ($\lesssim1\,\msun$). In our model these massive stars form from accretion from gas on a larger scale than the central gas reservoir, possibly avoiding pollution by \vems\ winds.

Finally,  in Section~\ref{ssec:anticorrs} we explained that the final Na and Al abundances are (for a given $\Tc$ and $\Delta Y$ of the polluter) proportional to the initial Ne and Mg abundances, respectively. YMCs in the Local Universe mostly form from gas that is not $\alpha$ enhanced (that is, $\NeFe=\MgFe\simeq0$), such that the maximum Na and Al are  a factor of 2.5-3 lower than that of GCs that form from $\alpha$-enhanced gas (for the same $\Tc$ and $\Delta Y$). This may explain why the N-rich stars in the 2\,\gyr old star cluster NGC\,1978 in the LMC show only a small/negligible Na enhancement  \citep{2020MNRAS.498.4472S}.

%______________
\subsection{Stellar ejections}
\label{ssec:ejections}
In clusters with short relaxation times (high density, low mass),  massive stars are dynamically ejected with high velocities \citep[so-called `runaway stars',][]{2023A&A...679A.109C, 2024Natur.634..809S}. If collisional dynamics becomes important already during GC formation, a fraction of the most massive stars can be ejected before they get a chance to pollute the GC. We here estimate how important this effect is.  Runaways are created dynamically after core collapse, which happens after a timescale comparable to the half-mass relaxation time ($\trh$), which for massive GCs is much longer than $\tnuc$. However, if clusters form hierarchically from merging sub-clusters, or the cluster forms with a high central concentration, core collapse can occur significantly earlier. We therefore consider the extreme case in which core collapse happens during GC formation, and a dynamically active binary starts ejecting massive stars while the GC is still forming. The binary components and the stars that are  ejected are typically the most massive stars in the cluster \citep{1975MNRAS.173..729H}.  The mass-loss rate as the result of dynamical ejections is \citep{2013MNRAS.432.2779B}
\begin{equation}
\dot{M}_{\rm ej} \simeq-0.006 \frac{M_0}{\trh}\simeq560\,\msun\,\myr^{-1}\left(\frac{\rhoh}{10^5\,\msun\,\pc^{-3}}\right)^{1/2}.
\label{eq:Mdot_ej}
\end{equation}
Here we used the expression for $\trh\propto M_0\rhoh^{-1/2}$ from \citet{1971ApJ...164..399S}, with $\psi=30$ to include the effect of the mass function \citep[see the discussion in][]{2020MNRAS.492.2936A,2020PhRvD.102l3016A}. From this we see that even for the case where dynamical ejections start immediately and for high -- but realistic \citep{2024Natur.632..513A} -- densities, only of order one EMS will be ejected during GC formation (few Myrs). For clusters in the Local Universe, with lower densities ($\rhoh\simeq10^4\,\msun\,\pc^{-3}$) and lower mass stars ($\sim50\,\msun$), this rough estimate predicts a handful of runaways before supernovae occur, consistent with was is found in detailed $N$-body calculations \citep{2011Sci...334.1380F}.

%______________
\subsection{The role of stellar collisions}
\label{ssec:collisions}
\citet{2018MNRAS.478.2461G} proposed that SMSs ($\gtrsim10^4\,\msun$) are the  origin of the abundance anomalies. In their SMS model, each GC forms  a SMS  via stellar collisions during GC formation. Some ingredients in the SMS model are similar to the ones of the model presented in this work:  the model relies on gas flowing into the proto-GC and accretion onto stellar seeds and a conveyer belt mechanism to produce hot-hydrogen burning products at low $\Delta Y$. 
In the SMS  model, gas accretion
leads to an increase of the stellar density and  SMS formation via stellar collisions. The resulting -- and required -- stellar density  to form a SMS is $\sim10^9\,\msun\,\pc^{-3}$ and this extreme density is in tension with results from recent observations of strongly-lensed young GCs ($\sim10\,\myr$) in the Cosmic Gems arc (redshift 10.2) with {\it JWST}, which have stellar surface densities of $\sim10^5\,\msun\,\pc^{-2}$ \citep{2024Natur.632..513A}. The SMS model assumes that a single SMS grows from collisions with low-mass stars in an unsegregated cluster. When considering mass segregation, collisions among stars of all masses and the formation of multiple VMSs leading eventually to a single SMS, the required density may be an order of magnitude lower, but this is still two orders of magnitude higher than what is observed. 

In addition, a caveat in the SMS model is that the gas accretion was assumed to be 100 per cent efficient, leading to the strongest increase of density \citep*{1998MNRAS.298...93B}. If not all inflowing gas is accreted -- more realistic and assumed here -- then the increase of density with mass is much weaker \citep[see section 3.1 of][]{1998MNRAS.298...93B} and the extreme densities may not be reached.

Also, SMSs may leave behind intermediate-mass black holes (IMBHs) with $\gtrsim10^4\,\msun$\footnote{The fate of metal-enriched $\sim10^4\,
\msun$ stars is uncertain and they may completely explode as the result of a general relativistic instability \citep{2023MNRAS.523.1629N,2024PhRvD.110c1301N}.}. 
Radio observations have now put strict upper limits of $\lesssim10^3\,\msun$ on the masses of putative IMBHs in 50 GCs \citep{2018ApJ...862...16T}, potentially a problem for the  SMS model. We discuss the role of IMBHs in the aEMS model in Section~\ref{ssec:imbh}. 

Finally, \citet{2018MNRAS.478.2461G} assumed that the collisions are mass conserving, but collisions can actually lead to a reduction of the mass because of  stripping of the outer envelope of the SMS \citep{2025arXiv250612132R}.
Stellar collisions are likely to occur at later stages of GC formation, but rather than efficiently growing the \vems, they may in fact be a promising  way to liberate slow moving processed material \citep{2025arXiv250612132R}, similar to the idea of interacting binaries \citep{2009A&A...507L...1D}, but earlier, hence with lower $\Delta Y$. To understand the role of stellar collisions in GC formation, detailed high-resolution hydrodynamical simulations are needed, which are currently out of reach. 
%______________
\subsection{Comparison with SMSs ($\gtrsim10^4\,\msun$) yields}
\label{ssec:compare}
In this paper we conclude that aEMSs with masses ${\rm few}\times10^3\,\msun$ and central temperatures of $\sim65-72\,\MK$ are able to reproduce the abundance trends of GCs. This appears to be in contradiction with the findings of \citet{2014MNRAS.437L..21D} who conclude that SMSs of $\gtrsim10^4\,\msun$ have {\it `the required central temperatures of $\sim74-78\,\MK$}' to reproduce observed abundance trends. The explanation for this apparent contradiction is twofold. Firstly, \citet{2014MNRAS.437L..21D} compare the SMS abundances to five GCs from \citet{2013ApJ...769....8D}. Of these five GCs, only $\omega$\,Cen (NGC\,5139) -- the most massive GC in the Milky Way ($\MGCi\simeq8\times10^6\,\msun$, Table~\ref{tab:mwgc}) -- shows the $\sim-0.3$ dex Mg depletion found in the SMS models. Another GC with a small Mg depletion ($\sim-0.1\,$dex)  is M\,13 (NGC\,6205). This cluster is not in Carretta's UVES sample. Based on its estimated initial mass and its metallicity (Table~\ref{tab:mwgc}), our aEMS model suggests that it had EMSs of $\sim2.5\times10^3\,\msun$ with central temperatures of $\sim69\,\MK$, that is, similar to the hottest EMSs in the UVES sample (NGC\,7078, $\sim72\,\MK$, see Fig.~\ref{fig:anticorrs1} and \ref{fig:anticorrs2}). This predicts a maximum Mg depletion of $-0.3\,$dex. The other three GCs do not show Mg depletion, like most clusters in the Carretta sample shown in Fig.~\ref{fig:anticorrs2}. Thus, SMSs overpredict Mg depletion, because this is only seen in a small fraction of massive, metal-poor GCs. In the aEMS model, Mg depletion is found only in the most massive, metal-poor GCs, where EMSs of $\gtrsim3\times10^3\,\msun$ are expected (see Fig.~\ref{fig:scaling3}). 

A second explanation is that Denissenkov and Hartwick determine their yields when their SMS reaches $\Delta Y = 0.15$. Although this is similar to the $\Delta Y$ in our model stars ($\sim0.1-0.2$, Fig.~\ref{fig:scaling1}), our accreting aEMSs undergo nucleosynthesis for a longer time because of continuous rejuvenation by gas accretion. In the absence of rejuvenation, our stars would have $\DYnor\simeq 0.2-0.3$ (see Figs.~\ref{fig:abundances_mass} and \ref{fig:scaling1}). From Fig.~\ref{fig:tc_m} we see that a star with $\Tc\simeq62\,\MK$ and $\DYnor\simeq0.4$ has a similar amount of Mg depletion as a star with $\Tc\simeq64\,\MK$ and $\DYnor\simeq0.1$. Hence, slightly lower mass, but more evolved aEMS are able to create a Mg depletion. For example, for $\omega$\,Cen the aEMS model predicts a Mg depletion of $-0.5\,$dex, comparable or even slightly more than the more massive SMS models of Denissenkov and Hartwick.

Finally, in the SMS model of Denissenkov and Hartwick, the MESA model is simply stopped at $\Delta Y = 0.15$ to determine the yields. The $\Delta Y$ in the aEMS model is the result of the combination of nucleosynthesis and rejuvenation. The relative abundances of all the light-elements are therefore a model prediction, and can be compared to observations, such as done by \citet{2024A&A...690A.199V}.
We conclude that SMSs ($\gtrsim10^4\,\msun$) are too hot to explain the abundances of the majority  of GCs and that the aEMSs are the preferred polluters of GCs. However, there is still sufficient freedom in the model predictions that we can not exclude that SMSs existed in the most massive, metal-poor GCs. In our aEMS model of M\,15 with $\Sigmagas=10^3(3\times10^3)\,\msun\,\pc^{-2}$ the most massive EMS is $6(8)\times10^3\,\msun$. Stellar collisions (see Section~\ref{ssec:collisions}) may increase this to $\gtrsim10^4\,\msun$.

%______________
\subsection{Black hole masses}
\label{ssec:imbh}
A consequence of GCs containing stars in excess of $10^3\,\msun$ is the  possibility that they leave behind massive black holes (BHs) that can be looked for. To quantify this, we look at the number of stars above $10^3\,\msun$ at the end of the accretion phase. For the most metal-poor and most massive GC in the data sample, NGC\,7078 (M\,15), we find 18 stars in the mass range $10^{3-4}\,\msun$ at the time the GC is fully formed ($t=\tdyn$). Their helium abundances are $Y\simeq0.42$. If we let the wind mass loss proceed until $Y\simeq1$, we find that these stars have final masses in the range $\sim800-3000\,\msun$. Assuming that 90\% of the final mass ends up in a BH \citep{2017MNRAS.470.4739S}, this predicts $18$ BHs in the mass range $720-2700\,\msun$ in M\,15. In total there are $\sim130$ stars with final masses $\gtrsim220\,\msun$, that is, creating BHs above the pair-instability gap \citep{2017MNRAS.470.4739S}, which are intermediate-mass BHs (IMBHs) with masses $\gtrsim120\,\msun$ \citep{2020A&A...640A..56R}. For a more typical GC, such as NGC\,6218, we predict $\sim40$ IMBHs with masses $\sim120-700\,\msun$.

There is no evidence that IMBHs with masses $\gtrsim10^4\,\msun$ are common in GCs \citep[possible exception being $\omega$\,Cen,][]{2024Natur.631..285H}, because radio observations have placed upper limits of $\sim10^3\,\msun$ in 50 nearby GCs \citep{2018ApJ...862...16T}. 
For the specific cases of M\,15 and 47\,Tucanae, pulsar accelerations and stellar kinematics place an upper limit of $\sim500\,\msun$ for the mass of a putative IMBH \citep[][respectively]{2014A&A...565A..43K, 2024A&A...682A..22D}. This  seems in tension with our prediction of numerous IMBHs $\gtrsim10^3\,\msun$, but a population of IMBHs with comparable masses is expected to  dynamically eject each other in a short time \citep[for example,][]{1993Natur.364..421K,2020MNRAS.492.2936A}. Dynamics leads to preferential ejection of the most massive IMBHs and this continues until a single IMBH (or a binary IMBH) remains. The corresponding timescale can be found from equation~(\ref{eq:Mdot_ej}). For a GC with $10^4\,\msun$ in IMBHs and an ejection rate of $\dot{M}_{\rm ej}\simeq -500\,\msun\,\myr^{-1}$, all IMBHs are ejected in $\sim 20\,\myr$. If we consider that both the cluster density and the mass function parameters $\psi$ in the expression for the relaxation time (see text under equation~\ref{eq:Mdot_ej}) decrease as the result of relaxation, then this timescale may be a few 100\,Myr. A final binary IMBH, or a single IMBH with a stellar-mass BH, may remain, because it is hard to dynamically eject a binary that is significantly more massive than the other cluster members. Because the most massive IMBHs are ejected first,  the last IMBH(s) to remain are `lite IMBHs' ($\sim100\,\msun$), which are below the upper limits discussed above.  These lite-IMBHs may also capture stellar companions that could be found with radial velocity surveys or with {\it Gaia} \citep{2024A&A...686L...2G}. 

While a binary IMBHs ejects other IMBHs, the binary can  become compact enough to spiral in by gravitational wave (GW) emission, before they are ejected in a dynamical recoil following a binary-single interaction  \citep{2006ApJ...637..937O}. GW inspiral is more likely to happen than ejection in GCs with high escape velocities \citep[for example,][]{2020MNRAS.492.2936A}. For mass ratios of $\sim0.1-1$ the merger remnant receives a  GW recoil kick of a few 100 km/s \citep{2012PhRvD..85h4015L} and is therefore ejected from clusters with escape velocities below $\sim300\,\msun$ (\citealt*{2019MNRAS.486.5008A};\,\citealt{2023MNRAS.526.4908C}). GCs with higher escape velocity are able to retain the first binary BH merger, and subsequent inspirals have smaller mass ratio and, therefore, receive (on average) a lower GW kick, and this could gradually grow the mass of the most massive IMBH. 
We therefore expect  GW signals of binary IMBHs and IMBH+stellar-mass BH at redshifts $z\gtrsim3$ (first $\sim1\,\gyr$ after GC formation, see above). These kind of binary BH mergers may also result from repeated mergers \citep{2016ApJ...831..187A}, but a fundamental difference is that these will have higher spins. Our model therefore predicts GW signals of IMBHs with the birth spin of the individual IMBHs, which could be low. These mergers will occur in the early phases of GC evolution, hence at redshifts of $z\gtrsim3$ and might  be found by Advanced LIGO in the next observing run (O5, starting late 2027).

%______________
\subsection{Observational evidence of star and cluster formation by inertial inflows} 
\label{sec:hub_filaments}
A key aspect of the aEMS model is that both the entire star cluster and its most massive stars form from converging flows within a larger turbulent region, over its dynamical timescale (Fig.\,\ref{fig:gcform}). Converging flow regions are an inevitable outcome of supersonic turbulence, even in the absence of gravitational collapse, as approximately one-third of the turbulent kinetic energy is stored in compressive motions across all scales. This scenario of accretion from large-scale filaments is strongly supported by observations of Galactic environments that form massive stars ($10-100\,\msun$). These environments were first described as hub-filament systems (HFSs) by \cite{Myers09_hubs}, where the `hub' represents the central region hosting massive stars, formed at the intersection of dense filaments that can extend several parsecs in length. The HFS morphology has since been found to be widespread in regions of massive star formation, as demonstrated by recent ALMA surveys of infrared dark clouds \citep[for example,][]{Zhou+22_ATOMS_hubs,Yang+23,Liu+23,Sharma+24,Rawat+24}. Kinematic studies of the filaments show clear evidence of mass accretion into the central hubs, with typical mass flow rates per filament of $\sim10^{-4}\,\msunyr$ \citep{Henshaw+14,Traficante+17,Yang+23,Wells+24,Alvarez-Gutierrez+24,2025A&A...696A.202S,Rawat+24}, supporting the growth of both individual massive stars and the entire cluster.

As a typical example, the G148.24+00.41 region studied by \cite{Rawat+24} can be viewed as a scaled-down version of our canonical GC model. It is a giant molecular cloud (GMC) with a mass $\Mgas\sim 6\times10^4\,\msun$ feeding a central star-forming clump of $2\times10^3\,\msun$ and 1.4\,pc size, through six main filaments of lengths from 14 to 38\,pc. The filaments provide the total inflow rate into the main hub $\Mdprist\sim700\,\msunmyr$. Assuming half of a reasonable GMC dynamical time, $\tdyn\sim6$\,Myr, has already passed, and accretion efficiency of 50 per cent, the mass of the central cluster can double in the next 3\,Myr to a value of $\MGCi\sim4\times10^3\,\msun$ (using the corresponding variables from the GC model). This example shows that the cluster formation is driven by converging flows from a larger cloud, yet only a fraction (7 per cent in this case, 10 per cent in our canonical GC model) of the cloud's mass is channeled into the cluster. 

High-resolution ALMA observations of regions forming massive stars have also ruled out the existence of prestellar cores with sufficient mass to directly collapse into massive stars. These observations show that prestellar cores have mass distributions that barely extend to $\sim10\,\msun$, and characteristic masses of $\sim1\,\msun$ \citep[for example,][]{2019ApJ...886..102S,Kong19cmf,Morii+23,Jiao+24,Cheng+24,Armante+24}, consistent with the numerical results in \cite{Padoan+20massive}. This supports the hypothesis that massive stars initially form with much lower masses and must undergo additional accretion from larger scales to reach their final mass. This conclusion is further reinforced by the observation that protostellar cores exhibit higher average masses and shallower mass distributions compared to prestellar cores within the same star-forming region \citep{Kong+21,Nony+23,Cheng+24,Armante+24}.

%______________
\subsection{Implications of a long formation time of the most massive stars}

The inertial-inflow model predicts that the most massive stars in any given environment are formed over a timescale of the order of that environment's dynamical time, $\tdyn$. In our canonical GC model $\tdyn\sim2\,\myr$, while in a typical nearby Galactic GMC of $\Mgas=10^4\,\msun$, $\tdyn\simeq4\,\myr$. As shown in Fig.\,\ref{fig:t0_mmax}, nearby GMCs are not expected to form stars in excess of $\sim30\,\msun$, because their $\tdyn$ values are in excess of the lifetime of more massive stars. This is consistent with the absence of massive stars in nearby star-forming regions, and the larger column density of more distant infrared dark clouds that harbour massive stars. Assuming a slightly larger column density for Galactic clouds forming massive stars, $\Sigmagas\sim100\,\msun\,\pc^{-2}$ and a cloud mass $\Mgas\sim10^5\,\msun$, $\tdyn\sim4.5\,\myr$ (equation~\ref{eq:tdyn}) and $\mfmax\sim250\,\msun$ (equation~\ref{eq:mfmax}). Because the star-formation time scales with the square root of the final stellar mass, within the constraint of the stellar lifetime, such a GMC can form stars up to $\sim100\,\msun$ in $\sim3\,\myr$, and a typical O-type star of $\sim50\,\msun$ in $\sim2\,\myr$, corresponding to an average accretion rate of $\sim3\times10^{-5}\,\msunyr$. This formation time is much longer than the standard literature value of $\sim0.1\,\myr$ (accretion rate of $\sim0.6\times10^{-3}\,\msunyr$), and it implies that O-type stars are kept heavily embedded by their accretion flows for a comparable timescale of $\sim2\,\myr$.

Since the work of \cite{Garmany+82}, it has been known that O-type stars are almost never found near their zero-age main sequence (ZAMS) with ages $\lesssim2\,\myr$ \citep[for example,][]{Garmany+82,Castro+14,Holgado+20,Schootemeijer+21}, the long-standing ZAMS problem. \cite{Holgado+20} have shown that this may be explained if massive stars are formed with accretion rates of $\sim10^{-5}\,\msunyr$ (the stars are colder, and hence detached from the ZAMS). However, \citet{Schootemeijer+21} argue that there is no evidence that massive stars are accreting for roughly half of their main-sequence time, so they conclude that the ZAMS problem is more likely to imply that O-type stars remain embedded in their birth clouds for $\sim2\,\myr$, as originally suggested by \cite{Garmany+82}. This is direct evidence of the long formation time predicted by the inertial-inflow model, because the expansion time of the HII region of an O-type star formed in just $\sim0.1\,\myr$ would be $\ll2\,\myr$, in the absence of a dense accretion flow. 

During its formation, a massive star is expected to be associated with a hot molecular core (HMC) first and a hyper/ultra compact HII (HCHII/UCHII) region later. Statistical estimates of the lifetime of UCHII regions yield a value of $\sim0.3\,\myr$ \citep{Wood+Churchwell89,Mottram+11}, and a comparable total duration of the embedded phase \citep[for example,][]{Mottram+11,Battersby+17,Nony+24}. This value is too long in the absence of a physical mechanism to confine the UCHII region, like continued accretion, but also too short to solve the ZAMS problem. In a followup paper, we plan to demonstrate that both the ZAMS problem and the lifetime of UCHII regions can be explained self-consistently in our inertial-inflow scenario, because the long formation time of O-type stars imply that the lifetime of UCHII regions was estimated incorrectly (too short) by almost an order of magnitude.
Interestingly, in their most detailed photoionization study of a UCHII region, \cite{Morisset+02} find that the ionizing star of G29.96-0.02 has an age of a few Myr. Another photoionization study of the W31 cluster \citep{Furness+10} concludes that the UCHII regions in W31 may have lifetimes in excess of 0.5\,Myr, if they are to be coeval with the four naked O-type stars in the cluster. 

%______________
\subsection{Implications for high redshift observations}
Nitrogen-rich, star forming  galaxies have been known for decades \citep[for example, Mrk 996,][]{2009MNRAS.398....2J}, and have received renewed attention thanks  to {\it JWST}. Well-known examples being the Sunburst  galaxy at a redshift $z\simeq2.4$ \citep{2023A&A...673A..50M} and GN-z11 discovered  at $z\simeq10.6$ \citep{2023A&A...677A..88B}.
The high nitrogen abundances
have been attributed to VMSs \citep{
2023A&A...679L...9V}, SMSs 
\citep{2023A&A...673L...7C} and AGB stars \citep{2023A&A...680L..19D}. We may also be looking at active galactic nuclei (AGNs), in which stars form in the accretion disc and simultaneously accrete and internally evolve  -- similar to what we propose here --  and produce large amounts of N \citep*{2021ApJ...910...94C}.
Without additional information, such as temperature-sensitive abundance measurements, such as Na, Mg and Al, it is not possible to confidently confirm the nature of the source. 
\citet{2024A&A...686A.185U} find that the majority of UV-bright galaxies at the peak of star formation ($z\sim2-3$) have prominent He\,II\,$\lambda$1640 emission lines (${\rm EW}\gtrsim3\,\r{A}$), which they attribute to VMSs. The high-incidence in galaxies with high star formation rate ($\sim10^{2-3}\,\msun\,\yr^{-1}$) is a logical outcome of our model, in which V/EMSs are forming in galaxies with high $\Sigmagas$. No detailed model atmospheres of EMSs exist and the structure of these stars depends sensitively on the uncertain convection in the outer layers. Models with efficient convection find that EMSs have surface temperatures of $\Teff\gtrsim 10^5 \,$K \citep{2025arXiv250612132R}. This may allow for photon energies in excess of 54.4 eV needed to get He\,II\,$\lambda$1640 and may even explain the observed P Cygni profile in CIV \citep{2024Natur.627...59M}, without the need for an AGN.

%%%%%%%%%%%%%%%%%%%%
\section{Summary and conclusions}
\label{sec:conclusions}
We present a parameterised model for the formation of GCs that explains the origin of their abundance anomalies, or multiple populations (MPs). It is based on  the inertial inflow model for massive star ($\sim10-100\,\msun$) formation, scaled up to the GC regime, where very massive stars (VMSs, $10^{2-3}\,\msun$) and extremely massive stars (EMSs, $10^{3-4}\,\msun$) are expected to form. 
The three dominant parameters in the model are the initial mass of the gas cloud, $\Mgas$, which sets the maximum stellar mass; the (surface) density of the gas, $\Sigmagas$, which combined with $\Mgas$ sets the maximum accretion rate
 and the metallicity of the gas, \feh, which sets the wind mass loss rates.
Wind mass loss becomes important already during the formation of \vemss\ and the final (maximum) masses of EMSs are set by a balance between accretion and wind mass loss and therefore dependent on both $\feh$ and $\Sigmagas$. 
GCs form in a single burst, on a dynamical timescale of the gas, and accretion of the most massive EMSs lasts for the full duration of GC formation. 
VMSs also contribute processed material, after accretion has stopped, but most of the wind mass loss ($\sim50-80$ per cent) 
is released by the most massive, accreting EMSs (aEMSs), which are expected to be fully convective, hence nucleosynthesis products are released into the intra-cluster medium where they are di

In this aEMS model, MPs form in a single star burst, but the duration of low-mass ($\lesssim1\,\msun$) star formation lasts $\sim1-2\,\myr$, hence towards the end of GC formation these low-mass stars  contain the burning products of the aEMSs. The formation mechanism of P1 and P2 stars is similar, which helps to understand the observed similarity of the stellar mass function of MPs \citep{2022ApJ...927..207D}. This differs from earlier models that relied on accretion of polluted material \citep{2013MNRAS.436.2398B, 2018MNRAS.478.2461G} onto low-mass stars, likely changing the mass function of P2 with respect to P1. In the model presented here, there is no first and second generation of stars, but there is a clear distinction in abundance patterns of low-mass stars forming early on (P1), and of those that form towards the end (P2), thereby explaining the observed discreteness of MP.

Abundances of Milky Way GCs can be well reproduced if GCs form from gas with a surface density of $\Sigmagas\simeq1.5\times10^3\,\msun\,\pc^{-2}$. Then typical GCs had EMSs in the mass range $\sim(1-5)\times10^3\,\msun$, which had  central temperatures in the range $\sim65-72\,\MK$. These temperatures are sufficiently high to lead to significant N enhancement and create the ubiquitous O-Na anticorrelation. The highest temperatures are found in the most massive and metal-poor GCs, where the model predicts the highest Al abundances and largest Mg depletion, as is observed. 

The criteria for forming MPs in star clusters is that (1) the cluster should be massive enough ($\gtrsim10^{5-6}\,\msun$), with the mass threshold being higher in high metallicity environment; and (2) the surface density of the gas should be high enough for MPs to form in time ($\gtrsim10^3\,\msun\,\pc^{-2}$). Although MPs have so far not been found in YMCs (Section~\ref{ssec:ymcs}), the conditions to form MPs can be met in the local Universe. 
It is therefore worthwhile to look for elements that are already enhanced at lower temperatures (stars of $\sim10^3\,\msun$), such as N and Na, in the most massive YMCs ($\gtrsim10^6\,\msun$) in low-metallicity galaxies, with high gas densities. Spectroscopy of individual low-mass stars may be feasible with {\it JWST} in Local Group galaxies.

In the aEMS model presented here, EMSs are a necessary outcome of GC formation at low metallicities and their ejecta are expected to pollute GC formation sites with hot hydrogen burning yields. This, therefore, provides a natural explanation \citep{2023A&A...679L...9V} for the N-rich galaxies observed in the early Universe by {\it JWST} \citep{2023A&A...677A..88B,2024ApJ...966...92S, 2024MNRAS.529.3301T}. From Galactic archeology studies in the Milky Way, it was inferred that the first star formation epoch of the Milky Way was dominated by GC formation \citep{2023MNRAS.525.4456B}. Galaxies at high redshift are biased towards galaxies that are actively forming GCs, because of their (temporarily) elevated light-to-mass ratio \citep{2018MNRAS.477..480Z}. For a stellar population with an upper cut-off in the MF of $\gtrsim10^3\,\msun$, the light-to-mass ratio is approximately an order of magnitude higher \citep{2025A&A...693A.271S} than that of a stellar population with an upper cut off at $\sim100\,\msun$. This may help interpreting  the overabundance of UV bright galaxies at high redshift \cite[for example,][]{2023ApJS..265....5H}.

Finally, EMSs of $\gtrsim10^3\,\msun$ with low metallicities could form black holes above the pair-instability gap \citep[$\sim45-150\,\msun$,][]{2017MNRAS.470.4739S}. Sequential mergers of stellar-mass black holes can also lead to such masses, with the difference being that they will have high spins. Gravitational wave signals of merging black holes of a few $100\,\msun$, and their spins, can soon be inferred with ongoing gravitational wave experiments and place constraints on this GC formation model. 

%%%%%%%%%%%%%%%%%%%%
\section*{Acknowledgements}
The authors thank the referee for constructive feedback and suggestions, that led to several important improvements in the presentation of the results. Nate Bastian and Eugenio Carretta are thanked for discussions and valuable comments on the pre-print.
The authors thank S\o ren Larsen and Carmela Lardo for discussions on abundances and Nikos Prantzos for discussions and sharing his nucleosynthesis results. 
MG acknowledges  financial support from grants PID2021-125485NB-C22 funded by MCIN/AEI/10.13039/501100011033 and SGR-2021-01069 (AGAUR).
MG and PP acknowledge financial support from grants CEX2019-000918-M, CEX2024-001451-M funded by MICIU/AEI/10.13039/501100011033.
 PP acknowledges support by the US National Science Foundation under Grant AST 2408023.
CC acknowledges support from the Swiss National Science Foundation (SNF; Project 200021-212160, PI CC). JSV is supported by STFC funding under grant number ST/V000233/1 (PI Vink),

%%%%%%%%%%%%%%%%%%%%
\appendix
%__________________________________
\section{Exact relations}
\label{app:exact}
Here we present the exact relations that are used in the model for the evolution of the IMF. When considering an initial seed mass of $\ms$, the expression for $\tf$ (equation~\ref{eq:tf_mf_mean}) becomes
\begin{equation}
\tf = \tfmax\left(\frac{\mf-\ms}{\mfmax-\ms}\right)^{1/2},
\label{eq:tf_exact}
\end{equation}
and the expression for $\mdotacc$ (equation~\ref{eq:mdotacc}) becomes
\begin{equation}
\mdotacc = \mdotaccmax\left(\frac{\mf-\ms}{\mfmax-\ms}\right)^{1/2}
\label{eq:mdotacc_exact}
\end{equation}
The exact result for the evolution of the stellar mass as a function of time since formation ($\tp=t-\tform$) considering both winds and accretion, $m(\tp)$ (equation~\ref{eq:mt}), then becomes
\begin{equation}
m(\tp) = \minf\tanh\left[\frac{\mdotacc \tp}{\minf}  + \arctanh\left(\frac{\ms}{\minf}\right) \right],
\label{eq:mt_exact}
\end{equation}
which is valid for $\tp\le\tf$.
After accretion ends ($\tp>\tf$), the mass evolves as the result of winds only as
\begin{equation}
m(\tp) = \frac{m(\tf)}{(\tp - \tf)m(\tf)\mdwhz/(100\,\msun)^2  + 1},
\label{eq:mtpost}
\end{equation}
where $m(\tf)$ follows from equation~(\ref{eq:mt_exact}) with $\tp=\tf$.

% Don't change these lines
\bsp	% typesetting comment
\label{lastpage}
\end{document}